\pdfoutput=1
\RequirePackage{ifpdf}
\ifpdf 
\documentclass[pdftex]{sigma}
\else
\documentclass{sigma}
\fi

\numberwithin{equation}{section}
\newtheorem{Lemma}{Lemma}[section]
\newtheorem{Theorem}[Lemma]{Theorem}
\newtheorem{Corollary}[Lemma]{Corollary}

\begin{document}

\newcommand{\arXivNumber}{1404.4392}

\allowdisplaybreaks

\renewcommand{\PaperNumber}{004}

\FirstPageHeading

\ShortArticleName{Hilbert--Schmidt Operators vs.~Integrable Systems~IV. The Relativistic Heun Case}

\ArticleName{Hilbert--Schmidt Operators vs.~Integrable Systems\\
of Elliptic Calogero--Moser Type~IV.\\
The Relativistic Heun (van Diejen) Case}

\Author{Simon N.M.~RUIJSENAARS}

\AuthorNameForHeading{S.N.M.~Ruijsenaars}

\Address{School of Mathematics, University of Leeds, Leeds LS2 9JT, UK}
\Email{\href{mailto:siru@maths.leeds.ac.uk}{siru@maths.leeds.ac.uk}}
\URLaddress{\url{http://www1.maths.leeds.ac.uk/~siru/}}

\ArticleDates{Received April 19, 2014, in f\/inal form January 10, 2015; Published online January 13, 2015}

\Abstract{The `relativistic' Heun equation is an 8-coupling dif\/ference equation that ge\-ne\-ralizes the 4-coupling Heun
dif\/ferential equation. It can be viewed as the time-independent Schr\"odinger equation for an analytic dif\/ference
operator introduced by van Diejen. We study Hilbert space features of this operator and its `modular partner', based on
an in-depth analysis of the eigenvectors of a~Hilbert--Schmidt integral operator whose integral kernel has a~previously
known relation to the two dif\/ference operators. With suitable restrictions on the parameters, we show that the commuting
dif\/ference operators can be promoted to a~modular pair of self-adjoint commuting operators, which share their
eigenvectors with the integral operator. Various remarkable spectral symmetries and commutativity properties follow from
this correspondence. In particular, with couplings varying over a~suitable ball in~${\mathbb R}^8$, the discrete spectra
of the operator pair are invariant under the $E_8$ Weyl group. The asymptotic behavior of an 8-parameter family of
orthonormal polynomials is shown to be shared by the joint eigenvectors.}


\Classification{33E05; 33E30; 39A45; 45C05; 47B39; 81Q05; 81Q10; 81Q80}

\tableofcontents

\section{Introduction and summary}\label{section1}

This paper is a~continuation of a~series of papers concerned with the commuting quantum Hamiltonians corresponding to
elliptic Calogero-Moser type systems. In Part~I~\cite{RI09}, we presented a~detailed study of the algebraic and
function-theoretic properties of kernel functions for these operators, which we introduced in~\cite{Rui04}. In
Part~II~\cite{RII09}, we obtained preliminary results pertaining to associated Hilbert space versions for the
case~$A_{N-1}$ and an explicit diagonalization of the pertinent Hilbert--Schmidt operator in the free case. In
Part~III~\cite{RIII09} we treated the nonrelativis\-tic~$BC_1$ (Heun) case, with the exploitation of the Hilbert--Schmidt
operator leading in particular to a~remarkable spectral invariance of the Heun operator under $S_4$-permutations of
a~reparametrized coupling vector.

The present Part~IV deals in particular with the dif\/ference operator generalization of the Heun dif\/ferential operator
introduced by van~Diejen~\cite{vDie94}. For our purposes, however, it is crucial to work with a~renormalized version of
this operator for which various symmetry properties become manifest~\cite{Rui02}. This operator leads to a~natural
`modular' partner commuting with it, and both operators are manifestly invariant under permutations of a~coupling
vector~$\gamma$ varying over~${\mathbb C}^8$. The operators are associated with two elliptic curves that share one
period. (The relation of the van Diejen operators to the dif\/ference operators arising from representations of the
Sklyanin algebra~\cite{Skl82,Skl83} was recently clarif\/ied in~\cite{RR13}.)

Since we are dealing with Hilbert space aspects in this paper, we assume throughout that the common period of the
elliptic curves is a~positive number given by~$\pi/r$, whereas we choose the two remaining (generally distinct) periods
to be imaginary and given by~$ia_{+}$ and~$ia_{-}$. (The latter play the role of shift parameters for the modular pair
of dif\/ference operators.) More specif\/ically, we let
\begin{gather}
\label{parp}
r,a_{+},a_{-}>0,
\end{gather}
and we often use the period average parameter
\begin{gather}
a\equiv (a_++a_-)/2,
\end{gather}
and the asymmetric shift parameters
\begin{gather}
\label{as}
a_s\equiv \min(a_+,a_-), \qquad a_l\equiv \max(a_+,a_-).
\end{gather}

We proceed with some remarks of a~general nature concerning commuting dif\/ference operators of the type occurring in this
paper, so as to convey some initial understanding for the problems we solve here. Consider f\/irst a~second order
dif\/ference operator of the form
\begin{gather}
\label{cAp}
{\cal A}_{+}=\exp(-ia_{-}d/dx)+V_{a,+}(x)\exp (ia_{-}d/dx) +V_{b,+}(x),
\end{gather}
where the coef\/f\/icients are elliptic functions with periods $\pi/r$ and $ia_{+}$. Thus, ${\cal A}_{+}$ leaves the space
of meromorphic functions invariant. Clearly, for~${\cal A}_{+}$ to be formally self-adjoint on the Hilbert space
\begin{gather}
\label{cH}
{\cal H} \equiv L^2([0,\pi/2r],dx),
\end{gather}
it suf\/f\/ices to require real-valuedness of $V_{b,+}(x)$ on the real line, and the conjugacy relation
\begin{gather}
\label{fsa}
V^{*}_{a,+}(x+ia_{-})=V_{a,+}(x),\qquad V^{*}_{a,+}(x):=\overline{V_{a,+}(\overline{x})},\qquad x\in{\mathbb C}.
\end{gather}
Now meromorphic eigenfunctions with arbitrary eigenvalues of such a~dif\/ference operator do exist, as follows for
instance from~\cite{Hor66}. However, the existence proofs involve non-constructive arguments (such as the Runge
approximation theorem), precluding any explicit knowledge about such eigenfunctions. Note also that one gets
inf\/inite-dimensional eigenspaces for any eigenvalue, since one is free to multiply by meromorphic functions with period~$ia_{-}$ (`quasi-constants').

In particular, no results are known that would entail the existence of eigenfunctions with real eigenvalues whose
restrictions to~$[0,\pi/2r]$ yield an orthogonal base for~${\cal H}$, hence leading to a~natural association of
a~self-adjoint operator~$\hat{{\cal A}_{+}}$ to the analytic dif\/ference operator~${\cal A}_{+}$. Therefore, even for
a~single dif\/ference operator of this type, it is a~wide open problem whether \emph{any} such reinterpretation exists.
(The uniqueness question is equally open; note that the symbols of the operators are exponentially increasing, so that
they do not belong to any of the well-studied classes of pseudo-dif\/ferential operators.)

Next, consider a~dif\/ference operator
\begin{gather}
\label{cAm}
{\cal A}_{-}=\exp(-ia_{+}d/dx)+V_{a,-}(x)\exp (ia_{+}d/dx) +V_{b,-}(x),
\end{gather}
for which the roles of the positive parameters $a_{+}$ and $a_{-}$ are swapped. Thus, $V_{a,-}$ and $V_{b,-}$ are now
elliptic functions with periods $\pi/r$ and $ia_{-}$, which are only restricted such that ${\cal A}_{-}$ is also
formally self-adjoint on ${\cal H}$ (i.e., $V_{b,-}$ is real-valued for real~$x$, and $V_{a,-}$ satisf\/ies the analog
of~\eqref{fsa}). Then, the coef\/f\/icients of ${\cal A}_{-}$ are quasi-constants for ${\cal A}_{+}$, and vice versa.
Accordingly, the dif\/ference operators ${\cal A}_{+}$ and ${\cal A}_{-}$ {\it commute}.

However, even when the numbers $a_{+}$ and $a_{-}$ have a~rational quotient, it seems utterly unlikely that in this
generality common meromorphic eigenfunctions of this commuting pair exist at all, since one is dealing with four
arbitrary elliptic coef\/f\/icients. Indeed, no such existence results are known. A~fortiori, this is the case for a~ratio
$a_{+}/a_{-}$ that is irrational. Under the latter restriction, it is actually easy to f\/ind commuting pairs of f\/irst
order dif\/ference operators admitting no joint eigenfunctions at all.

So when we are faced with a~{\it special} pair of commuting dif\/ference operators ${\cal A}_{\pm}$ of the above elliptic
type (as the van Diejen operators are), how can we ever show that any meromorphic joint eigenfunctions exist? And if
they are in addition formally self-adjoint on ${\cal H}$ (as the van Diejen operators are for restricted couplings), how
can we ever show that they have meromorphic joint eigenfunctions with real eigenvalues, whose restrictions to
$[0,\pi/2r]$ yield an orthogonal base for ${\cal H}$, hence leading to a~reinterpretation of the operators as bona f\/ide
commuting self-adjoint operators on ${\cal H}$? Assuming we can solve these problems by hook or by crook, how can we
ever get more explicit information about the eigenvalues and eigenfunctions, such as non-degeneracy and asymptotic
behavior?

These questions have been open for decades. We answer the f\/irst two in this paper, by enlisting the help of the
enigmatic kernel functions, and shed considerable light on the third with the aid of an equally unexpected ingredient,
namely an auxiliary 8-parameter family of orthonormal polynomials.

For a~mathematical physicist who views Hilbert space as the arena to reinterpret a~quantum integrable system (as we do),
these questions are the most pressing ones. Here, we use the expression `quantum integrable system' in the (nowadays
customary) loose sense of a~collection of commuting dif\/ferential or dif\/ference operators, which are often studied
without regard for Hilbert space features. We hasten to add that we have great appreciation for the study of special
function solutions to single or joint equations without addressing Hilbert space issues. For the case at hand, though,
we have nothing to report on {\it general} solutions to the two time-independent Schr\"odinger equations associated with
the commuting van Diejen operators. Indeed, our f\/indings pertain solely to the existence and properties of meromorphic
joint eigenfunctions yielding orthogonal bases in ${\cal H}$. As a~consequence, the associated eigenvalue pairs yield
a~countable discrete set in ${\mathbb R}^2$.

The kernel function from Part~I is def\/ined solely in terms of the elliptic gamma function~$G(r,a_+,a_-;x)$~\cite{Rui97},
cf.~\eqref{cS} below, and as such modular invariant (i.e., invariant under interchange of $a_+$ and~$a_-$). Closely
related kernel functions, which use instead the elliptic gamma function multiplied by quadratic exponentials, were
introduced from a~dif\/ferent perspective by~Komori, Noumi and Shiraishi~\cite{KNS09}. Since their focus is on one of the
dif\/ference operators, the absence of modular invariance for the latter kernel functions is of no consequence.

As will become clear below, the HS (Hilbert--Schmidt) operators associated with our modular invariant kernel function
yield the key tool to promote the modular pairs of commuting dif\/ference operators (given by~\eqref{cApmform}
and~\eqref{Ve}--\eqref{Varel}) to commuting self-adjoint operators that act on the Hilbert space~${\cal H}$~\eqref{cH}
and have solely pure point spectrum, provided the shift ratio $a_+/a_-\in(0,\infty)$ and coupling
vectors~$\gamma\in{\mathbb C}^8$ are suitably restricted. Moreover, for the case of a~real coupling vector, the HS
operator setting enables us to reveal a~striking invariance of the two discrete spectra under transformations from the
Weyl group associated with the $E_8$ Lie algebra. This suggests that the linear van Diejen dif\/ference equation may have
a~relation to the nonlinear elliptic $E_8$-symmetric Sakai dif\/ference equation~\cite{Sa01}, as a~generalization of the
relation between the linear Heun and nonlinear PVI Painlev\'e ODEs~\cite{Taka01,Take04, ZZ12}.

Unfortunately, we were unable to arrive at complete proofs of our results without confronting a~great many
technicalities. Moreover, a~plethora of non-obvious identities arises, whose verif\/ication sometimes involves substantial
calculations. Another (inevitable) dif\/f\/iculty occurs in digesting the def\/inition of the analytic dif\/ference operators.
Indeed, this is elaborate to an extent that in this introduction we have opted for describing the A$\Delta$Os (analytic
dif\/ference operators) in general terms, leaving their precise def\/inition to the main text (cf.~Section~\ref{section3}). For similar
expository reasons, we have chosen to describe our results in two stages, the f\/irst one yielding more context and
a~sketchy overview, whereas the second one (beginning with the paragraph containing~\eqref{defg}) is aimed at detailing
the f\/low chart and key results, introducing various important quantities and main ideas along the line.

At the outset, we would like to stress that to date no general methods exist to handle the Hilbert space theory of
analytic dif\/ference operators. From quite special cases with hyperbolic coef\/f\/icients, where explicit eigenfunctions are
known, it transpires that even when formal self-adjointness and unitary eigenfunction asymptotics are present, there may
be a~breakdown of self-adjointness and unitarity of the eigenfunction transform~\cite{Rui00}. The same phenomenon
happens for the `$g=2$' relativistic Lam\'e case analyzed in~\cite{Rui99g=2}.

In the relativistic elliptic $BC_1$ case at hand, there are only eigenfunction formulae available for an ${\mathbb
R}^8$-lattice of coupling constants. Indeed, for this discrete set of couplings eigenfunctions of Floquet--Bloch type for
one of the A$\Delta$Os were found by Chalykh~\cite{Cha07}. They involve a~transcendental system of `Bethe Ansatz'
equations whose solutions are quite inaccessible.

For the relativistic elliptic $A_1$ (Lam\'e) case, we have previously analyzed self-adjoint Hilbert space versions of
the modular pair of dif\/ference operators for couplings that are dense in a~certain bounded interval~\cite{Rui03}. For
these couplings there also exist eigenfunctions that can be tied in with eigenfunctions of Floquet--Bloch type,
generalizing Hermite's Lam\'e eigenfunctions~\cite{Rui99GL}. The question whether these Hilbert space results can be
extended to the whole pertinent interval has remained wide open until now. Indeed, one spin-of\/f of the present paper is
that this is indeed possible. This is because the relativistic Lam\'e operators arise by suitable 1-coupling
specializations of the 8-coupling relativistic Heun operators. To be sure, a~detailed treatment of these special cases
is beyond our present scope; we intend to return to the relativistic Lam\'e case in future work.

The perspective on the Hilbert space issues developed in the present paper is quite novel. Indeed, without the help of
the explicit integral operators of Hilbert--Schmidt type (a type of operator whose Hilbert space structure was already
clarif\/ied a~century ago), we would have no clue how to obtain bona f\/ide self-adjoint commuting operators for general
couplings. To elaborate on this, we emphasize that until now no arbitrary-coupling eigenfunctions of the elliptic $BC_1$
A$\Delta$Os were known at all, by contrast to the hyperbolic~$BC_1$ case~\cite{Rui99AW,BRS07}.

On the other hand, as we shall show in Section~\ref{section3}, it is possible to associate to both A$\Delta$Os at issue semi-bounded
symmetric operators \emph{without} invoking the relevant integral operators. For a~further study of these Hilbert space
operators, however, the latter are indispensable. As shall transpire, the operator with shift parameter~$a_s$ is
essentially self-adjoint, and its closure has an orthonormal base of eigenvectors. On the other hand, for generic
couplings none of the self-adjoint extensions of the operator with shift~$a_l$ shares these eigenvectors.

We are only able to elucidate this state of af\/fairs (in Section~\ref{section5}) because we can invoke detailed features of the
eigenvectors of the pertinent integral operators, assembled in Section~\ref{section2}. This additional information gives rise to
a~dif\/ferent domain def\/inition for the operator with shift~$a_l$, which does lead to the conclusion that the eigenvectors
are shared by the latter. It should be stressed that this second domain def\/inition would come out of the blue without
knowing the salient eigenvector properties beforehand. In point of fact, it would not even be clear that the subspace at
issue is \emph{dense} in~${\cal H}$.

In this connection we add that it is not widely appreciated just how dif\/ferently Hilbert space versions of A$\Delta$Os
behave compared to dif\/ferential operators. A~further quite simple example may help to convince readers who are familiar
with self-adjoint extension theory (which can be found, e.g., in~\cite{RS72}) that we are not out to kill phantoms while
exercising great care with Hilbert space domains for A$\Delta$Os.

Consider the Dirichlet and Neumann ONBs for ${\cal H}$, given by the well-known sine and cosine functions. Of course,
these can be used to turn the free A$\Delta$O $\exp(iad/dx)+\exp(-iad/dx)$ with $a>0$ into two distinct self-adjoint
operators with the obvious eigenvalues, just as the dif\/ferential operator $-d^2/dx^2$. The two resulting domains for the
Dirichlet and Neumann Laplacians thus obtained have a~dense intersection, containing in particular the smooth functions
with compact support in~$(0,\pi/2r)$. By contrast, the intersection of the two domains for the self-adjoint ope\-ra\-tors
associated with the free A$\Delta$O is given by the zero vector. This is because both domains consist of functions that
admit an analytic continuation to the strip $|\operatorname{Im} x|<a$, and functions in the `sine-domain' are odd, whereas
functions in the `cosine-domain' are even.

Having provided a~skeleton context, we proceed by putting some f\/lesh on the bones. The pair of commuting A$\Delta$Os was
already def\/ined in Subsection~3.1 of~\cite{RI09}. Here, however, we use slightly dif\/ferent notation and a~dif\/ferent
convention for the additive constant. The pair is of the form
\begin{gather}
\label{Apm}
A_{\delta}(\gamma;x)=
V_{\delta}(\gamma;x)\exp(-ia_{-\delta}d/dx)+V_{\delta}(\gamma;-x)\exp(ia_{-\delta}d/dx)+V_{b,\delta}(\gamma;x),\\
\delta=+,-.\nonumber
\end{gather}
As already announced, we postpone a~detailed def\/inition of the coef\/f\/icient functions (it is given
by~\eqref{Ve}--\eqref{cEt}). At this point we only mention that the shift coef\/f\/icient~$V_{\delta}(\gamma;x)$ and
additive `potential' $V_{b,\delta}(\gamma;x)$ are meromorphic functions of~$x$, the quantity $\gamma\in{\mathbb C}^8$ is
a~coupling vector, and the dependence on the period parameters~$r$ and $a_{\pm}$ is suppressed; moreover, we have
\begin{gather}
V_{b,\delta}\big(\gamma;x\!+\!\tfrac{\pi}{r}\big)=V_{b,\delta}(\gamma;x),\!\qquad V_{b,\delta}(\gamma;x\!+\!ia_{\delta})= V_{b,\delta}(\gamma;x),\!\qquad  V_{b,\delta}(\gamma;-x)=V_{b,\delta}(\gamma;x),  \!\!\! \!
\\
V_{\delta}\big(\gamma;x\!+\!\tfrac{\pi}{r}\big)=V_{\delta}(\gamma;x),\qquad V_{\delta}(\gamma;x\!+\!ia_{\delta})=\exp(-2r[4a+\langle
\zeta,\gamma\rangle])V_{\delta}(\gamma;x).
\end{gather}
Here, $\langle\cdot,\cdot\rangle$ is the canonical inner product on~${\mathbb C}^8$ (linear in the second slot),
and~$\zeta$ denotes the vector with coordinates~$\zeta_0=\cdots =\zeta_7=1$. Thus, $V_{b,\delta}(\gamma;x)$ is even and
elliptic with respect to the torus with periods $\pi/r$ and $ia_{\delta}$, whereas~$V_{\delta}(\gamma;x)$ is
quasi-periodic. Moreover, all of the coef\/f\/icients are entire functions of~$\gamma$ and invariant under shifting any
component~$\gamma_0,\ldots,\gamma_7$ over multiples of~$i\pi/r$.

From this description it is already clear that the two operators~$A_{\pm}(\gamma;x)$ leave the vector space of
meromorphic and even functions invariant; furthermore, it is readily verif\/ied that for two arbitrary coupling vectors
$\gamma^{(1)}$,  $\gamma^{(2)}$, we need only require equality of component sums to obtain their commutativity:
\begin{gather}
\label{Acom}
\big[A_+(\gamma^{(1)};x),A_-(\gamma^{(2)};x)\big]=0,\qquad \big\langle \zeta,\gamma^{(1)}\big\rangle=\big\langle \zeta,\gamma^{(2)}\big\rangle.
\end{gather}
We repeat, however, that when~$a_+/a_-$ is irrational, the commutativity relation~\eqref{Acom} does not even ensure that
\emph{any} joint eigenfunction exists.

The additive constant in~$V_{b,\delta}(\gamma;x)$ is chosen such that the kernel identity encoded in equation~(3.36) of
Part~I~\cite{RI09} becomes
\begin{gather}
\label{kid}
A_{\delta}(\gamma;x){\cal S}(\sigma(\gamma);x,y)=A_{\delta}(\gamma';y){\cal S}(\sigma(\gamma);x,y).
\end{gather}
Here we have
\begin{gather}
\label{gampr}
\gamma'\equiv -J\gamma,
\\
\label{sg}
\sigma(\gamma)\equiv -\tfrac14 \langle \zeta,\gamma\rangle,
\end{gather}
and~$J$ can be viewed as the ref\/lection associated with the highest $E_8$ root~$\zeta/2$, i.e.,
\begin{gather}
\label{J}
J\equiv {\bf 1}_8-\tfrac14 \zeta\otimes\zeta.
\end{gather}
(We recall pertinent Lie-algebraic information below Theorem~\ref{theorem6.3}.) Furthermore, ${\cal S}$ is the kernel
function
\begin{gather}
\label{cS}
{\cal S}(t;x,y)\equiv \prod\limits_{\delta_1,\delta_2=+,-}G(\delta_1x+\delta_2y-ia +it).
\end{gather}
Here and below, the dependence of the elliptic gamma function~$G(z)$ on~$r$ and~$a_{\pm}$ is suppressed and we use
properties of~$G(z)$ and the closely related (renormalized theta) functions $R_{\pm}(z)$ and~$s_{\pm}(z)$ reviewed
in~Appendix~\ref{appendixA}. Moreover, since we have
\begin{gather}
\sigma(\gamma)=\sigma(\gamma'),
\end{gather}
we may and shall abbreviate~$\sigma(\gamma)$ as~$\sigma$ without causing ambiguity.

We aim to associate two commuting self-adjoint operators~$\hat{A}_{\pm}(\gamma)$ with the A$\Delta$Os~\eqref{Apm} under
certain restrictions on the parameters $a_{+}$, $a_-$, and coupling vector~$\gamma$. However, for the A$\Delta$Os
$A_{\pm}(\gamma;x)$ the relevant Hilbert space is not given by~\eqref{cH}, but by the weighted $L^2$ space
\begin{gather}
\label{cHw}
{\cal H}_w\equiv L^2([0,\pi/2r], w(\gamma;x)dx),
\end{gather}
where the weight function is given~by
\begin{gather}
\label{w}
w(\gamma;x)\equiv 1/c(\gamma;x)c(\gamma;-x),
\end{gather}
with $c(\gamma;x)$ the generalized Harish-Chandra function
\begin{gather}
\label{c}
c(\gamma;x)\equiv \frac{1}{G(2x+ia)}\prod\limits_{\mu=0}^7G(x-i\gamma_{\mu}).
\end{gather}
(The vector $\gamma\in{\mathbb C}^8$ shall be restricted such that~$w(\gamma;x)$ is positive for~$x\in(0,\pi/2r)$.)
Accor\-ding\-ly, the similarity transformed Hamiltonians
\begin{gather}
H_{\delta}(\gamma;x)\equiv w(\gamma;x)^{1/2} A_{\delta}(\gamma;x)w(\gamma;x)^{-1/2},\qquad \delta=+,-,
\end{gather}
(with positive square roots taken for~$x\in(0,\pi/2r)$) give rise to commuting self-adjoint operators on~${\cal
H}$~\eqref{cH} whenever the A$\Delta$Os $A_{\pm}(\gamma;x)$ yield commuting self-adjoint operators on~${\cal
H}_w$~\eqref{cHw}.

Actually, it is more convenient to start from the A$\Delta$Os
\begin{gather}
\label{cApm}
{\cal A}_{\delta}(\gamma;x)\equiv c(\gamma;x)^{-1} A_{\delta}(\gamma;x)c(\gamma;x),\qquad \delta=+,-.
\end{gather}
They are of the above-mentioned form~\eqref{cAp},~\eqref{cAm}, and with suitable restrictions on the pa\-ra\-me\-ters formally
self-adjoint on~${\cal H}$:
\begin{gather}
\label{cApmform}
{\cal A}_{\delta}(\gamma;x)=
\exp(-ia_{-\delta}d/dx)+V_{a,\delta}(\gamma;x)\exp(ia_{-\delta}d/dx)+V_{b,\delta}(\gamma;x),\qquad \delta=+,-.
\end{gather}
Here, the coef\/f\/icients $V_{a,\delta}(\gamma;x)$ are elliptic in~$x$ with periods~$\pi/r$ and~$ia_{\delta}$, but not
even; just as~$V_{\delta}(\gamma;x)$ and~$V_{b,\delta}(\gamma;x)$, they are entire as functions of~$\gamma$, as opposed
to the meromorphy of~$c(\gamma;x)$.

The A$\Delta$Os~\eqref{cApm} satisfy the kernel identity
\begin{gather}
\label{cAid}
{\cal A}_{\delta}(\gamma;x){\cal K}(\gamma;x,y)={\cal A}_{\delta}(\gamma';-y){\cal K}(\gamma;x,y),
\end{gather}
where
\begin{gather}
\label{cK}
{\cal K}(\gamma;x,y)\equiv \frac{{\cal S}(\sigma;x,y)}{c(\gamma;x)c(\gamma';-y)},
\end{gather}
cf.~\eqref{kid}. A~crucial feature of these A$\Delta$Os is that their coef\/f\/icients are not only elliptic in~$x$, but
also manifestly invariant under~$\gamma$-transformations from the Weyl group of the Lie algebra~$D_8$,
cf.~\eqref{cApmform},~\eqref{Va} and~\eqref{Vbe}--\eqref{cEt}; we recall that this group is the semi-direct product of
the permutation group~$S_8$ and the group~$\Phi$ of even sign f\/lips:
\begin{gather}
W(D_8)=S_8 \ltimes \Phi.
\end{gather}
More comments on the pros and cons of the three avatars of the modular pair of operators can be found in~\cite{Rui02}.

At this point it should be mentioned that the kernel functions ${\cal S}(\sigma;x,y)$ and ${\cal K}(\gamma;x,y)$
are~\mbox{$S_8$-}, but not~$D_8$-invariant. (Indeed, for generic~$\gamma$ the sum parameter~$\sigma(\gamma)$~\eqref{sg} is not
invariant under any of the 127 nontrivial sign f\/lips $\phi\in\Phi$.) As will transpire, this lack of $D_8$-invariance
gives rise to (up to) 64 distinct HS operators that pairwise commute, cf.~Theorem~\ref{theorem6.2}.

Next, we recall that the ellipticity of the coef\/f\/icients of the A$\Delta$Os~${\cal A}_{\pm}(\gamma;x)$ entails
\begin{gather}
\label{cAcom}
\big[{\cal A}_+(\gamma^{(1)};x),{\cal A}_-(\gamma^{(2)};x)\big]=0,
\end{gather}
without the sum restriction necessary for~\eqref{Acom}. Now the A$\Delta$Os~${\cal A}_{\pm}(\gamma;x)$ are
\emph{formally} self-adjoint on the Hilbert space~\eqref{cH} for all $\gamma\in{\mathbb R}^8$. Since~\eqref{cAcom}
yields a~huge family of commuting A$\Delta$O pairs, it is clearly too optimistic (and indeed quite wrong) to expect that
there exist self-adjoint Hilbert space versions $\hat{{\cal A}}_+(\gamma^{(1)})$ and $\hat{{\cal A}}_-(\gamma^{(2)})$
that commute without restricting~$\gamma^{(1)}$ and~$\gamma^{(2)}$. As we shall show, there do exist commuting
self-adjoint Hilbert space versions when the couplings~$\gamma^{(1)}$ and~$\gamma^{(2)}$ are \emph{equal} (and satisfy
further restrictions), but there is no a~priori guarantee that even this is feasible.

We associate commuting self-adjoint operators~$\hat{{\cal A}}_{\pm}(\gamma)$ on~${\cal H}$ to the
A$\Delta$Os~\eqref{cApmform} in several stages. In Section~\ref{section2} we f\/irst assemble detailed information about the Hilbert
space eigenvectors of certain HS operators. This is the key to the remainder of the paper, inasmuch as we are going to
show that these eigenvectors give rise to joint eigenfunctions of the above A$\Delta$Os, thus answering the existence
question af\/f\/irmatively for a~countable set of real eigenvalues. In particular, we shall see in Sections~\ref{section3} and~\ref{section5} that
there is a~crucial dif\/ference in the def\/inition of dense domains in~${\cal H}$ for the A$\Delta$O ${\cal A}_s(\gamma;x)$
with the smallest shift parameter compared to the A$\Delta$O ${\cal A}_l(\gamma;x)$ with the largest one.

In more detail, the Hilbert space action of the former is def\/ined via the action of ${\cal A}_s(\gamma;x)$ on certain
spaces of functions that are meromorphic in a~suf\/f\/iciently large strip around the real axis (cf.~\eqref{St}
and~\eqref{Dtgam} below), yielding symmetric operators. As we shall show, this gives rise to a~self-adjoint
operator~$\hat{{\cal A}}_s(\gamma)$ that is essentially self-adjoint on these domains, provided certain restrictions on
the parameters are imposed. The point is now that a~similar domain def\/inition might also be used for the case of~${\cal
A}_l(\gamma;x)$, again giving rise to a~symmetric operator, but that this domain def\/inition is {\it not} the one leading
to a~commuting self-adjoint operator. Instead, a~more involved domain def\/inition also yields a~symmetric
operator~$\hat{{\cal A}}_l(\gamma)$, and this def\/inition does give rise to an essentially self-adjoint operator whose
closure commutes with~$\hat{{\cal A}}_s(\gamma)$.

More specif\/ically, the choice of domain for~$\hat{{\cal A}}_l(\gamma)$ is governed by certain remarkable properties of
the ONB (orthonormal base) of eigenvectors of the pertinent HS operator, and ensures that these eigenvectors are shared
by both operators~$\hat{{\cal A}}_{\pm}(\gamma)$. Now for real couplings spectral invariance of~$\hat{{\cal
A}}_{\pm}(\gamma)$ under $D_8$-transformations follows from the operators themselves being $D_8$-invariant. By contrast,
other~$E_8$ transformations on the coupling vector~$\gamma$ lead to 135 (generically) distinct operators, and their
remarkable isospectrality can only be shown (in Theorem~\ref{theorem6.3}) by invoking a~great many previous results.

To complete this rough sketch of our main results, we mention that the $n\to\infty$ asymptotics of an 8-parameter family
of orthonormal polynomials $p_n(\gamma;\cos(2rx))$ (given by~\eqref{pn}) is shown to coincide with that of the joint
eigenvectors. This state of af\/fairs has a~specialization to the relativistic Lam\'e case that was already introduced and
exploited in~\cite{Rui03}.

We continue with a~more detailed summary of the contents and organization of this paper, introducing various relevant
objects along the way. Section~\ref{section2} is devoted to a~comprehensive analysis of four HS integral operators~${\cal I}$, ${\cal T}$, $I$ and~$T$ on~${\cal H}$. (These operators are actually trace class, but we have no occasion to use this stronger
property.) In this study there is no need to restrict the period parameters~$r$ and~$a_{\pm}$ (save for our standing
assumption~\eqref{parp}), but the vector
\begin{gather}
\label{defg}
g\equiv \operatorname{Re} \gamma,\qquad \gamma=(\gamma_0,\dots,\gamma_7)\in{\mathbb C}^8,
\end{gather}
should vary over a~polytope, whereas~$\operatorname{Im} \gamma_{\mu}$, $\mu=0,\ldots,7$, is required to equal $0,\pi/2r$, or~$-
\pi/2r$. Since the kernel function~${\cal S}(\sigma(\gamma);x,y)$ has primitive period $4i\pi/r$ in~$\gamma_{\mu}$ (by
contrast to the $i\pi/r$-periodicity of~$c(\gamma;x)$, cf.~\eqref{c}), the choice of sign for~$\operatorname{Im} \gamma_{\mu}$
is important.

We restrict attention to two~$\gamma$-regimes for the A$\Delta$Os, which are def\/ined in terms of the choice of imaginary
parts. The f\/irst one is the case where all $\gamma_{\mu}$ are real. This regime is the one with maximal symmetry. The
second one is def\/ined~by
\begin{gather}
\label{imgam}
\operatorname{Im} \gamma_{\mu}=0,\quad \mu=0,1,2,3,\qquad \operatorname{Im} \gamma_{\mu}\in \{\pm \pi/2r\},\quad \mu=4,5,6,7.
\end{gather}
The second regime is the one that can be specialized to the relativistic Lam\'e case and that is needed for taking the
nonrelativistic (Heun) limit, cf.~Subsection~3.2 in~\cite{RI09}.

In Section~\ref{section2} we require in addition
\begin{gather}
\label{sumres}
\sum\limits_{\mu=0}^7\operatorname{Im} \gamma_{\mu}=0.
\end{gather}
Thus the sum parameter $\sigma(\gamma)$ is real in both regimes. In fact, we can reduce the cases where the $\operatorname{Im}
\gamma_{\mu}$-sum equals~$\pm 2 \pi/r$ to the case where it vanishes (cf.~Lemma~\ref{lemma2.1}), but we have no
comparable information for other sum values. (This is why~\eqref{sumres} is imposed in Section~\ref{section2}.)

The pertinent~$g$-polytope arises by a~two-step restriction of~$g$-polytopes that are important in later sections. The
largest one is
\begin{gather}
\label{Pit}
\tilde{\Pi}\equiv \{g\in{\mathbb R}^8\,|\, |g_{\mu}|<a,\mu=0,\ldots,7\},
\end{gather}
with the assumption~$\operatorname{Re} \gamma=g\in\tilde{\Pi}$ ensuring that the weight function~\eqref{w} is real-analytic
on~${\mathbb R}$ and positive for~$x\in(0,\pi/2r)$ for both~$\gamma$-regimes. (Using the def\/inition~\eqref{Gell}
of~$G(z)$, this assertion is easily verif\/ied.) In Section~\ref{section6} we also invoke the smaller parameter space
\begin{gather}
\label{Pi}
\Pi\equiv \{g \in\tilde{\Pi}\,|\, g'\in\tilde{\Pi}\},\qquad g'=-Jg,
\end{gather}
but in Section~\ref{section2} we need to restrict attention to
\begin{gather}
\label{Pir}
\Pi_r\equiv \{g\in\Pi\,|\, \sigma(g) \in(0,a)\},\qquad \sigma(g)=-\tfrac14\sum\limits_{\mu=0}^7 g_{\mu}.
\end{gather}
Throughout Section~\ref{section2}, it is understood that~$\operatorname{Im} \gamma$ either vanishes or is given
by~\eqref{imgam},~\eqref{sumres}, so that~$\sigma(g)=\sigma(\gamma)$.

The constraint~$\sigma\in(0,a)$ ensures that the pertinent integral operators are not only HS, but also have trivial
null space and dense range. We shall call a~(bounded) operator~\emph{complete} if\/f it has the latter property. The f\/irst
integral operator is def\/ined~by
\begin{gather}
\label{cI}
({\cal I}(\gamma)f)(x)\equiv \int_{0}^{\pi/2r} {\cal K}(\gamma;x,y)f(y)dy,\qquad g\in\Pi_r,\qquad f\in{\cal H},
\end{gather}
and its completeness follows from general completenes results obtained in~\cite{Rui12}, cf.~Lemma~\ref{lemma2.1}.
(From~\eqref{cK} and~\eqref{cS} it is easy to see that this completeness is violated at the interval ends. Indeed, we
have~${\cal S}(a;x,y)=1$, cf.~\eqref{cS} and~\eqref{refl}, so that ${\cal I}(\gamma)$ reduces to a~rank-one operator
for~$\sigma=a$; also, for $\sigma=0$ its kernel has poles at~$x=\pm y$, so then~${\cal I}(\gamma)$ is not even well
def\/ined as a~Hilbert space operator, let alone HS.) Using~\eqref{cK} and the identity
\begin{gather}
\overline{c(\gamma;x)}=c(\gamma;-x),\qquad g\in\Pi_r,\qquad x\in{\mathbb R},
\end{gather}
(cf.~\eqref{c} and the~$G$-conjugacy relation given by~\eqref{Mconj},~\eqref{cons1}), we see that its adjoint is given~by
\begin{gather}
{\cal I}(\gamma)^*={\cal I}(\gamma').
\end{gather}

We also have occasion to employ the positive trace class operator
\begin{gather}
{\cal T}(\gamma)\equiv {\cal I}(\gamma){\cal I}(\gamma').
\end{gather}
We recall that general arguments yield polar decompositions
\begin{gather}
{\cal I}(\gamma)={\cal T}(\gamma)^{1/2}{\cal U}(\gamma),\qquad {\cal I}(\gamma')={\cal U}(\gamma)^*{\cal
T}(\gamma)^{1/2},
\end{gather}
where~${\cal U}(\gamma)$ is a~unitary operator on~${\cal H}$; also, there exist two ONBs~$\{f_n(\gamma)\}_{n=0}^\infty$
and~$\{f_n(\gamma')\}_{n=0}^\infty$ such that we have a~singular value decomposition
\begin{gather}
\label{sing}
{\cal I}(\gamma)=\sum\limits_{n=0}^{\infty}\lambda_n(f_n(\gamma'),\cdot) f_n(\gamma),
\end{gather}
with{\samepage
\begin{gather}
\label{lamb}
\lambda_n=\lambda_n(\gamma)=\lambda_n(\gamma'),\qquad \lambda_0\ge \lambda_1\ge \lambda_2 \ge \dots >0,\qquad
\sum\limits_{n=0}^{\infty}\lambda_n^2<\infty,
\end{gather}
see, e.g.,~\cite[Chapter~VI]{RS72}.}

For our initial purposes, however, it is more convenient to work with the integral operator
\begin{gather}
\label{I}
(I(\gamma)f)(x)\equiv w(\gamma;x)^{1/2}\int_{0}^{\pi/2r} {\cal S}(\sigma;x,y)w(\gamma';y)^{1/2}f(y)dy,\qquad g\in\Pi_r,\qquad f\in{\cal H}.
\end{gather}
The weight functions occurring here are real-analytic on~${\mathbb R}$, even, $\pi/r$-periodic, and positive
on~$(0,\pi/2r)$. We are taking positive square roots in~\eqref{I}, and this entails that the kernel of~$I(\gamma)$ is
positive on~$(0,\pi/2r)^2 $. (The weight function~$w(\gamma;x)$ has double zeros for~$x\equiv 0 \pmod{\pi/2r}$,
cf.~\eqref{w} and~\eqref{c}.)

Since we have
\begin{gather}
I(\gamma)=M(\gamma) {\cal I}(\gamma)M(\gamma')^{-1},
\end{gather}
where~$M(\gamma)$ denotes the unitary operator of multiplication~by
\begin{gather}
\label{m}
m(\gamma;x)\equiv w(\gamma;x)^{1/2}c(\gamma;x),\qquad g\in\tilde{\Pi},\qquad x\in(0,\pi/2r),
\end{gather}
(with the positive square root understood), $I(\gamma)$ is also a~complete HS operator on~${\cal H}$. Its adjoint equals
$I(\gamma')$, and in Section~\ref{section2} the eigenvectors~$e_n(\gamma)$ of the positive trace class operator
\begin{gather}
\label{T}
T(\gamma)\equiv I(\gamma)I(\gamma'),\qquad g\in\Pi_r,
\end{gather}
are studied in considerable detail. In this enterprise we cannot avoid various case distinctions that depend on the six
possible orderings of the numbers~$\sigma(\gamma)$~\eqref{sg}, $d(\gamma)$~\eqref{dgam} and~$a_s$~\eqref{as}, which can
vary over $(0,a)$; moreover, there is an elaborate dependence on the maximal integer~$L$ such that~$La_s<a_l$.
Complications also arise from the contingency that the operator $T(\gamma)$ has degenerate eigenvalues; this prevents us
from appealing to continuity (let alone real-analyticity) of its eigenvectors
\begin{gather}
e_n(\gamma;\cdot)=m(\gamma;\cdot)f_n(\gamma;\cdot),\qquad n\in{\mathbb N}\equiv \{0,1,2,\ldots,\},
\end{gather}
in their dependence on the parameters~$a_+$, $a_-$ and~$\gamma$.

In Lemma~\ref{lemma2.2} we obtain information on the range of~$T(\gamma)$ and its powers. As a~consequence, we can infer
that all of the functions
\begin{gather}
\label{Fn}
F_n(\gamma;x)\equiv w(\gamma;x)^{-1/2}e_n(\gamma;x)=c(\gamma;x)f_n(\gamma;x),\qquad n\in{\mathbb N},\qquad x\in(0,\pi/2r),
\end{gather}
extend to functions that are even, $\pi/r$-periodic and holomorphic in the cut plane~${\mathbb C}_{\sigma}$, where
\begin{gather}
\label{Ccut}
{\mathbb C}_t\equiv {\mathbb C} \setminus \{x=m\pi/2r \pm ic\,|\, c\in [t,\infty),m\in{\mathbb Z}\},\qquad t>0.
\end{gather}
Lemma~\ref{lemma2.3} concerns analyticity properties of the functions
\begin{gather}
\label{gnF}
g_n(\gamma;x)\equiv \int_{0}^{\pi/2r} {\cal S}(\sigma;x,y)w(\gamma;y)F_n(\gamma;y)dy,\qquad n\in{\mathbb N}.
\end{gather}
(With suitable conventions, these functions amount to~$\lambda_n F_n(\gamma';x)$.) From the def\/inition~\eqref{cS} of the
kernel function~${\cal S}$ it is immediate that they are holomorphic for~$|\operatorname{Im} x|<\sigma$, even and
$\pi/r$-periodic. In Lemma~\ref{lemma2.3} we prove that these properties hold true for~$|\operatorname{Im}
x|<\sigma+m(\gamma)$, with~$m(\gamma)$ given by~\eqref{mgam}.

Next, we introduce
\begin{gather}
\label{Hn}
H_n(\gamma;x)\equiv P(\gamma;x) F_n(\gamma;x),\qquad g\in \Pi_r,
\end{gather}
where $P(\gamma;x)$ is the product function
\begin{gather}
\label{Pgam}
P(\gamma;x)\equiv \prod\limits_{\mu=0}^7 E(\pm x+i\gamma_\mu),
\end{gather}
and $E(z)$ is the entire function~\eqref{E}. Moreover, in~\eqref{Pgam} (and often below as well), we use the notation
\begin{gather}
f(\pm z+ p)\equiv f(z+p)f(-z+p).
\end{gather}

The importance of the functions~$H_n(\gamma;x)$ is due to their being holomorphic in all of the complex plane. More
precisely, we are only able to prove this for period ratios restricted~by
\begin{gather}
\label{aa1}
a_s/a_l\ne 1, 1/2.
\end{gather}
From entireness of~$H_n(\gamma;x)$ it is immediate that the functions~$F_n(\gamma;x)$ are meromorphic in~${\mathbb C}$,
with poles that can only occur at the zeros of~$P(\gamma;x)$. We prove entireness in several stages. As a~preparation,
we switch after Lemma~\ref{lemma2.3} to the functions~$H_n$. In view of~\eqref{gnF}, we need the relation
\begin{gather}
\label{HF}
w(\gamma;x)F_n(\gamma;x)=m_H(\gamma;x)H_n(\gamma;x),
\end{gather}
where~$m_H$ is the multiplier function
\begin{gather}
\label{mH}
m_H(\gamma;x)\equiv w(\gamma;x)/P(\gamma;x)= G(\pm 2x+ia)/P(-\gamma;x),
\end{gather}
cf.~\eqref{w},~\eqref{c}, and~\eqref{GE}.

The switch from $F_n$ to $H_n$ has another consequence. Indeed, it is readily verif\/ied that
\begin{gather}
w_H(\gamma;x)\equiv m_H(\gamma;x)/P(\gamma;x)=G(\pm 2x+ia)/P(\pm\gamma;x),\qquad g\in\Pi_r,
\end{gather}
is a~weight function with the same features as~$w(\gamma;x)$. (Specif\/ically, it is also real-analytic on~${\mathbb R}$,
even, $\pi/r$-periodic, positive on~$(0,\pi/2r)$, and has double zeros for~$x\equiv 0 \pmod{\pi/2r}$.) Hence we arrive
at a~fourth incarnation of the modular pair of A$\Delta$Os, namely,
\begin{gather}
\label{AHpm}
A^H_{\delta}(\gamma;x)\equiv P(\gamma;x)A_{\delta}(\gamma;x)P(\gamma;x)^{-1},\qquad \delta=+,-.
\end{gather}
These A$\Delta$Os are formally self-adjoint on the weighted $L^2$ space
\begin{gather}
\label{cHwH}
{\cal H}_{w_H}\equiv L^2([0,\pi/2r], w_H(\gamma;x)dx).
\end{gather}
Their importance hinges on the entire functions~$H_n(\gamma;x)$ being joint eigenfunctions, as will become clear in
Section~\ref{section5}.

The main result of Subsection~\ref{section2.2} is Lemma~\ref{lemma2.4}, which is the f\/irst step towards showing entireness
of~$H_n(\gamma;x)$. It can be rephrased as stating that the functions~$H_n(\gamma;x)$ are holomorphic in the cut
plane~${\mathbb C}_{\sigma +a_s}$. As will become clear in~Section~\ref{section4}, this result already suf\/f\/ices for deducing that the
functions $H_n(\gamma;x)$ are eigenfunctions of the A$\Delta$O $A_s^H(\gamma;x)$.

\looseness=-1
The derivation of the $H_n$-identity~\eqref{HHsp} for the special case $a_s=a_l$ at the end of~Subsection~\ref{section2.2}
foreshadows our reasoning to obtain the crucial generalizations~\eqref{Hnident} for $a_s<a_l$. At f\/irst reading, it may
be advisable to jump from this point to Section~\ref{section3}, inasmuch as the next two Subsections~\ref{section2.3} and~\ref{section2.4} are quite technical and their results are not necessary for Sections~\ref{section3} and~\ref{section4}.

\looseness=-1
To arrive in Section~\ref{section5} at the `correct' Hilbert space def\/inition of the A$\Delta$O with the largest shift
parameter~$a_l$ (namely, such that it shares its eigenvectors with ${\cal T}(\gamma)$ and~$\hat{{\cal A}}_s(\gamma)$),
we must improve on Lemma~\ref{lemma2.4} in two respects. Specif\/ically, in Subsections~\ref{section2.3} and~\ref{section2.4} we show not only that the
functions~$H_n(\gamma;x)$ are holomorphic in the larger cut plane~${\mathbb C}_{\sigma +a_l}$, but in the process also
uncover the remarkable identities~\eqref{Hnident} satisf\/ied by these functions. These principal results of Section~\ref{section2} are
encoded in~Theorem~\ref{theorem2.9}, and reveal the way to the desired domain def\/inition for~$\hat{{\cal A}}_l(\gamma)$.

To prevent getting lost whilst obviating too many snags at once, we have opted for treating the case~$a_l\in (a_s,2a_s]$
in~Subsection~\ref{section2.3}, using however the device of proving the auxiliary Lemmas~\ref{lemma2.5}--\ref{lemma2.8} for the
general case at the point where the need for these results becomes apparent for the special case under consideration.

It may well be possible to extend the arguments in Section~\ref{section2} to obtain holomorphy of~$H_n(\gamma;x)$ in all of~${\mathbb
C}$, but as it is now, our proof of holomorphy in~${\mathbb C}_{\sigma+a_l}$ is already quite long and elaborate.
Therefore we show entireness of the functions~$H_n(\gamma;x)$ in Section~\ref{section4}, by using the A$\Delta$O~$A_s^H(\gamma;x)$.
Unfortunately, we need the restriction~\eqref{aa1} in our proof. (We believe entireness still holds when~$a_s$ is equal
to~$a_l$ or~$a_l/2$.)

We begin Section~\ref{section3} by def\/ining the coef\/f\/icient functions in the A$\Delta$Os $A_{\pm}(\gamma;x)$~\eqref{Apm},
cf.~\eqref{Ve}--\eqref{cEt}; this involves solely the functions~$R_{\pm}$ from Appendix~\ref{appendixA}. As a~consequence, we get
explicit formulas for the coef\/f\/icients $V_{a,\pm}(\gamma;x)$ of ${\cal A}_{\pm}(\gamma;x)$~\eqref{cApmform}
and~$V_{\pm}^H(\gamma;x)$ of $A_{\pm}^H(\gamma;x)$~\eqref{AHpm}, namely,~\eqref{Va} and~\eqref{VH}--\eqref{VHfin}, resp.

In order to study Hilbert space aspects, we must make suitable assumptions on the pa\-ra\-meters. To begin with, we need to
restrict~$g$ to~$\tilde{\Pi}$~\eqref{Pit}. For ${\cal A}_s(\gamma;x)$ it suf\/f\/ices to require in addition
that~\eqref{aa1} hold true. For~${\cal A}_l(\gamma;x)$, however, we need the stronger requirement
\begin{gather}
\label{aa2}
a_s/a_l\notin \{1/k\,|\, k \in{\mathbb N}^*:={\mathbb N}\setminus \{0\} \}.
\end{gather}
With the constraint~\eqref{aa1} in force, we prove in Theorem~\ref{theorem3.1} that the A$\Delta$O~${\cal
A}_s(\gamma;x)$ gives rise to a~symmetric operator~$\hat{{\cal A}}_s(t,\gamma)$ on the dense subspace
$D_t(\gamma)$~\eqref{Dtgam} of~${\cal H}$~\eqref{cH}, provided $t>a_s$. This initial result involves a~key new
ingredient, namely, the auxiliary Harish-Chandra function
\begin{gather}
\label{cPol}
c_P(\gamma;x)\equiv \frac{\prod\limits_{\mu=0}^7E(x\pm i\gamma_{\mu})}{(1-\exp(-4irx))E(2x\pm i(a_+-a_-)/2)},\qquad
g\in\tilde{\Pi},
\end{gather}
where $E(z)$ is the entire function~\eqref{E}. Here, the subscript stands for `Polynomial': In Section~\ref{section7} we have
occasion to exploit the orthonormal polynomials spanning the Hilbert space
\begin{gather}
\label{cHP}
{\cal H}_P\equiv L^2([0,\pi/2r],(r/\pi)w_P(\gamma;x)dx),\qquad g\in\tilde{\Pi},
\end{gather}
where
\begin{gather}
\label{wP}
w_P(\gamma;x)\equiv 1/c_P(\gamma;\pm x).
\end{gather}

With the stronger constraint~\eqref{aa2} in ef\/fect, there is little dif\/f\/iculty in extending the symmetry result of
Theorem~\ref{theorem3.1} to the A$\Delta$O~${\cal A}_l(\gamma;x)$, but as already mentioned above, for
generic~$\gamma$'s the pertinent simple domain choices (namely, $D_t(\gamma)$ for $t>a_l$) do not lead to a~self-adjoint
operator commuting with~$\hat{{\cal A}}_s(\gamma)$.

In Section~\ref{section4} we start from the functions (cf.~the def\/inition~\eqref{Fn} of~$F_n$)
\begin{gather}
\label{fn}
f_n(\gamma;x)=\frac{1}{c(\gamma;x)}F_n(\gamma;x),\qquad n\in{\mathbb N},\qquad g\in\Pi_r.
\end{gather}
In view of the decomposition~\eqref{sing}, they yield an ONB of ${\cal T}(\gamma)$-eigenvectors, whereas the
functions~$e_n(\gamma;x)=m(\gamma;x)f_n(\gamma;x) $ give rise to an ONB of $T(\gamma)$-eigenvectors. In
Lemma~\ref{lemma4.1} we prove that the vectors~$f_n(\gamma;\cdot)$ belong to the def\/inition domain~$D_t(\gamma)$ of the
operator~$\hat{{\cal A}}_s(t,\gamma)$ for any $t\in(a_s,a]$. Hence we can def\/ine a~symmetric operator $\hat{{\cal
A}}_s(\gamma)$ by restricting the def\/inition domain to the span~${\cal C}(\gamma)$ of the ${\cal H}$-ONB
$\{f_n(\gamma;\cdot)\}_{n\in{\mathbb N}}$.

Lemma~\ref{lemma4.2} is a~pivotal auxiliary result, showing that~$\hat{{\cal A}}_s(\gamma'){\cal I}(\gamma')$
equals~${\cal I}(\gamma') \hat{{\cal A}}_{s}(\gamma)$ on~${\cal C}(\gamma)$. Its proof is relegated to~Appendix~\ref{appendixB}. On
the other hand, it is an easy consequence of this lemma that the operator~$\hat{{\cal A}}_s(\gamma)$ is essentially
self-adjoint on~${\cal C}(\gamma)$ and that the vectors~$f_n(\gamma;\cdot)$ can be redef\/ined so that they yield joint
eigenvectors of~${\cal T}(\gamma)$ and~$\hat{{\cal A}}_s(\gamma)$, cf.~Theorem~\ref{theorem4.3}.

Rewriting the resulting A$\Delta$O-eigenvalue equation
\begin{gather}
{\cal A}_s(\gamma;x)f_n(\gamma;x)=E_{n,s}(\gamma)f_n(\gamma;x),
\end{gather}
as
\begin{gather}
\label{AsHade}
A_s^H(\gamma;x)H_n(\gamma;x)=E_{n,s}(\gamma)H_n(\gamma;x),
\end{gather}
it now follows by using special features of the coef\/f\/icients~$V_s^H(\gamma;\pm x)$ that $H_n(\gamma;x)$ is an entire
function. This is the content of Theorem~\ref{theorem4.4}, which also lists relations between $H_n(\gamma;x)$-values
that encode the residue cancellations following from~\eqref{AsHade} and entireness. The last result of Section~\ref{section4} is
Lemma~\ref{lemma4.5}, which concerns the issue of degeneracy of the eigenspaces of~$\hat{{\cal A}}_s(\gamma)$. Def\/ining
spaces of elliptic multipliers~by
\begin{gather}
\label{cP}
{\cal P}(p):=\{\zeta(x)\ \mathrm{elliptic\ with\ periods}\ \pi/r, ip\},\qquad p>0,
\end{gather}
it states that whenever eigenvalue degeneracy occurs, the pertinent eigenfunctions must be related by multipliers
in~${\cal P}(a_s)$.

In Section~\ref{section5} we focus on the A$\Delta$O~${\cal A}_l(\gamma;x)$, assuming~\eqref{aa2}. We f\/irst show that the
(restrictions to $[0,\pi/2r]$ of the) functions ${\cal A}_l(\gamma;x)f_n(\gamma;x)$ belong to ${\cal H}$. In
Lemma~\ref{lemma5.1} we then prove that the Hilbert space operator~$\hat{{\cal A}}_l(\gamma)$ thus def\/ined on the
span~${\cal C}(\gamma)$ of the $f_n$'s is symmetric. The identities~\eqref{Hnident} are indispensable in this
enterprise: They ensure the vanishing of a~residue sum arising from simple poles that are met when shifting a~contour
over~$a_l$.

\looseness=1
The presence of these poles for generic $g\in\Pi_r$ is the reason why the vectors $f_n(\gamma;\cdot)\in{\cal H}$ cannot
be eigenvectors of any self-adjoint extension of the symmetric operator~$\hat{{\cal A}}_l^w(t,\gamma)$ from Section~\ref{section3}.
This is shown in more detail after Lemma~\ref{lemma5.1}, but we have not tried to isolate the nongeneric~$\gamma$'s for
which (a self-adjoint extension of)~$\hat{{\cal A}}_l^w(t,\gamma)$ coincides with~$\hat{{\cal A}}_l(\gamma)$ on~${\cal
C}(\gamma)$.

The crucial Lemma~\ref{lemma5.2} is the counterpart of Lemma~\ref{lemma4.2}. Its proof is relegated to Appendix~\ref{appendixC}. Once
more, the $H_n$-identities~\eqref{Hnident} play a~key role in our proof, which involves substantial ef\/fort. With this
arduous lemma out of the way, the remainder of Section~\ref{section5} amounts to an easy adaptation of the results in Section~\ref{section4}. In
Theorem~\ref{theorem5.3} we show that the ONB-vectors~$f_n(\gamma;\cdot)$ can be redef\/ined (if need be) so that they
become joint eigenvectors of~${\cal T}(\gamma)$ and $\hat{{\cal A}}_{\pm}(\gamma)$. The resulting eigenvalue A$\Delta$E
\begin{gather}
\label{AlHade}
A_l^H(\gamma;x)H_n(\gamma;x)=E_{n,l}(\gamma)H_n(\gamma;x),
\end{gather}
and the entireness of $H_n$ proved in Theorem~\ref{theorem4.4} can then be exploited to obtain additional
$H_n$-identities in Corollary~\ref{corollary5.4}.

Lemma~\ref{lemma5.5} is the analog of Lemma~\ref{lemma4.5}: It states that eigenvalue degeneracy in~\eqref{AlHade} can
only occur when the relevant eigenfunctions are related by a~multiplier in~${\cal P}(a_l)$~\eqref{cP}. Assuming the
quotient $a_s/a_l$ is irrational, it readily follows that there is no degeneracy for the joint eigenvalues
$(E_{n,s},E_{n,l})$, cf.~Theorem~\ref{theorem5.6}.

The latter theorem is of pivotal importance in Section~\ref{section6}. We use it to prove that the positive trace class operators
${\cal T}(\gamma)$, $g\in\Pi_r$, whose~$\gamma$'s are related by sign f\/lips, form a~commutative family,
cf.~Theorem~\ref{theorem6.2}. We refer to such~$\gamma$'s as~$\gamma$-clusters. Theorem~\ref{theorem6.2} also shows that
the def\/inition of~$\hat{{\cal A}}_l(\gamma)$ yields the same operator for all~$\gamma$'s in a~cluster. In this
connection we stress that in spite of the $D_8$-invariance of the A$\Delta$O ${\cal A}_l(\gamma;x)$, this is by no means
obvious, since the initial def\/inition domain of~$\hat{{\cal A}}_l(\gamma)$ consists of the span of the ${\cal
T}(\gamma)$-eigenvectors, and ${\cal T}(\gamma)$ is not invariant under sign f\/lips. By contrast, to see this invariance
for~$\hat{{\cal A}}_s(\gamma)$, it suf\/f\/ices to invoke well-known lore concerning symmetric vs.~essentially self-adjoint
operators~\cite{RS72}.

\looseness=-1
Armed with these insights, we can f\/inally address the isospectrality issue for the commuting Hilbert space
operators~$\hat{{\cal A}}_{\pm}(\gamma)$. In Theorem~\ref{theorem6.2} this issue is trivialized for the permutations and
sign f\/lips relevant to the two regimes, inasmuch as the \emph{operators} are shown to be the same. In general, however,
the operators~$\hat{{\cal A}}_{\pm}(\gamma')$ do not coincide with~$\hat{{\cal A}}_{\pm}(\gamma)$. In particular, it can
happen that the latter have~$x$-independent coef\/f\/icients, whereas the former do not. Even so, Lemma~\ref{lemma6.1} shows
that their spectra are the same. These issues are illustrated by explicit examples below Theorem~\ref{theorem6.2}.

As we prove in Theorem~\ref{theorem6.3}, for the f\/irst regime the couplings obtained by combining $D_8$-transformations
with the ref\/lection~$J$~\eqref{J} yield equal $\hat{{\cal A}}_{\pm}(\gamma)$-spectra. More specif\/ically, to ensure that
these orbits of the $E_8$ Weyl group belong to $\tilde{\Pi}$~\eqref{Pit}, we restrict the Euclidean length of~$\gamma$
by requiring $\|\gamma\|_2<a$. (We use the suf\/f\/ix 2, since the $\ell^{\infty}$- and $\ell^1$-norms of~$\gamma$ also play
a~role.)

On the other hand, this $\ell^2$-restriction seems somewhat unnatural from a~Lie-algebraic perspective. We collect some
salient aspects of the Lie algebra $E_8$ below Theorem~\ref{theorem6.3}, and ask a~geometric question concerning the
more `natural' orbit space~${\cal O}(E_8)$~\eqref{Gorbit}, but leave the answer open.

For the second regime we must restrict attention to a~subgroup of $D_8$-transformations, so we obtain
isospectral~$\gamma$-orbits under a~subgroup of the $E_8$ Weyl group, cf.~Theorem~\ref{theorem6.4}.

\looseness=-1
The f\/inal Section~\ref{section7} is concerned with the $n\to\infty$ behavior of the eigenvalues $\lambda_n(\gamma)^2$ of the positive
trace class operators ${\cal T}(\gamma)$, the eigenvalues $E_{n,\pm}$ of the self-adjoint operators $\hat{{\cal
A}}_{\pm}(\gamma)$, and their joint eigenvectors. Our main tool to obtain information about these issues is the ONB of
polynomials $p_n(\gamma;\cos(2rx))$ for the Hilbert space ${\cal H}_P$~\eqref{cHP}. The point is that we can compare the
large-$n$ asymptotics of the ${\cal T}(\gamma)$ eigenfunctions to that of the polynomials, which is explicitly known.
More precisely, we only need to invoke their $L^2$ asymptotics, which we established in~\cite{Rui05MR} by using solely
elementary Hilbert space lore. (Possibly, more information can be derived from the very detailed $L^{\infty}$ asymptotic
behavior obtained in~\cite{KMAV03} by elaborate Riemann--Hilbert problem techniques.)

In more detail, the starting point for our asymptotic analysis is Lemma~\ref{lemma7.1}, which shows that the action of
${\cal I}(\gamma')$ on the function
\begin{gather}
\psi_n(\gamma;x)\equiv \sqrt{\frac{r}{\pi}}p_n(\gamma;\cos(2rx))/c_P(\gamma;x),\qquad g\in\tilde{\Pi},
\end{gather}
yields a~function of the form $Ce^{-2nr\sigma}\psi_n(\gamma';x)$, up to an error term that also has exponential decay as
$n\to\infty$. (Surprisingly, the constant~$C$ depends only on~$\sigma$, cf.~\eqref{kapn} below.) This already suggests
that $\lambda_n(\gamma)$ becomes asymptotically equal to $Ce^{-2nr\sigma}$, and that the ${\cal
T}(\gamma)$-eigenfunc\-tion~$f_n(\gamma;x)$ is asymptotically proportional to~$\psi_n(\gamma;x)$.

In Theorems~\ref{theorem7.3},~\ref{theorem7.5} and~\ref{theorem7.7} these expectations are corroborated, but only at the expense of a~further restriction
on~$\sigma$. The general state of af\/fairs is analyzed in Lemmas~\ref{lemma7.2},~\ref{lemma7.4} and~\ref{lemma7.6}, which are of some interest in their
own right.

Even though we are unable to prove the above expectations without extra restrictions on~$\sigma$, we do give some solid
answers for these natural questions. By contrast, we only have circumstantial evidence for two further conjectures
regarding eigenvalue asymptotics, namely,
\begin{alignat}{3}
\label{conjs}
& E_{n,s}(\gamma)\sim \exp (2nra_s),\qquad && n\to \infty, \ \ (?)   &
\\
\label{conjl}
& E_{n,l}(\gamma)\sim \exp (2nra_l),\qquad && n\to \infty. \ \ (?)  &
\end{alignat}
Worse, even though we can prove that $\hat{{\cal A}}_s(\gamma)$ is bounded below, but unbounded above, we have not found
a~full proof of either property for $\hat{{\cal A}}_l(\gamma)$. Ironically enough, both properties follow in the same
way for the `wrong' symmetric operator $\hat{{\cal A}}_l^w(\gamma)$ as for $\hat{{\cal A}}_s(\gamma)$, but the
dif\/f\/iculty with handling $\hat{{\cal A}}_l(\gamma)$ is that its initial domain is the span of the joint eigenvectors
$f_n(\gamma;\cdot)$ of $\hat{{\cal A}}_s(\gamma)$ and ${\cal T}(\gamma)$. To be sure, in this paper we obtain
a~considerable amount of explicit information on the joint eigenfunctions $f_n(\gamma;x)$ of the \emph{analytic
difference operators} ${\cal A}_{\pm}(\gamma;x)$, but most of this seems `too far removed' from the Hilbert space ${\cal
H}$ to resolve domain closure issues.

Section~\ref{section7} is concluded with some remarks on the relativistic Lam\'e case; as already mentioned above, we hope to
elaborate on this special case elsewhere.

In Appendix~\ref{appendixA} we review salient features of the building block functions used in this paper. Proofs and further details
can be found in~\cite{Rui97}. Appendixes~\ref{appendixB} and~\ref{appendixC} contain the proofs of Lemmas~\ref{lemma4.2} and~\ref{lemma5.2}, resp.

\section[The HS operators~${\cal I} (\gamma)$, $I(\gamma)$, $T(\gamma)$ and ${\cal T}(\gamma)$, $g\in\Pi_r$]{The HS operators~$\boldsymbol{{\cal I} (\gamma)}$, $\boldsymbol{I(\gamma)}$, $\boldsymbol{T(\gamma)}$ and $\boldsymbol{{\cal T}(\gamma)}$, $\boldsymbol{g\in\Pi_r}$}\label{section2}

\subsection{Preliminaries}\label{section2.1}

We begin this section by taking a~closer look at the integral operator~${\cal I}(\gamma)$ with kernel~${\cal
K}(\gamma;x,y)$, cf.~\eqref{cK} and~\eqref{cS}. Choosing f\/irst~$\sigma=\sigma(\gamma) \in(0,2a)$, we can invoke the
series representation~\eqref{Gg},~\eqref{g} of the elliptic gamma function to write
\begin{gather}
\label{cSnew}
{\cal
S}(\sigma;x,y)=\exp\left(\sum\limits_{n=1}^{\infty}\frac{2\cos(2nrx)\cos(2nry)\sinh(2nr(a-\sigma))}{n\sinh(nra_+)\sinh(nra_-)}\right),
\qquad x,y\in{\mathbb R}.
\end{gather}
Viewing~${\cal S}(\sigma;x,y)$ as the kernel of an integral operator on~${\cal H}$~\eqref{cH} in its own right, it is
clear that its rank equals one for~$\sigma=a$. It is not hard to see that it has rank two for $\sigma=a+ a_{-\delta}/2$,
$\delta=+,-$. (Indeed, from~\eqref{Rrep} it follows that this~$\sigma$-choice yields the rank-two kernel
$R_{\delta}(x\pm y)$.) For~$\sigma\in(0,a)$, however, the right-hand side is of the form
\begin{gather}
\exp\left(\sum\limits_{n=1}^{\infty}c_n\cos(2nrx)\cos(2nry)\right),\qquad c_n>0,\quad\forall\, n\in{\mathbb N}^{*}\equiv
{\mathbb N}\setminus \{0\}, \quad \sum\limits_{n=1}^{\infty}c_n<\infty. \!\!\!
\end{gather}
It follows from~\cite[Lemma~1.1]{Rui12} that a~kernel with these properties yields a~positive HS operator. In
particular, it is complete in the sense def\/ined above~\eqref{cI}. We are now prepared for the following lemma.

\begin{Lemma}
\label{lemma2.1}
Assume $g=\operatorname{Re} \gamma$ belongs to $\Pi_r$~\eqref{Pir} and $\operatorname{Im} \gamma$ either vanishes or is given
by~\eqref{imgam} with
\begin{gather}
\sum\limits_{\mu=4}^7\operatorname{Im} \gamma_{\mu}\in \{0, \pm 2 \pi/r\}.
\end{gather}
Then the integral operator
\begin{gather}
\label{cIg}
({\cal I}(\gamma)f)(x)= \int_0^{\pi/2r}\frac{{\cal S}(\sigma;x,y)}{c(\gamma;x) c(\gamma';-y)}f(y)dy, \qquad f\in{\cal H},
\end{gather}
is HS with trivial null space and dense range. Moreover, it is real-analytic in~$g$ on $\Pi_r$ and real-analytic in
$a_{\pm}$ on~$(0,\infty)^2$ in the Hilbert--Schmidt norm topology. Finally, assuming~$\gamma$ satisfies
\begin{gather}
\label{gampos}
\sum\limits_{\mu=4}^7\operatorname{Im} \gamma_{\mu}= 0,\qquad c(\gamma;x)=c(\gamma';x),
\end{gather}
it is a~self-adjoint operator with positive eigenvalues.
\end{Lemma}
\begin{proof}
Assuming f\/irst~\eqref{sumres}, we have~$\sigma\in(0,a)$, so that the kernel~\eqref{cSnew} def\/ines a~positive HS operator
on~${\cal H}$. Furthermore, the functions~$1/c(\gamma;x)$ and~$1/c(\gamma';-x) $ are real-analytic on~${\mathbb R}$ and
nonzero on~$(0,\pi/2r)$, since~$|g_{\mu}|,|g'_\mu|<a$, cf.~\eqref{c}. Therefore, they yield bounded multiplication
operators on~${\cal H}$ with trivial null space. Hence completeness of~${\cal I}(\gamma)$ readily follows.

Next, we assume~$ \operatorname{Im} \gamma_{\mu}= \pi/2r$, $\mu=4,5,6,7$. Then we are dealing with a~kernel of the form
\begin{gather}
\exp\left(\sum\limits_{n=1}^{\infty}\frac{2(-)^n\cos(2nrx)\cos(2nry)\sinh(2nr\eta)}{n\sinh(nra_+)\sinh(nra_-)}\right),\qquad
  \eta \in (0,a).
\end{gather}
As it stands, this type of kernel is not studied in~\cite{Rui12}. However, it is easy to verify that the corresponding
integral operator is the product of a~positive HS operator of the previous form and the unitary parity operator
\begin{gather}
(Pf)(x):=f(\pi/2r-x),\qquad f\in{\cal H}.
\end{gather}
From this completenes of~${\cal I}(\gamma)$ is clear, and taking $\operatorname{Im} \gamma\to -\operatorname{Im} \gamma$ leaves ${\cal
I}(\gamma)$ invariant. (Recall in this connection that~$c(\gamma;x)$ is $i\pi/r$-periodic in~$\gamma_{\mu}$.)

Analyticity of ${\cal I}(\gamma)$ in~$g$ with respect to the HS norm amounts to analyticity of its kernel ${\cal
K}(\gamma;x,y)$ in~$g$ in the strong topology of the Hilbert space $ L^2([0,\pi/2r]^2, dxdy)$ (i.e., the $L^2$-norm
topology). From the explicit kernel formula it is readily verif\/ied that the kernel is strongly analytic in~$g$ in
a~complex neighborhood of $\Pi_r$, proving the real-analyticity assertion. In the same way, real-analyticity in~$a_+$
and~$a_-$ can be verif\/ied.

The additional~$\gamma$-assumptions~\eqref{gampos} entail $\operatorname{Im} \sigma(\gamma)=0$ and self-adjointness of ${\cal
I}(\gamma)$. Therefore, positivity of~${\cal I}(\gamma)$ follows from the kernel~\eqref{cSnew} yielding a~positive HS
operator on~${\cal H}$ for $\sigma\in (0,a)$.
\end{proof}

Since the HS operator ${\cal I}(\gamma)$ is real-analytic in~$g$ on the connected set~$\Pi_r$, we need only exhibit one
special~$\gamma$ for which all of its singular values~$\lambda_n$~\eqref{lamb} are nondegenerate to infer generic
nondegeneracy. Unfortunately we were unable to f\/ind any such `explicitly solvable'~$\gamma$ for the f\/irst regime. For
the second~$\gamma$-regime we shall exhibit such a~choice later on, cf.~\eqref{gfree0}--\eqref{lamfree}. (Recall the two
regimes were def\/ined in the paragraph containing~\eqref{imgam}.) Even so, it seems hard to make good use of that. The
dif\/f\/iculty is that we cannot exclude isolated degeneracies as~$\gamma$ varies, so that we cannot even make
a~\emph{continuous} choice for the ONBs in~\eqref{sing}.

We mention in passing that~${\cal I}(\gamma)$, $g\in\Pi_r$, is in fact a~trace class operator, but we have no occasion
to use this feature. (It follows from the integral operator with kernel~\eqref{cSnew} being trace class, cf.~the lemma
on p.~65 of~\cite{RS79}.)

For the remainder of this section we assume~\eqref{sumres} and
\begin{gather}
g\in\Pi_r,
\end{gather}
which entails
\begin{gather}
g'\in\Pi_r,\qquad \sigma\in(0,a).
\end{gather}
Also, since
\begin{gather}
g_{\mu}'=-g_\mu -\sigma>-a,
\end{gather}
we obtain an upper bound
\begin{gather}
g_\mu,g_{\mu}'<a-\sigma,\qquad \mu=0,\ldots,7.
\end{gather}

Next, we introduce
\begin{gather}
\label{dgam}
d(\gamma)\equiv\min_{\mu=0,\ldots,7}(g_{\mu}+a),
\\
\label{mgam}
m(\gamma)\equiv \min(a_s, d(\gamma),\sigma(\gamma)),
\end{gather}
where $a_s$ is the smallest shift parameter, cf.~\eqref{as}. We have
\begin{gather}
\label{dga}
d(\gamma)\in(0,a),
\end{gather}
since at least one $g_{\mu}$ must be negative on account of $\langle\zeta,\gamma\rangle<0$, cf.~\eqref{sg}. Note also
that the assumptions made thus far are compatible with each one of the three numbers~$a_s, d(\gamma),\sigma(\gamma)$ in
the interval~$(0,a)$ being smaller than the other ones, and that~$d(\gamma)$ in general dif\/fers from~$d(\gamma')$.

We proceed to determine for which~$\gamma$ we have an equality
\begin{gather}
\label{selfd}
c(\gamma;x)=c(\gamma';x),
\end{gather}
implying positivity of ${\cal I}(\gamma)$, cf.~Lemma~\ref{lemma2.1}. Of course, this is the case when~$\gamma$ is
a~multiple of~$\zeta$, since then $\gamma'$ is equal to~$\gamma$, cf.~\eqref{gampr} and~\eqref{J}. But this is not the
only choice that guarantees the \emph{self-duality} relation~\eqref{selfd}. Indeed, for the f\/irst regime it is plain
that all of the quantities of interest (in particular $c(\gamma;x)$) are~$S_8$-invariant. For a~given $S_8$-orbit we can
f\/ix an ordering choice by requiring
\begin{gather}
\gamma_0\le \gamma_1\le \cdots \le \gamma_7.
\end{gather}
Then the coordinates of the corresponding~$\gamma'$ are oppositely ordered. Consider now the numbers
\begin{gather}
s_\mu\equiv\gamma_\mu -\gamma_{7-\mu}'=\gamma_\mu +\gamma_{7-\mu}+\sigma,\qquad \mu=0,\dots,7.
\end{gather}
They all vanish if and only if $\gamma'$ is related to~$\gamma$ by the reversal permutation~$r_8$. From this we readily
conclude that~${\cal I}(\gamma)$ is positive for a~four-dimensional subset~$\Pi_r^s(1)\subset\Pi_r$ of
self-dual~$\gamma$'s. Indeed, we can freely choose four numbers constrained~by
\begin{gather}
-a<\gamma_0\le \gamma_1\le\gamma_2\le\gamma_3\le -\sigma/2\in(-a/2,0),
\end{gather}
and set
\begin{gather}
\gamma_{7-\mu}\equiv -\gamma_{\mu}-\sigma,\qquad \mu=0,1,2,3.
\end{gather}
Then $\Pi_r^s(1)$ is given by the union of the $S_8$-orbits of the~$\gamma$'s thus obtained.

Turning to the second~$\gamma$-regime, we still have symmetry under $S_4\times S_4$. Fixing the ordering by requiring
\begin{gather}
\label{gord2}
g_0\le g_1\le g_2 \le g_3,\qquad g_4\le g_5\le g_6 \le g_7,
\end{gather}
we introduce
\begin{gather}
t_{\mu}\equiv g_{\mu}-g_{3-\mu}',\qquad t_{\mu +4}\equiv g_{\mu+4}-g_{7-\mu}',\qquad \mu=0,1,2,3.
\end{gather}
These numbers are all zero if\/f~$g$ is related to $g'$ by the permutation $r_4\times r_4$. Hence we deduce again
that~${\cal I}(\gamma)$ is positive for a~four-dimensional subset~$\Pi_r^s(2)\subset\Pi_r$ of self-dual~$\gamma$'s: we
can choose~$g_0$, $g_1$, $g_4$, $g_5$ satisfying
\begin{gather}
-a<g_{\mu}\le g_{\mu +1}\le -\sigma/2\in(-a/2,0),\qquad \mu=0,4,
\end{gather}
and let
\begin{gather}
\label{g2}
g_{3-\mu}\equiv -g_{\mu}-\sigma,\qquad g_{7-\mu}\equiv -g_{\mu+4}-\sigma,\qquad \mu=0,1.
\end{gather}
Then $\Pi_r^s(2)$ is given by the union of the $S_4\times S_4$-orbits of the resulting~$g$'s.

We now focus on the operator~$I(\gamma)$~\eqref{I}. Clearly, it is also a~complete HS operator, and for
$x,y\in(0,\pi/2r)$ its kernel is positive. Thus the same is true for the trace class operator~$T(\gamma)$~\eqref{T},
whose kernel equals
\begin{gather}
\label{Tgker}
T(\gamma;x,y)=w(\gamma;x)^{1/2}\int_{0}^{\pi/2r} dz{\cal S}(\sigma;x,z)w(\gamma';z){\cal
S}(\sigma;z,y)w(\gamma;y)^{1/2}.
\end{gather}
Hence we may and will assume that the $T(\gamma)$-eigenvectors~$e(\gamma)$ are real-valued functions~$e(\gamma;x)$
on~$[0,\pi/2r]$. Moreover, since they belong to the range of~$T(\gamma)$, it is clear that these functions are
restrictions of functions that are real-analytic on~${\mathbb R}$, odd and $\pi/r$-periodic. Furthermore, they have
zeros for~$x\equiv 0 \pmod{\pi/2r}$. (The latter zeros and oddness arise from the double zeros of~$w(\gamma;x)$
for~$x=k\pi/2r,k\in{\mathbb Z}$.)

It follows from the above that the kernel~\eqref{Tgker} can be rewritten as
\begin{gather}
T(x,y)=\sum\limits_{n=0}^{\infty}\lambda_n^2e_n(x)e_n(y),
\end{gather}
where $\{e_n(\gamma;x)\}_{n\in{\mathbb N}}$ is an ONB consisting of $T(\gamma)$-eigenfunctions and
the~$\gamma$-dependence is suppressed. From now on we f\/ix the sign of the real-valued functions~$e_n(\gamma;x)$~by
requiring
\begin{gather}
\label{signen}
e_n(x)=p_nx^{k_n}(1+O(x)),\qquad p_n>0,\qquad x\to 0.
\end{gather}
Then the only ambiguity left in the choice of the eigenfunctions~$e_n(\gamma;x)$ arises from eventual degeneracies in
the spectrum of~$T(\gamma)$. When~$\gamma$ and $\gamma'$ are not related by a~permutation, then~$I(\gamma)$ need not be
self-adjoint. However, in case degeneracy occurs, we may and shall relate the choice of bases by requiring
\begin{gather}
\label{bases}
I(\gamma)=\sum\limits_{n\in{\mathbb N}}\lambda_n (e_n(\gamma'),\cdot)e_n(\gamma),\qquad \lambda_0\ge \lambda_1\ge
\lambda_2\ge \cdots >0.
\end{gather}
(To show that this singular value decomposition does not clash with the sign convention~\eqref{signen}, we can appeal to
the HS-continuity on the connected set $\Pi_r$ following from Lemma~\ref{lemma2.1}, combined with positivity
of~$I(\gamma)$ for~$\gamma$ in the subsets~$\Pi_r^s(j)$, $j=1,2$.)

As mentioned before, we are unable to rule out degenerate $T(\gamma)$-eigenvalues in general. However, the largest
eigenvalue~$\lambda_0$ \emph{is} nondegenerate. This follows from the Krein--Rutman (gene\-ra\-lized Perron--Frobenius)
theorem~\cite{Hel13}, which also implies
\begin{gather}
\label{e0pos}
e_0(\gamma;x)>0,\qquad \forall\, x\in(0,\pi/2r).
\end{gather}

We proceed to scrutinize the $T(\gamma)$-eigenfunctions. To this end we observe that we have
\begin{gather}
\label{enk}
e_n=\lambda_n^{-2k} T^ke_n, \qquad k\in{\mathbb N}.
\end{gather}
Thus the eigenfunctions belong to the range of~$T^k$ for any $k\in{\mathbb N}$. To exploit this property, we begin~by
characterizing various operator ranges. First we consider functions of the form
\begin{gather}
\label{gform}
g(x)\equiv(w(\gamma';\cdot)^{-1/2}I(\gamma')f)(x)=\int_{0}^{\pi/2r} {\cal S}(\sigma;x,y)w(\gamma;y)^{1/2}f(y)dy,\qquad
f\in{\cal H}.
\end{gather}
The function $w(\gamma;x)$ (given by~\eqref{w},~\eqref{c}) is a~meromorphic, even and $\pi/r$-periodic function with
pole locations
\begin{gather}
p_{\delta,k,l}(\gamma_{\mu})\equiv i\delta (\gamma_{\mu}+a+ka_++la_-) \pmod{\pi/r},\\
 \delta=+,-,\qquad \mu=0,\ldots,7,\qquad k,l\in{\mathbb N},\nonumber
\end{gather}
at a~minimal distance $d(\gamma)\in(0,a) $ from the real axis, cf.~\eqref{dga}. Thus the
function~$w(\gamma;y)^{1/2}f(y)$ belongs to~$L^1([0,\pi/2r],dy)$. In view of~\eqref{cSnew}, this entails that~$g(x)$
extends to a~holomorphic function in the space~$S_{\sigma}$, where
\begin{gather}
\label{St}
S_t\equiv \{F(x)\; {\rm holomorphic\; for}\; |\operatorname{Im} x|<t\,|\, F(x)=F({-}x),F(x\!+\!\pi/r)=F(x)\},\ \
 t>0.   \!\!\!\!
\end{gather}

Consider next
\begin{gather}
h(x)\equiv (w(\gamma;\cdot)^{-1/2}T(\gamma)f)(x)= \int_{0}^{\pi/2r}{\cal S}(\sigma;x,y)w(\gamma';y)g(y)dy,\qquad |\operatorname{Im} x|<\sigma.
\end{gather}
Choosing $ x\in(0,\pi/2r)$, we get two simple poles of~${\cal S}(\sigma;x,y)$ at distance~$\sigma$ from the integration
contour~$[0,\pi/2r]$ in the~$y$-plane, located at (cf.~\eqref{cS})
\begin{gather}
y_{\delta}=x +i\delta\sigma,\qquad \delta =+,-.
\end{gather}
If we now choose~$\operatorname{Re} x$ equal to a~f\/ixed~$p\in(0,\pi/2r)$, then we may let~$\operatorname{Im} x$ increase from 0
to~$\sigma$, so that the pole at~$y=y_-$ converges to~$ p$. But we are free to indent the integration contour upwards
at~$p$, so long as this indentation does not lead out of the analyticity region of~$w(\gamma';y)g(y)$. (Note the contour
gets pinched for $p=0$ and $p=\pi/2r$, which is why we exclude the interval endpoints.) Hence we can let $\operatorname{Im} x$
increase to $2\sigma$. Likewise, we can let $\operatorname{Im} x$ decrease beyond $-\sigma$, as long as~$\operatorname{Im}
x>-2\sigma$.

As a~result, we see that~$h(x)$ is holomorphic in the rectangle~$\operatorname{Re} x\in(0,\pi/2r),\operatorname{Im} x\in
(-2\sigma,2\sigma)$. As before, $h(x)$ is also holomorphic in the strip~$|\operatorname{Im} x|<\sigma$, even and
$\pi/r$-periodic, so it follows that~$h(x)$ is holomorphic in the rectangles~$\operatorname{Im} x\in(-2\sigma,2\sigma),\operatorname{Re} x\in(m\pi/2r,(m+1)\pi/2r), m\in{\mathbb Z}$.

A moment's thought now shows that the above reasoning can be iterated to obtain the following lemma.

\begin{Lemma}
\label{lemma2.2}
The range of~$T(\gamma)^k$, $\gamma\in\Pi_r$, $k\in{\mathbb N}^*\!\equiv{\mathbb N}\setminus\{0\}$, consists of
functions~$w(\gamma;x)^{1/2}F(x)$, where~$F(x)$ belongs to the space~$S_{\sigma}$~\eqref{St} and is holomorphic in the
rectangles
\begin{gather}
|\operatorname{Im} x|<2k\sigma,\qquad \operatorname{Re} x\in(m\pi/2r,(m+1)\pi/2r),\qquad m\in{\mathbb Z}.
\end{gather}
\end{Lemma}

\subsection[Holomorphy of $H_n(\gamma;x)$ for~$|\operatorname{Im}  x|<\sigma+a_s$]{Holomorphy of $\boldsymbol{H_n(\gamma;x)}$ for~$\boldsymbol{|\operatorname{Im}  x|<\sigma+a_s}$}\label{section2.2}

By virtue of~\eqref{enk}, Lemma~\ref{lemma2.2} leads to the conclusion that the functions~$F_n(\gamma;x)$ def\/ined
by~\eqref{Fn} (and with phases and ordering determined via~\eqref{signen},~\eqref{bases}) are even, $\pi/r$-periodic and
holomorphic in the cut plane~${\mathbb C}_{\sigma}$ given by~\eqref{Ccut}.

We continue to study their behavior at the cuts. By evenness and $\pi/r$-periodicity it suf\/f\/ices to consider the two
cuts~$x=it, x=it+\pi/2r,t\ge\sigma$. To this end we choose $f(y)=w(\gamma;y)^{1/2}F_n(\gamma;y)$ in~\eqref{gform}. From
the base f\/ixing~\eqref{signen},~\eqref{bases} we then deduce that we have
\begin{gather}
\label{FnF}
\lambda_n F_n(\gamma';x)= \int_{0}^{\pi/2r} {\cal S}(\sigma;x,y)w(\gamma;y)F_n(\gamma;y)dy,\qquad |\operatorname{Im} x|<\sigma,
\end{gather}
and that this equation is also valid when we interchange~$\gamma$ and $\gamma'$.

To study the two cuts, where contour pinching occurs, it is expedient to switch to a~dif\/ferent integration interval.
Specif\/ically, using evenness and $\pi/r$-periodicity of the integrand, we may and shall start from
\begin{gather}
\label{gn}
\lambda_n F_n(\gamma';x)= \frac12 \int_{I(s)} {\cal S}(\sigma;x,y)w(\gamma;y)F_n(\gamma;y)dy,\qquad |\operatorname{Im}
x|<\sigma,
\end{gather}
where
\begin{gather}
I(s)\equiv [-\pi/4r +s,3\pi/4r+s],\qquad s\in[0,\pi/4r],
\end{gather}
and the shift parameter~$s$ is at our disposal. The latter will be chosen equal to 0 or $\pi/4r$ depending on $\operatorname{Re} x$, so as to avoid poles crossing the contour ends. (To be sure, this minor nuisance might be discarded~by
changing to circle integrals via the substitution $z=\exp (2iry)$. However, this setting gives rise to new bookkeeping
trouble we prefer to avoid.)

To clarify the analytic behavior in~$x$ at the cuts, the obvious choice is~$s=0$. Choosing f\/irst~$x=it$,
$t\in(0,\sigma)$, and letting~$t$ increase to~$\sigma$, the two poles at
\begin{gather}
y=\pm(x-i\sigma),
\end{gather}
collide at the origin. But~$w(\gamma;y)$ has a~double zero for~$y=0$, so that~$F_n(\gamma';it)$ has a~f\/inite limit
for~$t\uparrow \sigma$. In particular, it follows that~$F_n(\gamma';x)$ cannot have a~pole at~$x=i\sigma$. Choosing
now~$x=\pi/2r+it$, $t\in(0,\sigma)$, and letting~$t$ increase to~$\sigma$, the two poles at
\begin{gather}
y=x-i\sigma,\qquad y=\pi/r -x+i\sigma,
\end{gather}
collide at~$y=\pi/2r$. Since~$w(\gamma;y)$ has a~double zero for~$y=\pi/2r$, too, we deduce that~$F_n(\gamma';x)$ cannot
have a~pole at~$x=\pi/2r+i\sigma$ either.

Next, we intend to show that~$F_n(\gamma';x)$ is actually holomorphic for~$|\operatorname{Im} x|<\sigma +m(\gamma)$,
cf.~\eqref{mgam}. Recalling~\eqref{St}, we can reformulate this aim as the following lemma.

\begin{Lemma}
\label{lemma2.3}
The function~$F_n(\gamma';x)$ continues analytically to a~function in~$S_{\sigma+m(\gamma)}$.
\end{Lemma}
\begin{proof}
We use~$m:= m(\gamma)$ in this proof. By evenness and $\pi/r$-periodicity in~$x$ it suf\/f\/ices to study what happens when
we f\/ix~$\operatorname{Re} x\in [0,\pi/2r]$ and let $\operatorname{Im} x$ increase, beginning with~$\operatorname{Im} x-\sigma\in(-m/4,0)$. To
ensure that not more than two poles cross the contour, we choose~$s=0$ for $\operatorname{Re} x\in[0,\pi/6r]\cup
[\pi/3r,\pi/2r]$ and~$s=\pi/4r$ for $\operatorname{Re} x\in(\pi/6r,\pi/3r)$ (say). Indeed, with this choice only two poles of
the integrand are at a~distance less than $m/4$ from the contour.

We now shift up the contour by~$m/2$. Then we only pass the pertinent simple pole at~$y\equiv i\sigma-x \pmod {\pi/r}$,
picking up a~residue term. (The vertical parts of the rectangular closed contour at issue cancel
by~$\pi/r$-periodicity.) As a~result we obtain a~new representation
\begin{gather}
\label{gns}
\lambda_nF_n(\gamma';x)=\rho_n^{(1)}(\gamma';x)+ \frac12 \int_{C^+(s)}{\cal S}(\sigma;x,y)w(\gamma;y)F_n(\gamma;y)dy,\\
 \operatorname{Im} x-\sigma\in(-m/4,0),\nonumber
\\
\rho_n^{(1)}(\gamma';x):= -i\pi r_0 G(2i\sigma -ia)G(2x-ia)G(-2x+2i\sigma-ia)\nonumber\\
\hphantom{\rho_n^{(1)}(\gamma';x):=}{}\times w(\gamma;x-i\sigma)F_n(\gamma;x-i\sigma),\label{res1}
\end{gather}
where~$r_0$ denotes the residue of~$G(z)$ at the pole~$z=-ia$, and where $C^+(s)$ denotes the interval~$I(s)$ shifted up
by~$m/2$. Using~\eqref{w} and~\eqref{c}, the residue term can be rewritten as
\begin{gather}
\rho_n^{(1)}(\gamma';x)   =   -i\pi r_0 G(2i\sigma -ia)G(2x-ia)G(2i\sigma-2x+ia)
\nonumber
\\
\hphantom{\rho_n^{(1)}(\gamma';x)   =}{}
\times F_n(\gamma;x-i\sigma) \prod\limits_{\mu=0}^7G(x-i\sigma+i\gamma_{\mu})G(-x-i\gamma_{\mu}').
\end{gather}
From this we see that it gives rise to a~function that is holomorphic in the strip~$\operatorname{Im} x-\sigma\in(-m,\sigma)$,
save for those poles at~$x\equiv i(a-\gamma_{\mu}')\pmod{\pi/r}$ that belong to this strip. (They arise from the
factors~$G(-x-i\gamma_{\mu}')$, all other factors being holomorphic in the strip.) Since we have
\begin{gather}
a-g_{\mu}'=a+g_{\mu}+\sigma\ge d(\gamma)+\sigma,
\end{gather}
it follows that~$\rho_n^{(1)}(\gamma';x)$ is holomorphic for~$\operatorname{Im} x-\sigma\in(-m,m)$.

Next, we consider the contour integral in~\eqref{gns}. It extends to a~function that is holomorphic in the strip~$\operatorname{Im} x-\sigma\in(-m/2,m/2)$, since this restriction ensures that the relevant pole at $i\sigma-x+ia_s \pmod{\pi/r}$
stays above and the pole at~$x-i\sigma$ stays below the contour. Therefore, we have now shown that~$F_n(\gamma';x)$
extends to a~function in~$S_{\sigma+m/2}$.

To prove the stronger result~$F_n(\gamma';x)\in S_{\sigma+m}$, we f\/irst choose~$\operatorname{Im} x-\sigma\in (m/4, m/2)$ in
the representation~\eqref{gns}. Then we push the contour down by~$m/2$, picking up a~residue term at the
pole~$x-i\sigma$ that is again given by~\eqref{res1} (by virtue of evenness and $\pi/r$-periodicity in~$y$). Thus we
wind up with
\begin{gather}
\label{gns2}
\lambda_nF_n(\gamma';x)=2\rho_n^{(1)}(\gamma';x)+ \frac12 \int_{I(s)}{\cal S}(\sigma;x,y)w(\gamma;y)F_n(\gamma;y)dy,\\
  \operatorname{Im} x-\sigma\in(m/4,m/2). \nonumber
\end{gather}
The integral now continues analytically to~$\operatorname{Im} x-\sigma\in(0,a_s)$. Since~$a_s\ge m$
and~$\rho_n^{(1)}(\gamma';x)$ is holomorphic for~$\operatorname{Im} x-\sigma\in(-m,m)$ (as already shown), the lemma follows.
\end{proof}

Note that we need only interchange~$\gamma$ and $\gamma'$ to arrive at the conclusion
\begin{gather}
F_n(\gamma;x)\in S_{\sigma+m(\gamma')}.
\end{gather}
In order to improve this result, it is convenient to get rid of~$\gamma$-dependent poles. This is the reason we now turn
to the functions~$H_n(\gamma;x)$ def\/ined by~\eqref{Hn}. From~\eqref{E} we see that~$P(\gamma;x)$ is an entire, even and
$\pi/r$-periodic function whose zeros are at a~minimal distance~$\sigma+d(\gamma')$ from the real line
(recall~$\gamma'_\mu$ equals~$-\gamma_\mu-\sigma$). At this stage, therefore, we can only conclude that~$H_n(\gamma;x)$
is holomorphic for~$\operatorname{Re} x\ne k\pi/2r$, $k\in{\mathbb Z}$, and satisf\/ies
\begin{gather}
H_n(\gamma;x)\in S_{\sigma+m(\gamma')}.
\end{gather}

We proceed to prove $H_n(\gamma';x)$ belongs to $S_{\sigma+a_s}$. To this end, we multiply~\eqref{gns2}
by~$2P(\gamma';x)$ and write the result as (cf.~\eqref{HF})
\begin{gather}
\label{HH}
2 \lambda_nH_n(\gamma';x)=\mu_0(\gamma;x)H_n(\gamma;x-i\sigma)+ P(\gamma';x) \int_I{\cal
S}(\sigma;x,y)m_H(\gamma;y)H_n(\gamma;y)dy.
\end{gather}
Here we may take $\operatorname{Im} x-\sigma\in(0,m(\gamma))$, and from now on we shall use the integration intervals
\begin{gather}
\label{intI}
I=I(0)=[-\pi/4r,3\pi/4r],\qquad C^+=C^+(0),
\end{gather}
it being clear from context when we need a~shift by~$\pi/4r$ to avoid multiple pole crossings. Also, the residue term
now involves the multiplier
\begin{gather}
\mu_0(\gamma;x)\equiv -4i\pi r_0G(2i\sigma -ia)G(2x-ia) G(-2x+2i\sigma-ia)P(\gamma';x)m_H(\gamma;x-i\sigma)
\nonumber\\
\hphantom{\mu_0(\gamma;x)}{}=-4i\pi r_0G(2i\sigma
-ia)G(2x-ia)G(-2x+2i\sigma+ia)\prod\limits_{\mu}\frac{E(-x+i\gamma_{\mu}')}{E(-x+i\sigma-i\gamma_\mu)}.\label{mu0}
\end{gather}
This meromorphic function is holomorphic in the strip~$\operatorname{Im} x\in(\sigma,\sigma+a)$, whereas the integral is
holomorphic for~$\operatorname{Im} x\in(\sigma, \sigma+a_s)$. Moreover, the factor~$H_n(\gamma;x-i\sigma)$ is holomorphic
for~$\operatorname{Im} x\in(0,2\sigma)$. As a~result,~\eqref{HH} entails that~$H_n(\gamma';x)$ belongs
to~$S_{\min(\sigma+a_s,2\sigma)}$. For the case~$\sigma\ge a_s$, therefore, we obtain the announced
relation~$H(\gamma';x)\in S_{\sigma +a_s}$.

Turning to the remaining case~$\sigma<a_s$, there exists a~unique $N\in{\mathbb N}^*$ such that
\begin{gather}
\label{N}
N\sigma<a_s,\qquad (N+1)\sigma \ge a_s.
\end{gather}
Now we already know~$H(\gamma';x)\in S_{2\sigma}$, so that also~$H(\gamma;x)\in S_{2\sigma}$. Therefore the
factor~$H_n(\gamma;x-i\sigma)$ in~\eqref{HH} is holomorphic for~$\operatorname{Im} x\in(-\sigma,3\sigma)$. Hence~\eqref{HH}
entails that~$H_n(\gamma';x)$ belongs to~$S_{\min(\sigma+a_s,3\sigma)}$. Repeating this reasoning~$N-1$ times, we obtain
the following lemma.

\begin{Lemma}
\label{lemma2.4}
The function~$H_n(\gamma;x)$~\eqref{Hn} is holomorphic for~$\operatorname{Re} x\ne k\pi/2r$, $k\in{\mathbb Z}$, and satisfies
\begin{gather}
\label{Hnas}
H_n(\gamma;x)\in S_{\sigma+a_s}.
\end{gather}
\end{Lemma}

In the next two subsections we assume $a_s<a_l$, and we shall prove that we may replace~$a_s$ by~$a_l$ in~\eqref{Hnas}.
This involves a~major additional ef\/fort. We conclude this subsection by deriving one result for the special
case~$a_s=a_l=a$, namely, the identity
\begin{gather}
\label{Hnidsp}
H_n(\gamma';ia+\tau\pi/2r)=e^{4r(\sigma -a)}H_n(\gamma';\tau\pi/2r)
\prod\limits_{\mu=0}^7\big(1-(-)^{\tau}\exp(2r\gamma_{\mu}')\big),\qquad \tau=0,1. 
\end{gather}
The derivation yields the simplest example for the reasoning we shall use to obtain the general
identities~\eqref{Hnident} of Theorem~\ref{theorem2.9}, unencumbered by the profusion of technicalities we are unable to
avoid for the general case.

We begin by taking $x=ia+\tau\pi/2r$ in~\eqref{HH}. From~\eqref{mu0} we read of\/f that~$\mu_0$ vanishes for these values
of~$x$. A~key point is now that we have an identity
\begin{gather}
{\cal S}(\sigma;ia+\tau\pi/2r,y)=e^{4r(\sigma -a)}{\cal S}(\sigma;\tau\pi/2r,y),\qquad \tau=0,1,\qquad(a_s=a_l).
\end{gather}
(This can be checked from the def\/inition~\eqref{cS} of~${\cal S}$ by using the A$\Delta$Es satisf\/ied by the~$G$- and
$R_\delta$-functions, cf.~\eqref{Gades},~\eqref{Rade}.) Moreover, from~\eqref{FnF},~\eqref{Hn} and~\eqref{HF} we obtain
\begin{gather}
\label{HHsp}
2 \lambda_nH_n(\gamma';\tau\pi/2r)= P(\gamma';\tau\pi/2r) \int_I{\cal
S}(\sigma;\tau\pi/2r,y)m_H(\gamma;y)H_n(\gamma;y)dy.
\end{gather}
Combining these formulas, we get
\begin{gather}
H_n(\gamma';ia+\tau\pi/2r)=e^{4r(\sigma
-a)}H_n(\gamma';\tau\pi/2r)\frac{P(\gamma';ia+\tau\pi/2r)}{P(\gamma';\tau\pi/2r)},\qquad (a_s=a_l).
\end{gather}
Finally, recalling the def\/inition~\eqref{Pgam} of~$P(\gamma;x)$, we obtain~\eqref{Hnidsp} by using the A$\Delta$Es
obeyed by the~$E$- and $G_t$-function, cf.~\eqref{Eades} and~\eqref{Gtade}, resp.

\subsection[Holomorphy of $H_n(\gamma;x)$ for~$|\operatorname{Im} x|<\sigma+a_l$ and $\protect{a_l\in(a_s,2a_s]}$]{Holomorphy of $\boldsymbol{H_n(\gamma;x)}$ for~$\boldsymbol{|\operatorname{Im} x|<\sigma+a_l}$ and $\boldsymbol{a_l\in(a_s,2a_s]}$}\label{section2.3}

In Section~\ref{section4} we shall see that Lemma~\ref{lemma2.4} already suf\/f\/ices to get a~grip on the Hilbert space
version~$\hat{{\cal A}}_s(\gamma)$ of the A$\Delta$O ${\cal A}_s(\gamma;x)$. To handle~${\cal A}_l(\gamma;x)$, our next
aim is to show that~$H_n(\gamma;x)$ belongs to~$S_{\sigma+a_l}$. This involves similar ingredients, but also quite a~few
new ones, in particular various non-trivial identities. In order to retain a~clear view on our line of reasoning, we
have opted for a~separate treatment of the special case~$a_l\in(a_s, 2a_s]$ in this subsection, before handling the
general case in the next one. In the same vein, however, we already state and prove the general version of the four
auxiliary Lemmas~\ref{lemma2.5}--\ref{lemma2.8} at the point where the need arises for the special case at hand.

Since we assume~$a_s<a_l$ in the remainder of this section, there is a~unique~$L\in{\mathbb N}^*$ satisfying
\begin{gather}
\label{Ldef}
a_l>La_s,\qquad a_l\le (L+1)a_s.
\end{gather}
As we have already seen, the representation~\eqref{HH} holds true for~$\operatorname{Im} x\in(\sigma,\sigma+a_s)$. Just as in
the proof of Lemma~\ref{lemma2.3}, we can now show
\begin{gather}
\label{Haux}
H_n(\gamma';x)\in S_{\sigma +a_s+\min (m(\gamma),a_l-a_s)},
\end{gather}
by letting the simple poles at the two relevant locations among
\begin{gather}
y\equiv\pm (x-i\sigma-ia_s) \pmod{\pi/r},
\end{gather}
pass the contour in several stages. This results in a~representation
\begin{gather}
2 \lambda_nH_n(\gamma';x)=\sum\limits_{\ell =0}^1\mu_{\ell}(\gamma;x)H_n(\gamma;x-i\sigma-i\ell a_s)
\nonumber\\
\hphantom{2 \lambda_nH_n(\gamma';x)=}{} + P(\gamma';x) \int_I{\cal S}(\sigma;x,y)m_H(\gamma;y)H_n(\gamma;y)dy,\label{HH2}
\end{gather}
where we may take $\operatorname{Im} x-\sigma-a_s\in(0,\min (m(\gamma),a_l-a_s))$. The multipliers are given~by
\begin{gather}
\mu_{\ell}(\gamma;x) = -4i\pi r_{\ell}G(2i\sigma +i\ell a_s -ia)G(2x-i\ell a_s-ia) \nonumber\\
\hphantom{\mu_{\ell}(\gamma;x) =}{}
\times G(-2(x-i\sigma)+i\ell
a_s-ia)\xi_{\ell}(\gamma;x),\label{muell}
\end{gather}
with
\begin{gather}
\label{xiell}
\xi_{\ell}(\gamma;x):= P(\gamma';x)m_H(\gamma;x-i\sigma -i\ell a_s).
\end{gather}

As will become clear shortly, we need detailed information about the poles and zeros of the multipliers
$\mu_{\ell}(\gamma;x)$, not only for $\ell=0,1$, but also for $\ell =2,\ldots,L$, with~$L$ given by~\eqref{Ldef}. More
specif\/ically, as we let $\operatorname{Im} x$ increase, the multiplier~$\mu_{\ell}$ arises after the simple poles of ${\cal
S}(\sigma;x,y)$ at~$\pm y\equiv x-i\sigma -i\ell a_s\pmod{\pi/r}$ pass the real line. Therefore, the only relevant poles
and zeros of~$\mu_{\ell}$ are those satisfying~$\operatorname{Im} x -\sigma \ge \ell a_s$. Furthermore, in this section we let
$\operatorname{Im} x$ increase until we reach $\sigma +a_l$, so only f\/initely many of these locations matter. It is expedient
to digress at this point, so as to obtain the pertinent information.

\begin{Lemma}
\label{lemma2.5}
Assuming $a_l>a_s$, define $L\in{\mathbb N}^*$ by~\eqref{Ldef}. The poles of the multipliers
\begin{gather}
\label{multip}
\mu_{\ell}(\gamma;x),\qquad \ell=0,1,\ldots,L,
\end{gather}
given by~\eqref{muell}, \eqref{xiell}, with
\begin{gather}
\label{imres}
\operatorname{Im} x-\sigma \in [\ell a_s, a_l),
\end{gather}
can only occur at the locations
\begin{gather}
\label{crit}
x\equiv i\sigma +i(\ell +m)a_s/2+ia_l/2 \pmod{\pi/2r},\qquad m=0,1,\ldots, L-\ell,\quad m\ne \ell,
\end{gather}
and they are at most simple poles. The multipliers~\eqref{multip} satisfy
\begin{gather}
\label{edgezero}
\mu_{\ell}(\gamma;i\sigma+i\ell a_s+k\pi/2r) =0, \qquad k\in{\mathbb Z},
\end{gather}
and have further relevant zeros at
\begin{gather}
\label{muzer0}
x\equiv i\sigma+i\gamma_{\mu}+ia_l/2+i(2j-1)a_s/2 \pmod{\pi/r},\qquad \mu=0,\ldots,7,\quad j=1,\ldots,\ell,
\\
\label{muzer1}
x\equiv ia+i(\ell+k_1)a_s/2 \pmod{\pi/2r}, \qquad k_1\in{\mathbb N},
\\
\label{muzer2}
x\equiv ia_l +i(\ell+k_2+1)a_s/2 \pmod{\pi/2r}, \qquad k_2\in{\mathbb N},
\end{gather}
provided that these locations obey~\eqref{imres} and are disjoint from~\eqref{crit}. Finally, for the
special~$\sigma$-values
\begin{gather}
\label{sigsp}
\sigma_K:=a-Ka_s/2,\qquad 1\le K\le L,
\end{gather}
we have
\begin{gather}
\label{muvan}
\mu_{\ell}(\gamma;x)= 0,\qquad K\le \ell\le L.
\end{gather}
\end{Lemma}
\begin{proof}
First, using the def\/inition~\eqref{mH} of the multiplier function $m_H$, the identities~\eqref{RR} and
the~$E$-A$\Delta$Es~\eqref{Eades}, we can rewrite $\xi_{\ell}$ as
\begin{gather}
\xi_{\ell}(\gamma;x)=R_s(2(x-i\sigma-i\ell a_s)+ia_s/2)R_l(2(x-i\sigma-i\ell a_s)-ia_l/2)
\nonumber\\
\hphantom{\xi_{\ell}(\gamma;x)=}{} \times\prod\limits_{\mu=0}^7\frac{E(-x-i\sigma -i\gamma_{\mu})}{E(-x+i\sigma+i\ell
a_s-i\gamma_{\mu})}\prod\limits_{j=1}^\ell \frac{1}{G_t(a_l;-x+i\sigma+i\gamma_{\mu}+i(j-1/2)a_s)}.
\end{gather}
Thus we get simple zeros at
\begin{gather}
\label{Rszeros}
x-i\sigma\equiv ija_s/2 \pmod{\pi/2r},\qquad j\in{\mathbb Z},
\end{gather}
due to the $R_s$-factor, and at
\begin{gather}
\label{Rlzeros}
x-i\sigma\equiv i\ell a_s+ija_l/2 \pmod{\pi/2r},\qquad j\in{\mathbb Z},
\end{gather}
due to the $R_l$-factor. Taking into account the zero locations of $E(z)$~\eqref{E} and pole locations of
$G_t(a_l;z)$~\eqref{Gt}, it is straightforward to verify that the product factor in~$\xi_{\ell}$ has no poles in the
half plane $\operatorname{Im} x-\sigma\ge \ell a_s$; for $\ell=0$ it has no zeros in the half plane either, whereas for
$\ell>0$ the zeros~\eqref{muzer0} arise.

Secondly, we note that the poles of the second~$G$-function in~\eqref{muell} are irrelevant, whereas the poles of the
third one are located at
\begin{gather}
x-i\sigma \equiv i\ell a_s/2 +ija_s/2+ika_l/2 \pmod{\pi/2r},\qquad j,k\in{\mathbb N}.
\end{gather}
The poles satisfying~\eqref{imres} are then given~by
\begin{gather}
x-i\sigma \equiv i(\ell +j_1/2)a_s \pmod{\pi/2r},\qquad j_1=0,1,\ldots, 2(L-\ell),
\nonumber\\
x-i\sigma\equiv i(L+1/2)a_s \pmod{\pi/2r}\qquad \big(a_l>(L+1/2)a_s\big),  \label{relpo1}
\end{gather}
and~by
\begin{gather}
\label{relpo2}
x-i\sigma \equiv i(\ell +j_2)a_s/2+ia_l/2 \pmod{\pi/2r},\qquad j_2=0,1,\ldots, L-\ell,
\end{gather}
and these poles are simple unless $(\ell +j_1-j_2)a_s=a_l$. In any case, when we take the zeros~\eqref{Rszeros} of the
$R_s$-factor into account, we can cancel all of the poles~\eqref{relpo1}, and then we are left with the simple
poles~\eqref{relpo2}. For the case~$\ell\le L/2$, however, the poles for $j_2=\ell$ are cancelled by the zeros
in~\eqref{Rlzeros} with $j=1$.

The upshot of this analysis is that the critical pole locations of the multipliers $\mu_{\ell}(\gamma;x)$ are indeed
given by~\eqref{crit}, so we need only verify the remaining zero assertions. The zeros~\eqref{edgezero} arise from the
zeros~\eqref{Rlzeros} by taking~$j=0$. The source of the zeros~\eqref{muzer1},~\eqref{muzer2} is the second~$G$-factor
in~\eqref{muell}. Finally, the~$\sigma$-choices~\eqref{sigsp} imply that the f\/irst~$G$-factor in~\eqref{muell} vanishes
for the~$\ell$-values in~\eqref{muvan}.
\end{proof}

Note that the lemma implies in particular that $\mu_{\ell}(\gamma;x)$ is always regular at $x\equiv i\sigma +i\ell
a_s/2+ia_l/2 \pmod{\pi/2r}$ and vanishes at $x\equiv i\sigma +i\ell a_s \pmod{\pi/2r}$; these locations correspond to
the midpoint and lower edge of the interval in~\eqref{imres}. Also, we are saying `at most simple poles' in the
statement of the lemma, since there may be (nongeneric) cancellation with the zeros~\eqref{muzer0}--\eqref{muzer2}.
Furthermore, for the choices $\sigma=\sigma_K$, the multipliers $\mu_K,\ldots,\mu_L$ vanish identically,
cf.~\eqref{sigsp},~\eqref{muvan}. We point out that the absence of the corresponding poles agrees with the kernel
function reducing to
\begin{gather}
{\cal S}(\sigma_K;x,y)=\prod\limits_{m=1}^K\frac{1}{R_l(x\pm y +i(K+1-2m)a_s/2)},\qquad 1\le K \le L.
\end{gather}
(This identity readily follows from the~$G$-A$\Delta$Es~\eqref{Gades}.)

After this digression, we are prepared to study the three terms on the right-hand side of~\eqref{HH2}. Clearly, the contour integral
term continues to a~function that is holomorphic for
\begin{gather}
\operatorname{Im} x-\sigma\in(a_s,a_l),\qquad (L=1),
\\
\operatorname{Im} x-\sigma\in(a_s,2a_s),\qquad (L>1).
\end{gather}
We proceed to supply full details for the case $L=1$ in this subsection, after which the case of general~$L$ in
Subsection~\ref{section2.4} can be more readily understood. In particular, to unburden our account in the next subsection, we shall
prove some more lemmas for general~$L$ when the need arises for the $L=1$ case at issue in the present subsection, just
as we did in the previous lemma.

For~$L=1$ both pertinent multipliers $\mu_0$ and~$\mu_1$ can have only critical poles for~$\operatorname{Im}
x-\sigma\in(a_s,a_l)$ that are located at $x\equiv i\sigma +ia \pmod{\pi/2r}$. From this and the
representation~\eqref{HH2} it is not hard to see that we have
\begin{gather}
\label{Hna}
H_n(\gamma';x)\in S_{\sigma +a},\qquad (L=1).
\end{gather}
Indeed, recalling~\eqref{Haux}, we can let $\operatorname{Im} x$ increase from $\sigma +a_s$ to $\sigma +a$ without meeting
poles. (If need be, this can be done in several `$\sigma$-steps', cf.~the argument in the paragraph
con\-tai\-ning~\eqref{N}.)

Even though we can now continue~\eqref{HH2} meromorphically to $\operatorname{Im} x-\sigma\in (a_s,a_l)$, it may seem that we
cannot avoid the simple poles at $x=i\sigma +ia$ and $x=\pi/2r+i\sigma +ia$, which are present in $\mu_0(\gamma;x)$ and
$\mu_1(\gamma;x)$ for generic parameters. Indeed, since~\eqref{Hna} is also valid for $H_n(\gamma;x)$, we do obtain
well-def\/ined function values~$H_n(\gamma;ia),H_n(\gamma;ia-ia_s)$ and~$H_n(\gamma;\pi/2r+ia),H_n(\gamma;\pi/2r+ia-ia_s)$
for these~$x$-choices in~\eqref{HH2}, but at f\/irst sight there is no indication that these values are related in such
a~way that the sum of the residues at the relevant pole vanishes.

In fact, however, the residue sum does vanish. This hinges on identities satisf\/ied by the kernel function, namely the
case $j=1$ of the remarkable identities collected in the next lemma. (They are needed to handle general~$L$.)

\begin{Lemma}
\label{lemma2.6}
Letting $j\in{\mathbb N}^*$, and introducing
\begin{gather}
\label{xtau}
x_0:=ia_l/2,\qquad x_1:=\pi/2r+ ia_l/2,
\\
\label{Dj}
D_j(\sigma;x,y):= {\cal S}(\sigma; x,y)-\exp(4jr(\sigma -a)) {\cal S}(\sigma; x-ija_s,y),
\end{gather}
we have
\begin{gather}
\label{cruxid}
D_j(\sigma;x_{\tau}+ija_s/2,y)=0,\qquad \tau=0,1.
\end{gather}
\end{Lemma}

\begin{proof}
To prove these identities, we invoke the def\/inition~\eqref{cS} of the kernel function and
the~$G$-A$\Delta$Es~\eqref{Gades}. They enable us to write
\begin{gather}
\frac{{\cal S}(\sigma; x_{\tau}+ija_s/2,y)}{{\cal S}(\sigma; x_{\tau}-ija_s/2,y)}
=\prod\limits_{m=1}^j\frac{R_l(x_{\tau}\pm y-ia +i\sigma+i(j+1)a_s/2 -ima_s)}{R_l(x_{\tau}\pm y +ia-i\sigma-i(j+1)a_s/2
+ima_s)}.
\end{gather}
When we now use evennness and~$\pi/r$-periodicity of $R_l$, and invoke~\eqref{Rade}, then we see this equals $\exp
(4jr(\sigma -a))$.
\end{proof}

Continuing with the special case $L=1$ under scrutiny, there are now three subcases, namely $\sigma>a- a_s$, $\sigma<a-
a_s$ and $\sigma=a-a_s $. In each of these we have
\begin{gather}
\label{mu0zero}
\mu_0(\gamma;ia)=\mu_0(\gamma;\pi/2r+ia)=0.
\end{gather}
(These zeros come from~\eqref{muzer1} with $k_1=0$; they are always relevant and disjoint from the
locations~\eqref{crit}.) Focusing on~$x$-values near $ia$ until further notice, we can invoke~\eqref{HH} in the f\/irst
subcase. We telescope this representation by using $D_1$~\eqref{Dj}:
\begin{gather}
2\lambda_nH_n(\gamma';x)=\mu_0(\gamma;x)H_n(\gamma;x-i\sigma) +P(\gamma';x)
\nonumber\\
\hphantom{2\lambda_nH_n(\gamma';x)=}{}
\times\left(e^{4r(\sigma -a)}\int_I{\cal S}(\sigma;x-ia_s,y)m_H(\gamma;y)H_n(\gamma;y)dy\right.\nonumber\\
\left.
\hphantom{2\lambda_nH_n(\gamma';x)=}{}
+\int_I
D_1(\sigma;x,y)m_H(\gamma;y)H_n(\gamma;y)dy\right).
\end{gather}
Next, we use an immediate consequence of~\eqref{gn}, viz.,
\begin{gather}
\int_I {\cal S}(\sigma;x-ia_s,y)m_H(\gamma;y)H_n(\gamma;y)dy=\frac{2\lambda_nH_n(\gamma';x-ia_s)}{P(\gamma';x-ia_s)},
\end{gather}
to deduce
\begin{gather}
2\lambda_nH_n(\gamma';x)=\mu_0(\gamma;x)H_n(\gamma;x-i\sigma) +2\lambda_nH_n(\gamma';x-ia_s)e^{4r(\sigma
-a)}\frac{P(\gamma';x)}{P(\gamma';x-ia_s)}
\nonumber\\
\hphantom{2\lambda_nH_n(\gamma';x)=}{}
+ P(\gamma';x)\int_I D_1(\sigma;x,y)m_H(\gamma;y)H_n(\gamma;y)dy.\label{Hia3}
\end{gather}

We are now prepared to let~$x$ converge to $ia$. The limit of the~$P$-ratio can be evaluated by using
the~$E$-A$\Delta$Es~\eqref{Eades} and then the $G_t$-A$\Delta$E~\eqref{Gtade}. Since we need limits of more
general~$P$-ratios shortly, we collect them in the following lemma.

\begin{Lemma}
\label{lemma2.7}
Letting $j\in{\mathbb N}^* $, we have
\begin{gather}
\lim_{x\to x_{\tau}}\frac{P(\gamma'; x+ija_s/2)}{P(\gamma'; x-ija_s/2)}\nonumber\\
\qquad{} =\prod\limits_{\mu=0}^7\prod\limits_{m=1}^j
\big(1-(-)^{\tau}\exp[2r(\gamma_{\mu}'+(j+1-2m)a_s/2)]\big),\qquad \tau=0,1.   \label{Pid}
\end{gather}
\end{Lemma}

\begin{proof}
Using the~$P$-def\/inition~\eqref{Pgam} and the~$E$-A$\Delta$Es~\eqref{Eades}, the left-hand side can be written as
\begin{gather}
\label{Gtrat}
\lim_{x\to x_{\tau}}\prod\limits_{\mu=0}^7\prod\limits_{m=1}^j
\frac{G_t(a_l;x-i\gamma_{\mu}'+i(j+1)a_s/2-ima_s)}{G_t(a_l;-x-i\gamma_{\mu}'+i(j+1)a_s/2-ima_s)}.
\end{gather}
Due to our standing assumption~$g\in\Pi_r$, the $G_t$-factors in the numerator have f\/inite limits. From~\eqref{Gtade} we
then deduce equality to the right-hand side.
\end{proof}

It should be noted that~\eqref{Pid} may be false when the~$g$-restriction $g\in\Pi_r$ is dropped. Indeed, an extra sign
appears for each $G_t$-ratio in~\eqref{Gtrat} with both $G_t$'s having a~pole.

Taking $x\to ia$, we obtain from~\eqref{mu0zero} and Lemmas~\ref{lemma2.6},~\ref{lemma2.7} (with $j=1$, $\tau=0$) the identity
\begin{gather}
\label{Hnrel}
H_n(\gamma';ia)=\exp(4r(\sigma -a))H_n(\gamma';ia-ia_s)\prod\limits_{\mu=0}^7(1-\exp(2r\gamma_{\mu}')).
\end{gather}
We proceed to show that this simple relation holds true in the remaining two subcases as well. For the second subcase
$\sigma<a-a_s $, we telescope~\eqref{HH2} for~$x$ near $ia$ as
\begin{gather}
2\lambda_nH_n(\gamma';x)=\sum\limits_{\ell=0}^1\mu_{\ell}(\gamma;x)H_n(\gamma;x-i\sigma-i\ell a_s) +P(\gamma';x)
\nonumber\\
\hphantom{2\lambda_nH_n(\gamma';x)=}{}
\times\left(e^{4r(\sigma -a)}\int_I{\cal S}(\sigma;x-ia_s,y)m_H(\gamma;y)H_n(\gamma;y)dy\right.
\nonumber\\
\left. \hphantom{2\lambda_nH_n(\gamma';x)=}{}
+\int_I
D_1(\sigma;x,y)m_H(\gamma;y)H_n(\gamma;y)dy\right).\label{HH2ia1}
\end{gather}
Using next~\eqref{HH}, this implies
\begin{gather}
2\lambda_nH_n(\gamma';x)=\sum\limits_{\ell=0}^1\mu_{\ell}(\gamma;x)H_n(\gamma;x-i\sigma-i\ell a_s)
\nonumber\\
\hphantom{2\lambda_nH_n(\gamma';x)=}{}
+\big(2\lambda_nH_n(\gamma';x-ia_s)-\mu_0(\gamma;x-ia_s)H_n(\gamma;x-ia_s-i\sigma)\big)
\\
\hphantom{2\lambda_nH_n(\gamma';x)=}{}
\times {} e^{4r(\sigma
-a)}\frac{P(\gamma';x)}{P(\gamma';x-ia_s)}
+ P(\gamma';x)\int_I D_1(\sigma;x,y)m_H(\gamma;y)H_n(\gamma;y)dy.\nonumber
\end{gather}
When we now let~$x$ converge to $ia$, we can use once more~\eqref{mu0zero} and Lemmas~\ref{lemma2.6},~\ref{lemma2.7} (with $j=1$, $\tau=0$),
together with the limit
\begin{gather}
\label{muqlim}
\lim_{x\to ia}\left(\mu_1(\gamma;x)-e^{4r(\sigma-a)}\mu_0(\gamma;x-ia_s)\frac{P(\gamma';x)}{P(\gamma';x-ia_s)}\right)=0.
\end{gather}
The latter shall be proved shortly (among more general limits), cf.~\eqref{muspqu} with $k=0,j=1$ and $\tau=0$. Taking
it for granted, it is clear that the relation~\eqref{Hnrel} follows again.

Continuing with the last subcase~$\sigma =a-a_s=\sigma_2$ (cf.~\eqref{sigsp}), we obtain from~\eqref{edgezero} the edge
zeros
\begin{gather}
\label{nongen}
\mu_1(\gamma;ia)=0,\qquad \mu_0(\gamma;ia-ia_s)=0,\qquad (\sigma =\sigma_2),
\end{gather}
in addition to~$\mu_0(\gamma;ia)=0$, cf.~\eqref{mu0zero}. Recalling the proof of Lemma~\ref{lemma2.3}, we deduce that
for~$x$ suf\/f\/iciently close to $ia$ we have a~representation
\begin{gather}
2\lambda_nH_n(\gamma';x)=\frac12 \mu_1(\gamma;x)H_n(\gamma;x-i\sigma-ia_s)+\mu_0(\gamma;x)H_n(\gamma;x-i\sigma)
\\
\hphantom{2\lambda_nH_n(\gamma';x)=}{}
+P(\gamma';x) \int_{C^+}\!\!\big(e^{4r(\sigma -a)} {\cal S}(\sigma_2;x-ia_s,y)+
D_1(\sigma;x,y)\big)m_H(\gamma;y)H_n(\gamma;y)dy.\nonumber
\end{gather}
Note that we need not worry any more about $\gamma_{\mu}$-dependent poles thanks to our switch from~$F_n$ to~$H_n$; now,
however, we should choose~$C^+$ equal to the interval~$I$ shifted up by a~distance ${\le}\min(a_s,\sigma)/2$ that
ensures~$C^+$ stays below the poles at~$y\equiv -x+i\sigma +ia_l\pmod{\pi/2r}$, cf.~\eqref{intI}. Choosing the same
contour, we also have
\begin{gather}
2\lambda_nH_n(\gamma';x-ia_s)=\frac12 \mu_0(\gamma;x-ia_s)H_n(\gamma;x-ia_s-i\sigma)
\nonumber\\
\hphantom{2\lambda_nH_n(\gamma';x-ia_s)=}{}
+P(\gamma';x-ia_s)\int_{C^+}{\cal S}(\sigma_2;x-ia_s,y)m_H(\gamma;y)H_n(\gamma;y)dy.
\end{gather}
Using the zeros~\eqref{nongen} and~\eqref{mu0zero}, we can now proceed as in the f\/irst subcase to reobtain the
relation~\eqref{Hnrel}.

The upshot is that we have now proved the relation~\eqref{Hnrel} for the case~$L=1$. We shall use this relation (with
$\gamma'\to\gamma$) to analyze the behavior of the sum on the right-hand side of~\eqref{HH2} as~$x$ converges
to~$i\sigma+ia$. Another key result for doing so is
\begin{gather}
\label{mur}
\lim_{x\to i\sigma +ia}\frac{\mu_1(\gamma;x)}{\mu_0(\gamma;x)}=- \exp(4r(\sigma -a))
\prod\limits_{\mu=0}^7(1-\exp(2r\gamma_{\mu})), \qquad (\sigma \ne \sigma_1=a_l/2).
\end{gather}
As before, we proceed to collect more general limits, needed for~$L>1$.

\begin{Lemma}
\label{lemma2.8}
Assume $k\in{\mathbb N}$, $j\in{\mathbb N}^*$ and $k+j\le L$, with~$L$ given by~\eqref{Ldef}. Provided
\begin{gather}
\label{sigcon}
\sigma<a-(k+j)a_s/2,
\end{gather}
we have
\begin{gather}
\lim_{x\to x_{\tau}}\bigg(\mu_{k+j}(\gamma;x+ija_s/2)\nonumber\\
\qquad{} -e^{4jr(\sigma-a)} \mu_k(\gamma;x-ija_s/2)\frac{P(\gamma';
x+ija_s/2)}{P(\gamma'; x-ija_s/2)}\bigg)=0,\qquad \tau=0,1.  \label{muspqu}
\end{gather}
Furthermore, restricting~$\sigma$~by
\begin{gather}
\label{sigres}
\sigma\notin \{\sigma_1,\ldots,\sigma_L\},\qquad \sigma_K=a_l/2-(K-1)a_s/2,\qquad K=1,\ldots,L,
\end{gather}
we have
\begin{gather}
\label{muqu}
\lim_{x\to x_{\tau}}\frac{\mu_{k+j}(\gamma;x+i\sigma +i(2k+j)a_s/2)}{\mu_k(\gamma;x+i\sigma
+i(2k+j)a_s/2)}=-\pi_{j,\tau}(\gamma),\qquad \tau=0,1,
\end{gather}
where
\begin{gather}
\label{pijtau}
\pi_{j,\tau}(\gamma):=\exp(4j r(\sigma -a)) \prod\limits_{\mu=0}^7\prod\limits_{m=1}^j
\big(1-(-)^{\tau}\exp[2r(\gamma_{\mu}+(j +1-2m)a_s/2)]\big).
\end{gather}
Finally, let $\sigma=\sigma_K,1\le K\le L$. Then~\eqref{muqu} holds for $k+j<K$, whereas for $k+j\ge K$ we have
\begin{gather}
\label{muspec}
\mu_{k+j}(\gamma;x)=0,\qquad \mu_k(\gamma;x_{\tau}+i\sigma_K+i(2k+j)a_s/2)\in{\mathbb C},\qquad \tau=0,1.
\end{gather}
\end{Lemma}

\begin{proof}
The constraint~\eqref{sigcon} is equivalent to the inequality $\sigma<\sigma_{k+j}$. Recalling~\eqref{muvan}, we see
that this implies that the residue functions~$\mu_{\ell}(\gamma;x)$ do not vanish identically for $\ell\le k+j$. It is
also easy to check that the imaginary parts of the limit~$x$-values in~\eqref{muspqu} are smaller than the imaginary
parts of the possible pole locations~\eqref{crit}. Hence the limits do not lead to poles in $\mu_{k+j}$ and~$\mu_k$, and
from Lemma~\ref{lemma2.7} we see that the~$P$-quotient has a~limit for $x\to x_{\tau}$. As a~consequence, the limit
in~\eqref{muspqu} involves the dif\/ference of two functions that are both regular at $x=x_{\tau}$. It therefore suf\/f\/ices
to show that the quotient of the two functions converges to 1 for $x\to x_{\tau}$. In view of the
def\/initions~\eqref{muell}, \eqref{xiell}, this is equivalent to
\begin{gather}
\label{Gqu}
\frac{r_{k+j}}{r_k}\frac{G(i(k+j)a_s+2i\sigma-ia)}{G(ika_s+2i\sigma-ia)}
\\
\qquad{}\times\lim_{x\to x_{\tau}}\frac{G(-ika_s+2x-ia)}{G(-i(k+j)a_s+2x-ia)}\frac{G(ika_s+2i\sigma
-2x-ia)}{G(i(k+j)a_s+2i\sigma -2x-ia)}=\exp(4jr(\sigma -a)).\nonumber
\end{gather}
Using~\eqref{rqu} and~\eqref{Gades}, we see the left-hand side equals
\begin{gather}
\prod\limits_{m=1}^j\frac{R_l(i(k+m-1)a_s-ia_l/2+2i\sigma)} {R_l(i(k+m)a_s+ia_l/2)}\lim_{x\to x_{\tau}}
\prod\limits_{m=1}^j\frac{R_l(-i(k+m)a_s+2x-ia_l/2)} {R_l(i(k+m)a_s+2i\sigma-2x-ia_l/2-ia_s)}
\nonumber\\
\qquad {} =\prod\limits_{m=1}^j\frac{R_l(i(k+m-1)a_s-ia_l/2+2i\sigma)}
{R_l(i(k+m-1)a_s-3ia_l/2+2i\sigma)}\frac{R_l(-i(k+m)a_s+ia_l/2)} {R_l(i(k+m)a_s+ia_l/2)}.\label{RRqu}
\end{gather}
Finally, using evenness of $R_l$ and~\eqref{Rade}, we infer that this equals the right-hand side of~\eqref{Gqu},
proving~\eqref{muspqu}.

The~$\sigma$-restriction~\eqref{sigres} is needed to prevent $\mu_{k+j}$ from vanishing identically, cf.~\eqref{muvan}.
The limits~\eqref{muqu} now follow from a~long, but straightforward calculation. Specif\/ically,
using~\eqref{xiell},~\eqref{mH} and~\eqref{RR}, we obtain
\begin{gather}
\frac{\xi_{k+j}(\gamma;x+i\sigma +i(2k+j)a_s/2)}{\xi_k(\gamma;x+i\sigma +i(2k+j)a_s/2)}   =
\frac{R_s(2x-ija_s+ia_s/2)R_l(2x-ija_s-ia_l/2)} {R_s(2x+ija_s+ia_s/2)R_l(2x+ija_s-ia_l/2)}
\nonumber
\\
\hphantom{\frac{\xi_{k+j}(\gamma;x+i\sigma +i(2k+j)a_s/2)}{\xi_k(\gamma;x+i\sigma +i(2k+j)a_s/2)}   =}{}
    \times\frac{P(-\gamma; x+ija_s/2)}{P(-\gamma; x-ija_s/2)}.
\end{gather}
Using next Lemma~\ref{lemma2.7} and the $R_{\delta}$-A$\Delta$Es~\eqref{Rade}, we see that the limit of
this~$\xi_{\ell}$-ratio for $x\to x_{\tau}$ equals
\begin{gather}
-e^{-8jra}\prod\limits_{\mu=0}^7\prod\limits_{m=1}^j \big(1-(-)^{\tau} \exp[2r(-\gamma_{\mu}+(j+1-2m)a_s/2)]\big)
=-e^{4j r(\sigma -a)}\pi_{j,\tau}(\gamma),
\end{gather}
cf.~\eqref{pijtau}. In view of~\eqref{muell}, it remains to verify the limit
\begin{gather}
\frac{r_{k+j}}{r_k}\frac{G(i(k+j)a_s+2i\sigma-ia)}{G(ika_s+2i\sigma-ia)}
\nonumber\\
\qquad{}\times\lim_{x\to x_{\tau}}\frac{G(-ika_s-2x-ia)}{G(-i(k+j)a_s-2x-ia)}\frac{G(ika_s+2i\sigma
+2x-ia)}{G(i(k+j)a_s+2i\sigma +2x-ia)}=e^{-4jr(\sigma -a)}.\!\!\!
\end{gather}
This can be done by slightly modifying~\eqref{Gqu},~\eqref{RRqu}.

For the special value $\sigma=\sigma_K$, the restriction~$k+j<K$ ensures that~$\mu_{k+j}$ is not identically zero,
cf.~\eqref{muvan}. Then the previous calculation applies with the same outcome. For~$k+j\ge K$, however, $\mu_{k+j}$
does vanish identically. (In particular,~\eqref{muqu} is \emph{false} for~$k+j\ge K$.) On the other hand, the
inequality~$k+j\ge K$ ensures that the argument of~$\mu_k$ belongs to the zero locations~\eqref{muzer2}. It may also
belong to the critical locations~\eqref{crit}, and we can have $k\ge K$ as well, but in all of these
cases~\eqref{muspec} follows, completing the proof of the lemma.
\end{proof}

Using~\eqref{muqu} for $k=0$, $j=1$ and~$\tau=0$, we obtain~\eqref{mur}. There are now several cases concerning the
occurrence of poles at~$x=i\sigma+ia$ for the multipliers~$\mu_0$,~$\mu_1$ in~\eqref{HH2}. For the special values $\sigma
=a_s/2,a_s$, both poles are cancelled by the zeros~\eqref{muzer1}. For the special value $\sigma=a_l/2=\sigma_1$, the
$\mu_0$-pole is cancelled by~\eqref{muzer2}, whereas~$\mu_1$ vanishes identically by~\eqref{muvan} with~$K=1$.
For~$\sigma$ not taking these three values, the pole of~$\mu_0$ is present, but the pole of~$\mu_1$ can still be
cancelled by a~zero coming from~\eqref{muzer0}, provided at least one $\gamma_{\mu}$ vanishes. For the latter case,
however, we obtain~$H_n(\gamma;ia)=0$ from~\eqref{Hnrel} (with $\gamma'\to\gamma$, of course), so again the limit of the
right-hand side of~\eqref{HH2} is f\/inite. Finally, for generic parameters, both multipliers have simple poles, and
then~\eqref{mur} can be combined with~\eqref{Hnrel} to deduce residue cancellation.

We have now shown that~$H_n(\gamma;x)$ is regular at $x=i\sigma +ia$, and in the process we have arrived at the
relation~\eqref{Hnrel}. We can repeat this analysis for the choice $x=i\sigma+\pi/2r+ia$ with obvious changes. This
results in the announced holomorphy property
\begin{gather}
\label{HnL1}
H_n(\gamma;x) \in S_{\sigma+a_l},\qquad (L=1),
\end{gather}
alongside with the relations
\begin{gather}
\label{HnrelL=1}
H_n(\gamma;x_{\tau}+ia_s/2)= \pi_{1,\tau}(\gamma)H_n(\gamma;x_{\tau}-ia_s/2),\qquad \tau=0,1,\qquad (L=1).
\end{gather}

\subsection[Holomorphy of $H_n(\gamma;x)$ for~$|\operatorname{Im} x|<\sigma+a_l$ and $a_l>a_s$]{Holomorphy of $\boldsymbol{H_n(\gamma;x)}$ for~$\boldsymbol{|\operatorname{Im} x|<\sigma+a_l}$ and $\boldsymbol{a_l>a_s}$}\label{section2.4}

With Lemmas~\ref{lemma2.5}--\ref{lemma2.8} at our disposal, we are suf\/f\/iciently prepared to tackle the general-$L$ case.

\begin{Theorem}
\label{theorem2.9}
Let $(n,a_{+},a_{-},g)\in {\mathbb N}\times (0,\infty)^2\times \Pi_r$. Then we have
\begin{gather}
\label{Hnal}
H_n(\gamma;x)\in S_{\sigma +a_l},
\end{gather}
and~$H_n(\gamma;x)$ is holomorphic in the cut plane ${\mathbb C}_{\sigma+a_l}$. $($Here, ${\mathbb C}_{t}$ and $S_t$,
$t>0$, are given by~\eqref{Ccut} and~\eqref{St}.$)$ Furthermore, with~$L$ defined by~\eqref{Ldef}, $x_{\tau}$
by~\eqref{xtau}, and $\pi_{\ell,\tau}(\gamma)$ by~\eqref{pijtau}, $H_n(\gamma;x)$ satisfies the identities
\begin{gather}
\label{Hnident}
H_n(\gamma;x_{\tau}+i\ell a_s/2)= \pi_{\ell,\tau}(\gamma) H_n(\gamma;x_{\tau}-i\ell a_s/2),\qquad \ell =1,\ldots,L,\qquad
\tau=0,1.
\end{gather}
\end{Theorem}
\begin{proof}
We have already proved~\eqref{Hnal} for the case $a_l=a_s$ (cf.~Lemma~\ref{lemma2.4}), and both~\eqref{Hnal}
and~\eqref{Hnident} for the case $L=1$, cf.~\eqref{HnL1} and~\eqref{HnrelL=1}. Also, holomorphy for~$x$ varying over the
set $\operatorname{Re} x \ne k\pi/2r, k\in{\mathbb Z}$, follows from~$F_n(\gamma;x)$ being holomorphic in that set (recall
Lemma~\ref{lemma2.2}) and~$P(\gamma;x)$ being entire. We proceed to study the case $L>1$.

To begin with, f\/ixing $\operatorname{Re} x\in [0,\pi/2r]$, we can let the two relevant simple poles at
\begin{gather}
y\equiv \pm (x-i\sigma-i\ell a_s)\pmod{\pi/2r},\qquad \ell=0,\ldots,L,
\end{gather}
cross the contour in a~by now familiar way. For~$\operatorname{Im} x-\sigma\in (\nu a_s,(\nu +1)a_s)$ with $\nu <L$, this
yields the representation
\begin{gather}
2 \lambda_nH_n(\gamma';x)=\sum\limits_{\ell =0}^\nu \mu_{\ell}(\gamma;x)H_n(\gamma;x-i\sigma-i\ell a_s)
\nonumber\\
\hphantom{2 \lambda_nH_n(\gamma';x)=}{} + P(\gamma';x) \int_I{\cal S}(\sigma;x,y)m_H(\gamma;y)H_n(\gamma;y)dy,\label{HHnu}
\end{gather}
whereas~\eqref{HHnu} with $\nu =L$ pertains to~$\operatorname{Im} x-\sigma\in (La_s,a_l)$. For the special cases~$\operatorname{Im}
x-\sigma =\nu a_s$, with $\nu =0,\ldots,L$, we can use instead the representations
\begin{gather}
2 \lambda_nH_n(\gamma';x)=\frac12 \mu_\nu (\gamma;x)H_n(\gamma;x-i\sigma-i\nu a_s)+\sum\limits_{\ell =0}^{\nu
-1}\mu_{\ell}(\gamma;x)H_n(\gamma;x-i\sigma-i\ell a_s)
\nonumber\\
\hphantom{2 \lambda_nH_n(\gamma';x)=}{}
+ P(\gamma';x) \int_{C^+}{\cal S}(\sigma;x,y)m_H(\gamma;y)H_n(\gamma;y)dy,\label{HHnusp}
\end{gather}
which are valid in a~suf\/f\/iciently small $\operatorname{Im} x$-interval centered at~$\sigma+\nu a_s$, cf.~the proof of
Lemma~\ref{lemma2.3}. (More precisely, for $\nu =0$ we can choose~$C^+$ equal to the interval~$I$~\eqref{intI} shifted
up by $\min (a_s,\sigma)/2$, for $\nu =1,\ldots,L-1$ we can shift up by $a_s/2$, whereas we may need to choose a~smaller
shift for $\nu =L$ to ensure~$C^+$ stays below the poles at the two pertinent locations among~$y\equiv -x+i\sigma
+ia_l\pmod{\pi/2r}$.)

These representations express equality of functions that are a~priori meromorphic in the pertinent strip. However, the
multipliers~$\mu_{\ell}(\gamma;x)$ have no relevant poles for $\operatorname{Im} x-\sigma<a$, cf.~Lemma~\ref{lemma2.5}, so we
deduce right away
\begin{gather}
H_n(\gamma;x)\in S_{\sigma +a}.
\end{gather}
Letting $\operatorname{Im} x$ increase further, we see from~\eqref{crit} that we meet (generic) multiplier poles at
$i\sigma+x_{\tau}+ij a_s/2$ for $j=1,\ldots,L$ and $\tau=0,1$, so we must show that residue cancellations occur. This
hinges on the identities~\eqref{Hnident}. Therefore, our aim is to prove their validity recursively in~$\ell$.

To start the recursion, we focus on the `lowest' pole for $x=i\sigma +ia$. Just as for the $L=1$ case already studied,
we need case distinctions. For $\sigma >a-a_s$, we arrive at the desired relation~\eqref{Hnrel}
via~\eqref{mu0zero}--\eqref{Hia3}, while for~$\sigma\in(a-2a_s,a-a_s)$ we can invoke~\eqref{HH2ia1}--\eqref{muqlim} to
deduce~\eqref{Hnrel}. More generally, for~$\sigma\in(a-(\nu +1)a_s,a-\nu a_s)$ with $\nu \le L/2+1$, we can
combine~\eqref{HHnu} with $x=ia$ and~\eqref{HHnu} with $x=ia-ia_s$ and $\nu \to \nu -1$ in a~similar way, yielding
\begin{gather}
2 \lambda_nH_n(\gamma';x)= \sum\limits_{\ell =0}^\nu \mu_{\ell}(\gamma;x)H_n(\gamma;x-i\sigma-i\ell a_s)
+e^{4r(\sigma-a)}\frac{P(\gamma';x)}{P(\gamma';x-ia_s)} \nonumber\\
\hphantom{2 \lambda_nH_n(\gamma';x)=}{}
\times\left(2\lambda_nH_n(\gamma';x-ia_s)- \sum\limits_{\ell =0}^{\nu
-1}\mu_{\ell}(\gamma;x-ia_s)H_n(\gamma;x-ia_s-i\sigma-i\ell a_s)\right)
\nonumber\\
\hphantom{2 \lambda_nH_n(\gamma';x)=}{}
+ P(\gamma';x)\int_I D_1(\sigma;x,y)m_H(\gamma;y)H_n(\gamma;y)dy.\label{HHiagen}
\end{gather}
Letting~$x$ converge to $ia$, we can once more invoke~\eqref{mu0zero},~\eqref{cruxid},~\eqref{Pid} and~\eqref{muspqu}
with $j=1$, $k=0,\ldots,\nu -1$ and~$\tau=0$ (note that~\eqref{sigcon} holds true), and so~\eqref{Hnrel} follows again.

It remains to study the special cases~$\sigma=a- \nu a_s=\sigma_{2\nu}$ for $\nu \ge 1$ and $\nu \le L/2+1$. Then the
representation~\eqref{HHnusp} applies suf\/f\/iciently close to $ia$, whereas near $ia-ia_s$ we need~\eqref{HHnusp} with
$\nu \to \nu -1$. Using these representations, we arrive in the same way at~\eqref{HHiagen}, but now with~$I$ replaced
by~$C^+$ and factors $1/2$ included in the `highest' residue terms of the two sums. Using the same equations as before,
we then obtain once more~\eqref{Hnrel} by letting~$x$ converge to~$ia$.

The upshot is that we have proved~\eqref{Hnident} for~$\ell =1$ and $\tau=0$. Our reasoning applies to the~$\tau=1$ case
with obvious changes, so~\eqref{Hnident} holds true for~$\ell =1$. We proceed to use this identity to show regularity of
$H_n(\gamma';x)$ at $x=i\sigma +ia$, assuming f\/irst~$\sigma$ is not equal to one of the exceptional
values~$\sigma_K$~\eqref{sigsp}.

For the case~$a \in (\nu a_s,(\nu +1)a_s)$, we need~\eqref{HHnu} with~$x$ taking values near
\begin{gather}
\label{ptau}
p_{\tau}:=i\sigma+x_{\tau}+ia_s/2.
\end{gather}
Since~$a>a_s$, we have $\nu \ge 1$. Now from~\eqref{crit} we see that~$\mu_{\ell}(\gamma;x)$ with $\ell>1$ is regular
for~$x=p_{\tau}$. Hence we need only study the behavior of the sum function
\begin{gather}
\label{Sn}
S_n(\gamma;x):=\mu_0(\gamma;x)H_n(\gamma;x-i\sigma)+ \mu_1(\gamma;x)H_n(\gamma;x-i\sigma-ia_s),
\end{gather}
as~$x$ converges to $p_{\tau}$.

Since $\sigma \notin \{\sigma_1,\ldots,\sigma_L\}$, we can invoke~\eqref{muqu} with $j=1$ and $k=0$. Now there are two
cases. Either~$\mu_0(\gamma;x)$ is regular at~$ p_{\tau}$ or it has a~simple pole. In the f\/irst case~\eqref{muqu}
entails~$\mu_1(\gamma;x)$ is regular, too, so~$S_n(\gamma;x)$ is regular. The second case has two subcases:
either~$\mu_1(\gamma;x)$ is regular or it has a~simple pole. In the f\/irst subcase the right-hand side of~\eqref{muqu}
vanishes, and then~\eqref{Hnident} with~$\ell=1$ entails~$H_n(\gamma;p_{\tau}-i\sigma)=0$. Hence~$S_n(\gamma;x)$ is once
more regular at~$p_{\tau}$. In the second subcase we can combine~\eqref{muqu} with $j=1,k=0$ and~\eqref{Hnident}
with~$\ell=1$ to deduce residue cancellation, hence regularity of~$S_n(\gamma;x)$. As a~consequence, we obtain
regularity of~$H_n(\gamma';x)$ at~$p_{\tau}$.

For the case~$a =\nu a_s$ we need~\eqref{HHnusp}, but since this case can only occur for~$\nu >1$, the fac\-tor~$1/2$
in~\eqref{HHnusp} is innocuous, and residue cancellation follows in the same way.

We are now left with the exceptional values $\sigma=\sigma_K$. From the f\/inal assertion of Lemma~\ref{lemma2.8} we see
that for~$K>1$ our previous reasoning applies. For $K=1$,~\eqref{muspec} says that $\mu_1$ vanishes identically, whereas
$\mu_0(\gamma;x)$ is regular for $x=i\sigma_1+x_{\tau}+ia_s/2$. Therefore $S_n(\gamma;x)$ and $H_n(\gamma';x)$ are once
more regular at $x=p_{\tau}$~\eqref{ptau}.

As a~result, we obtain
\begin{gather}
H_n(\gamma;x)\in S_{\sigma +a+a_s/2},
\end{gather}
which completes the f\/irst step of the recursion.

We now assume inductively that the function~$H_n(\gamma;x)$ satisf\/ies the identities~\eqref{Hnident} for $0<\ell\le J<L$
and that it belongs to~$S_{\sigma+a+Ja_s/2}$. (We have just proved this is true for $J=1$.) We proceed to
prove~\eqref{Hnident} for $\ell =J+1=:M$.

First we study the~$\sigma$-values for which we can use~\eqref{HHnu} with~$x=x_{\tau}+iMa_s/2$. (By assumption,
$H_n(\gamma';x)$ is holomorphic at these~$x$-values.) Thus we need
\begin{gather}
\label{Msig}
a_l/2+Ma_s/2\in(\sigma+\nu a_s,\sigma+(\nu +1)a_s).
\end{gather}
Since~$\sigma$ must satisfy $\sigma\in(0,a)$, the possible $\nu $-values are
\begin{gather}
\label{nures}
(M-1)/2\le \nu \le (L+M)/2,
\end{gather}
with `small' $\nu $-values corresponding to~$\sigma$ near~$a$, and `large' $\nu $-values to~$\sigma$ near~$0$.
Fixing~$\nu_0$ satisfying~\eqref{nures}, there are now two subcases, namely, $\nu_0\ge M$ and~$\nu_0< M$.

Assuming f\/irst~$\nu_0\ge M$, we arrive in a~now familiar way at
\begin{gather}
2 \lambda_nH_n(\gamma';x)= \sum\limits_{\ell =0}^{\nu_0}\mu_{\ell}(\gamma;x)H_n(\gamma;x-i\sigma-i\ell
a_s)+e^{4Mr(\sigma-a)}\frac{P(\gamma';x)}{P(\gamma';x-iMa_s)}
\nonumber\\
\qquad{}
\times \left(2\lambda_nH_n(\gamma';x-iMa_s)- \sum\limits_{\ell =0}^{\nu_0
-M}\mu_{\ell}(\gamma;x-iMa_s)H_n(\gamma;x-iMa_s-i\sigma-i\ell a_s)\right)
\nonumber\\
\qquad{}
+ P(\gamma';x)\int_I D_M(\sigma;x,y)m_H(\gamma;y)H_n(\gamma;y)dy.\label{HHM}
\end{gather}
The condition~\eqref{sigcon} is obeyed for the~$\ell$-values occurring here. Indeed, the largest one is~$\nu_0$,
and~\eqref{Msig} entails
\begin{gather}
\sigma<a_l/2+(M-2\nu_0)a_s/2\le a_l/2-\nu_0a_s/2.
\end{gather}
Therefore, letting~$x$ converge to $x_{\tau}+iMa_s/2$, we are allowed to use~\eqref{muspqu} with $j=M$
and~$k=0,\ldots,\nu_0-M$. It implies that the residue terms in the second sum cancel against the terms with
$\ell=M,\ldots,\nu_0$ in the f\/irst one. Also, the summand containing~$D_M$ vanishes by~\eqref{cruxid} with~$j=M$.
Invoking~\eqref{Pid} with~$j=M$, we see that $H_n(\gamma;x)$ satisf\/ies~\eqref{Hnident} with~$\ell=M$, provided the
remaining sum
\begin{gather}
\label{rsum}
\sum\limits_{\ell =0}^{M-1}\mu_{\ell}(\gamma;x_{\tau}+iMa_s/2)H_n(\gamma;x_{\tau}+iMa_s/2-i\sigma-i\ell a_s),
\end{gather}
vanishes.

In order to prove this, we need to distinguish two further subcases. First we assume
\begin{gather}
\label{sigex}
\sigma \notin {\cal N}:= \{ka_s/2 \,|\, k=1,\ldots, M-1 \}.
\end{gather}
Then we have
\begin{gather}
\label{muellv}
\mu_{\ell}(\gamma;x_{\tau}+iMa_s/2)=0,
\end{gather}
for $\ell=0,\ldots,M-1$, thanks to~\eqref{muzer1}. Indeed, our assumption entails that $x_{\tau}+iMa_s/2$ is not among
the critical locations~\eqref{crit}. Hence the sum~\eqref{rsum} does vanish in this subcase.

Next, let $\sigma\in{\cal N}$. Thus we can write~$\sigma$ as
\begin{gather}
\label{sigN}
\sigma=(M-N)a_s/2,\qquad 1\le N\le M-1.
\end{gather}
Then~\eqref{muellv} still holds true for~$\ell>N$, since this implies that $x_{\tau}+iMa_s/2$ is not of the
form~\eqref{crit}. For~$\ell\le N$, however, the zero location~$x_{\tau}+iMa_s/2$ following from~\eqref{muzer1}
coincides with a~pole location in~\eqref{crit}, unless~$2\ell =N$. Therefore we have
\begin{gather}
\label{mufin}
\mu_{\ell}(\gamma;x_{\tau}+iMa_s/2)\in{\mathbb C},\qquad \ell\le N,
\end{gather}
and the sum~\eqref{rsum} reduces to
\begin{gather}
\sum\limits_{0\le k<N/2}\Big(\mu_{k}(\gamma;x_{\tau}+iMa_s/2)H_n(\gamma;x_{\tau}+i(N-2k)a_s/2)
\nonumber\\
\qquad{} +\mu_{N-k}(\gamma;x_{\tau}+iMa_s/2)H_n(\gamma;x_{\tau}-i(N-2k)a_s/2)\Big).\label{rsumsp}
\end{gather}
Now for~$0\le k <N/2$ it follows not only from~\eqref{muqu} that
\begin{gather}
\mu_{N-k}(\gamma;x_{\tau}+iMa_s/2)=-\pi_{N-2k,\tau}(\gamma) \mu_{k}(\gamma;x_{\tau}+iMa_s/2),\qquad \tau=0,1,
\end{gather}
but also, using our induction assumption, that
\begin{gather}
\label{Hrel}
H_n(\gamma;x_{\tau}+i(N-2k)a_s/2)=\pi_{N-2k,\tau}(\gamma) H_n(\gamma;x_{\tau}-i(N-2k)a_s/2),\qquad \tau=0,1.
\end{gather}
From this we conclude that the terms in the sum~\eqref{rsumsp} cancel in pairs.

The upshot is that we have now proved~\eqref{Hnident} for $\ell =M$ in the subcase~$\nu_0\ge M$. For the subcase~$\nu_0<
M$ we need to start from~\eqref{HHM} with the second sum omitted. Reasoning as before, we see that~\eqref{Hnident} with
$\ell =M$ follows, provided the sum~\eqref{rsum} with $M-1$ replaced by~$\nu_0$ vanishes. Once again, the
assumption~\eqref{sigex} entails~\eqref{muellv} for~$\ell=0,\ldots,\nu_0$, so that the sum does vanish.

Next, assume~$\sigma$ is given by~\eqref{sigN}. From~\eqref{Msig} we then deduce
\begin{gather}
a_l<(2\nu_0+2-N)a_s.
\end{gather}
Since $a_l>La_s$, this entails
\begin{gather}
2\nu_0+2-N>L.
\end{gather}
Now we have $N\le M-1\le L-1$, so we infer~$\nu_0\ge N$. Thus we can again use~\eqref{mufin}--\eqref{Hrel} to prove that
the sum vanishes, so that~\eqref{Hnident} with $\ell =M$ follows once more.

We proceed to study the remaining special cases
\begin{gather}
a_l/2+Ma_s/2=\sigma+\nu_0a_s,\qquad (M-1)/2< \nu_0< (L+M)/2,
\end{gather}
yielding $\sigma=\sigma_{2\nu_0+1-M}$. Assuming f\/irst~$\nu_0\ge M$, we arrive at~\eqref{HHM} with $I\to C^+$ and factors~1/2 for the `highest' residue terms. We can now follow our previous reasoning with obvious changes to
infer~\eqref{Hnident} with $\ell =M$.

Finally, for~$\nu_0<M$ we need to omit the second sum in (the adaptation of)~\eqref{HHM}. Again, it is easy to adapt the
above arguments. In particular, the assumption~\eqref{sigN} now implies $a_l=(2\nu_0-N)a_s$, entailing~$\nu_0>N$.

We have now completed our induction proof of the identity~\eqref{Hnident} with $\ell =M$. Therefore it remains to show
\begin{gather}
H_n(\gamma;x)\in S_{\sigma +a+Ma_s/2}.
\end{gather}
This will follow, provided we can prove that~$H_n(\gamma';x)$ has no pole at
\begin{gather}
\label{ptauM}
p_{\tau,M}:=i\sigma+x_{\tau}+iMa_s/2.
\end{gather}

Just as for $M=1$, we need case distinctions to show this. For the case
\begin{gather}
\label{alnu}
a_l/2+Ma_s/2 \in (\nu a_s,(\nu +1)a_s),
\end{gather}
we can use~\eqref{HHnu} with~$x$ taking values near $p_{\tau,M}$. Now~\eqref{alnu} implies~$a_l<(2\nu+2-M)a_s$. Since
$a_l>La_s$, it follows that~$\nu \ge (L+M)/2$. Inspecting the critical locations~\eqref{crit}, we infer
that~$\mu_{\ell}(\gamma;x)$ with $\ell>M$ is regular for~$x=p_{\tau,M}$. But we have~$M\le(L+M)/2$, so it suf\/f\/ices to
show that the sum function
\begin{gather}
S_{n,M}(\gamma;x):=\sum\limits_{\ell=0}^M\mu_{\ell}(\gamma;x)H_n(\gamma;x-i\sigma-i\ell a_s),
\end{gather}
is regular at $p_{\tau,M}$.

To this end we mimic the reasoning below~\eqref{Sn}. Thus, assuming f\/irst $\sigma\ne \sigma_K$~\eqref{sigsp}, we
invoke~\eqref{muqu}, now with $0\le k<M/2$ and~$ j=M-2k$. Then we repeat the case analysis for the
pairs~$\mu_k,\mu_{M-k}$, now using the identities~\eqref{Hnident} with~$\ell=M,M-2,\ldots$, to deduce regularity
of~$S_{n,M}(\gamma;x)$ and~$H_n(\gamma';x)$ at~$p_{\tau,M}$.

For the case~$a_l/2+Ma_s/2 =\nu a_s$ we need~\eqref{HHnusp}. In this case we have~$\nu >(L+M)/2$, so the factor $1/2$
in~\eqref{HHnusp} multiplies a~residue function $\mu_{\nu}$ that is regular at~$p_{\tau,M}$. Hence residue cancellation
follows just as for the previous case.

Turning to the exceptional values $\sigma=\sigma_K$, our previous reasoning applies for the case~$K>M$. Finally,
consider the choices
\begin{gather}
\sigma=\sigma_{M-N},\qquad N=0,\ldots,M-1.
\end{gather}
From~\eqref{muspec} we then conclude that~$\mu_{\ell}$ vanishes identically for $\ell\ge M-N$, while~$\mu_{\ell}$ is
regular at~$p_{\tau,M}$ for~$0\le\ell<M-N$. Hence we deduce once again regularity of $S_{n,M}(\gamma;x)$ and
$H_n(\gamma';x)$ at $x=p_{\tau,M}$~\eqref{ptauM}. This completes the induction step, and so Theorem~\ref{theorem2.9}
follows.
\end{proof}

We hope that the energy of the patient reader has not been dissipated after this subcase orgy. The remainder of this
section is concerned with assertions that have more palatable proofs.

\begin{Corollary}
\label{corollary2.10}
The kernel of the integral operator~${\cal T}(\gamma)$, $\gamma\in\Pi_r$, is given~by
\begin{gather}
{\cal T}(\gamma;x,y)=\sum\limits_{n=0}^\infty \lambda_n(\gamma)^2f_n(\gamma;x)\overline{f_n(\gamma;y)},
\end{gather}
where $f_n$ is of the form
\begin{gather}
\label{fH}
f_n(\gamma;x)=\frac{H_n(\gamma;x)}{c(\gamma;x)P(\gamma;x)}=\frac{F_n(\gamma;x)}{c(\gamma;x)},
\end{gather}
with $H_n(\gamma;x)/F_n(\gamma;x)$ even, $\pi/r$-periodic and holomorphic in the cut plane ${\mathbb C}_{\sigma
+a_l}/{\mathbb C}_{\sigma +d(\gamma')}$.
\end{Corollary}
\begin{proof}
This follows from Theorem~\ref{theorem2.9} and the def\/initions~\eqref{Ccut},~\eqref{sing},~\eqref{Fn} and~\eqref{Hn}
of~${\mathbb C}_t$, $f_n$, $F_n$ and $H_n$, resp., noting also that the zeros of $P(\gamma;x)$ are at a~distance
\begin{gather}
\min_{\mu=0,\ldots,7}(a-g_{\mu})=\min_{\mu=0,\ldots,7}(a+g_{\mu}'+\sigma)=d(\gamma')+\sigma<\sigma+a_l,
\end{gather}
from the real line, cf.~\eqref{dgam}.
\end{proof}

\begin{Corollary}
Letting $\ell\in\{1,\ldots,L\}$, assume
\begin{gather}
\label{Pnonz}
P(\gamma;x_{\tau}+i\ell a_s/2)\ne 0.
\end{gather}
Then~$F_n(\gamma;x)$ is regular at the two~$x$-values $x_{\tau}\pm i\ell a_s/2$, and we have
\begin{gather}
\label{Fnident}
F_n(\gamma;x_{\tau}+ i\ell a_s/2)=\exp(4\ell r(\sigma -a)) F_n(\gamma;x_{\tau}- i\ell a_s/2),\qquad \forall\, n\in{\mathbb N}.
\end{gather}
\end{Corollary}
\begin{proof}\sloppy
The assumption implies~$P(\gamma;x_{\tau}-i\ell a_s/2)\ne 0$, cf.~\eqref{Pgam}. Thus the regularity claim is clear
from~\eqref{Hn}. Also, dividing~\eqref{Hnident} by~$P(\gamma;x_{\tau}+i\ell a_s/2)$, we deduce~\eqref{Fnident}
from Lem\-ma~\ref{lemma2.7}.
\end{proof}

It may seem plausible that~\eqref{Fnident} is valid without requiring~\eqref{Pnonz}. In fact, however, $F_n(\gamma;x)$
can be regular for~$x=x_{\tau}\pm i\ell a_s/2$, yet fail to fulf\/il~\eqref{Fnident}. We demonstrate this via an
explicitly solvable~$\gamma$-choice, which is of considerable interest in its own right. First, we introduce
\begin{gather}
\gamma^f_0=-a,\qquad \gamma^f_1=-a_-/2,\qquad \gamma^f_2=-a_+/2,\qquad \gamma^f_3=0,\nonumber
\\
\gamma^f_4=\gamma^f_0+i\pi/2r,\qquad \gamma^f_5=\gamma^f_1+i\pi/2r,\qquad \gamma^f_6=\gamma^f_2-i\pi/2r,\qquad \gamma^f_7=\gamma^f_3-i\pi/2r.\!\!\!\label{gfree0}
\end{gather}
Thus, $\gamma^f$ is on the boundary of~$\tilde{\Pi}$ (cf.~\eqref{Pit}), and~$\sigma(\gamma^f)=a$. Next, we def\/ine
\begin{gather}
\label{gLame}
\gamma_{\rm L}(b):=\gamma^f+b \zeta/2,\qquad b\in (0,a).
\end{gather}
We have
\begin{gather}
\sigma(\gamma_{\rm L}(b))=a-b,
\end{gather}
so the~$b$-restriction ensures~$\operatorname{Re} \gamma_{\rm L}(b)$ belongs to $\Pi_r$~\eqref{Pir}. (The subscript L stands
for Lam\'e, cf.~Section~\ref{section7}.) In fact, it is easily checked that $\gamma_{\rm L}(b)'$ is related to~$\gamma_{\rm L}(b)$ by
the permutation~$r_4\times r_4$, so that
\begin{gather}
\operatorname{Re} \gamma_{\rm L}(b)\in\Pi_r^s(2),
\end{gather}
cf.~the paragraph containing~\eqref{gord2}. Finally, the~$G$-duplication formula~\eqref{Gdup} entails
\begin{gather}
\label{clam}
c(\gamma_{\rm L}(b)^{(')};x)=G(2x+ia-ib)/G(2x+ia)=:c_{\rm L}(b;2x).
\end{gather}

Consider now the special case $b=a_s<a_l$. This yields
\begin{gather}
\label{cfree}
c_{\rm L}(a_s;z)=1/R_l(z+ia_l/2),
\end{gather}
so that the weight function
\begin{gather}
w_{\rm L}(b;z):=1/c_{\rm L}(b;\pm z),
\end{gather}
is given~by
\begin{gather}
w_{\rm L}(a_s;z)=p_l^2s_l(z)^2,
\end{gather}
cf.~\eqref{sR}. As a~consequence, the kernel of the integral operator~\eqref{I} is given~by
\begin{gather}
I(\gamma_{\rm
L}(a_s);x,y)=p_l^2s_l(2x)s_l(2y)\exp\left(\sum\limits_{n=1}^{\infty}\frac{4\cos(2nrx)\cos(2nry)\cosh(nra_s)}{n\sinh(nra_l)}\right).
\end{gather}

We are now in the position to invoke (a special case of) Theorem~2.2 in~\cite{Rui12}, yielding the identity
\begin{gather}
I(\gamma_{\rm L}(a_s);x,y)=\frac{8ir e^{ra_l}}{p_l^2s_l(ia_s)}
\sum\limits_{k=1}^{\infty}\frac{\sinh(kra_s)}{\sinh(kra_l)}\sin(2krx)\sin(2kry).
\end{gather}
This identity amounts to~$I(\gamma_{\rm L}(a_s))$ having an eigenfunction-ONB given~by
\begin{gather}
e_n(\gamma_{\rm L}(a_s);x)=\sqrt{4r/\pi}\sin2(n+1)rx,\qquad n\in{\mathbb N},
\end{gather}
with nondegenerate eigenvalues
\begin{gather}
\label{lamfree}
\lambda_n(\gamma_{\rm L}(a_s))=\frac{2\pi i e^{ra_l}}{p_l^2s_l(ia_s)}\frac{\sinh(n+1)ra_s}{\sinh(n+1)ra_l}.
\end{gather}
Next, we use the relation~\eqref{Fn} between $e_n$ and $F_n$ to obtain
\begin{gather}
\label{Fnfree}
F_n(\gamma_{\rm L}(a_s);x)=\sqrt{4r/\pi}\frac{\sin2(n+1)rx}{p_ls_l(2x)}.
\end{gather}
From this it is immediate that none of the $F_n$-identities~\eqref{Fnident} holds true. (Note also that~the poles
of~$F_n$ are at a~distance~$a_l/2$ from~${\mathbb R}$, in accordance with Corollary~\ref{corollary2.10}.)

By contrast, the product on the right-hand side of the $H_n$-identities~\eqref{Hnident} vanishes for $\gamma=\gamma_{\rm
L}(a_s)$. Thus these identities say that $H_n(\gamma_{\rm L}(a_s);x)$ vanishes for $x=x_{\tau}+i\ell a_s/2$. From the
duplication formula~\eqref{Edup} for the~$E$-function we have
\begin{gather}
P(\gamma_{\rm L}(b);x)=E(\pm 2x-ia+ib),
\end{gather}
implying
\begin{gather}
H_n(\gamma_{\rm L}(a_s);x)=\sqrt{4r/\pi}\frac{E(\pm 2x-ia+ia_s)}{p_ls_l(2x)}\sin2(n+1)rx.
\end{gather}
From this it is clear that $H_n$ is a~holomorphic function with the zeros just mentioned. Moreover, these zeros show
that the condition~\eqref{Pnonz} is violated.

On the other hand,~\eqref{Pnonz} is not a~necessary condition for~\eqref{Fnident} to be valid. We shall show this~by
analyzing a~special situation. First, we choose $a_l\in (a_s,2a_s]$, so that we need only consider the case~$\ell=1$.
For notational convenience we also take $\tau=0$ and study the relation between~$F_n(\gamma';ia)$ and
$F_n(\gamma';ia-ia_s)$. To this end we divide~\eqref{Hia3} by $P(\gamma';x)$, yielding
\begin{gather}
2\lambda_nF_n(\gamma';x)=\nu_0(\gamma;x)F_n(\gamma;x-i\sigma) +2\lambda_nF_n(\gamma';x-ia_s)e^{4r(\sigma -a)}
\nonumber\\
\hphantom{2\lambda_nF_n(\gamma';x)=}{}
+ \int_I D_1(\sigma;x,y)w(\gamma;y)F_n(\gamma;y)dy,
\end{gather}
where (cf.~\eqref{mH} and~\eqref{mu0})
\begin{gather}
\nu_0(\gamma;x)  :=  \frac{\mu_0(\gamma;x)w(\gamma;x-i\sigma)}{P(\gamma';x)m_H(\gamma;x-i\sigma)}
\nonumber\\
\hphantom{\nu_0(\gamma;x)~}{}
  =   -4i\pi
r_0\frac{G(2i\sigma-ia)G(2x-ia)G(-2x+2i\sigma+ia)}{\prod\limits_{\mu=0}^7G(x+i\gamma_{\mu}')G(-x+2i\sigma+i\gamma_{\mu}')}.
\end{gather}

Next, we assume
\begin{gather}
\label{gampass}
\gamma_0'=0,\qquad \gamma_{\mu}'\ne 0,a_s,\qquad \mu=1,\ldots,7.
\end{gather}
This ensures that the function~$P(\gamma';x)$ has a~simple zero for $x=ia$, whereas it is nonzero for $x=ia-ia_s$.
Now~\eqref{Hnrel} entails that~$H_n(\gamma';x)$ vanishes for $x=ia$. Therefore the function~$F_n(\gamma';x)$ is regular
for $x=ia$. Since it is also regular for $x=ia-ia_s$, it follows from Lemma~\ref{lemma2.6} that we have
\begin{gather}
F_n(\gamma';ia)-F_n(\gamma';ia-ia_s)e^{4r(\sigma -a)}=\delta_n(\gamma),
\end{gather}
where the dif\/ference term reads
\begin{gather}
\delta_n(\gamma):= (2\lambda_n)^{-1}\lim_{x\to ia}\nu_0(\gamma;x)F_n(\gamma;x-i\sigma)
\nonumber\\
\hphantom{\delta_n(\gamma)~}{}
= -\frac{4i\pi
r_0G(2i\sigma-ia)}{\lambda_n\prod\limits_{\mu=1}^7G(ia+i\gamma_{\mu}')G(2i\sigma-ia+i\gamma_{\mu}')}\lim_{x\to ia}
F_n(\gamma;x-i\sigma).\label{dn}
\end{gather}

From this we deduce that whenever the factor multiplying the $F_n$-limit is nonzero, $F_n(\gamma;x)$ is regular for
$x=ia-i\sigma$. It should be noted that this property also follows from Corollary~\ref{corollary2.10} when
$a-\sigma<\sigma +d(\gamma')$, but this inequality is not implied by~\eqref{gampass}. (Indeed, we can choose~$\sigma$
arbitrarily close to~$0$.) However, when we choose, e.g.,
\begin{gather}
\gamma_1'=-2\sigma,\qquad \sigma<a/2,
\end{gather}
then the prefactor vanishes. If we now require in addition
\begin{gather}
2\sigma >a-d(\gamma'),
\end{gather}
then $F_n(\gamma;x)$ is regular for $x=ia-i\sigma$ by virtue of Corollary~\ref{corollary2.10}. Hence we have
$\delta_n(\gamma)=0$ for all $n\in{\mathbb N}$, showing that the condition~\eqref{Pnonz} is not necessary
for~\eqref{Fnident} to hold true.

In case the prefactor does not vanish, we cannot rule out that $F_n(\gamma;ia-i\sigma)$ vanishes for all $n\in{\mathbb
N}$. But when we require $\sigma>a/2$, then we obtain from~\eqref{FnF}
\begin{gather}
\lambda_n F_n(\gamma;ia-i\sigma)= \int_{0}^{\pi/2r} {\cal S}(\sigma;ia-i\sigma,y)w(\gamma';y)F_n(\gamma';y)dy,\qquad
\sigma>a/2.
\end{gather}
It now follows from the completeness of $I(\gamma)$ that~$F_n(\gamma;ia-i\sigma)$ cannot vanish for
\emph{all}~$n\in{\mathbb N}$. In particular, by~\eqref{e0pos} the integrand is positive for $n=0$, so we have
$F_0(\gamma;ia-i\sigma)> 0$. Therefore, assuming~\eqref{gampass}, $\sigma>a/2$, and a~nonzero prefactor in~\eqref{dn},
we infer
\begin{gather}
F_0(\gamma';ia)\ne F_0(\gamma';ia-ia_s)e^{4r(\sigma -a)}.
\end{gather}
More generally, under the same assumptions, we have
\begin{gather}
F_n(\gamma;ia-i\sigma)\ne 0 \ \Rightarrow \ F_n(\gamma';ia)\ne F_n(\gamma';ia-ia_s)e^{4r(\sigma -a)}.
\end{gather}

\section[Symmetry domains for the A$\Delta$Os~${\cal A} _{\pm}(\gamma;x)$, $g\in\tilde{\Pi}$]{Symmetry domains for the A$\boldsymbol{\Delta}$Os~$\boldsymbol{{\cal A} _{\pm}(\gamma;x)}$, $\boldsymbol{g\in\tilde{\Pi}}$}\label{section3}

First of all, we need to def\/ine the coef\/f\/icient functions in the A$\Delta$Os $A_{\pm}(\gamma;x)$ given by~\eqref{Apm}
more precisely. This involves the building blocks~$R_{\pm}(x)$, whose salient features can be found in~Appendix~\ref{appendixA}. The
shift coef\/f\/icients~$V_{\pm}$ are given~by
\begin{gather}
\label{Ve}
V_{\delta}(\gamma;x)\equiv\frac{\prod\limits_{\mu=0}^7R_{\delta}(x-i\gamma_{\mu}-ia_{-\delta}/2)}
{R_{\delta}(2x+ia_{\delta}/2)R_{\delta}(2x-ia_{-\delta}+ia_{\delta}/2)},\qquad \delta=+,-,
\end{gather}
and the additive coef\/f\/icients~$V_{b,\pm}$ read
\begin{gather}
\label{Vbe}
V_{b,\delta}(\gamma;x)\equiv \frac{\sum\limits_{t=0}^3p_{t,\delta}(\gamma) [{\cal E}_{t,\delta}(\xi;x)-{\cal
E}_{t,\delta}(\xi;\omega_{t,\delta})]}{2R_{\delta}(\xi -ia_{\delta}/2)R_{\delta}(\xi -ia_{-\delta}-ia_{\delta}/2)},\qquad
\delta=+,-.
\end{gather}
Here we are using half-periods
\begin{gather}
\omega_{0,\delta}=0,\qquad \omega_{1,\delta}=\pi/2r,\qquad \omega_{2,\delta}=ia_{\delta}/2,\qquad
\omega_{3,\delta}=-\pi/2r-ia_{\delta}/2,
\end{gather}
the product functions are given~by
\begin{gather}
p_{0,\delta}(\gamma)\equiv \prod\limits_\mu R_{\delta}(i\gamma_{\mu}),\qquad p_{1,\delta}(\gamma)\equiv \prod\limits_\mu
R_{\delta}(i\gamma_{\mu}-\pi/2r),
\\
p_{2,\delta}(\gamma)\equiv \exp(-2r a_{\delta})\exp(r\langle \zeta,\gamma\rangle)\prod\limits_\mu
R_{\delta}(i\gamma_{\mu}-ia_{\delta}/2),
\\
\label{p3}
p_{3,\delta}(\gamma)\equiv \exp(-2r a_{\delta})\exp(-r\langle \zeta,\gamma\rangle)\prod\limits_\mu
R_{\delta}(i\gamma_{\mu}+\pi/2r+ia_{\delta}/2),
\end{gather}
and ${\cal E}_{t,\delta}$ reads
\begin{gather}
\label{cEt}
{\cal E}_{t,\delta}(\xi;x)\equiv \frac{R_{\delta}(x+\xi -ia-\omega_{t,\delta})R_{\delta}(x-\xi
+ia-\omega_{t,\delta})}{R_{\delta}(x-ia-\omega_{t,\delta})R_{\delta}(x+ia-\omega_{t,\delta})},\qquad t=0,1,2,3.
\end{gather}
Thus the functions ${\cal E}_{t,\delta}(\xi;x)$ and $V_{b,\delta}(\gamma;x)$ are elliptic in~$x$ with periods $\pi/r$
and $ia_{\delta}$, whereas $V_{\delta}(\gamma;x)$ has period $\pi/r$ and obeys the quasi-periodicity relation
\begin{gather}
\label{Vper}
V_{\delta}(\gamma;x+ia_{\delta})/V_{\delta}(\gamma;x)=\exp(8r(\sigma -a)).
\end{gather}

It is not clear by inspection, but true that the function $V_{b,\delta}(\gamma;x)$ does not depend on the parameter
$\xi\in{\mathbb C}$. This can be verif\/ied directly by noting that $V_{b,\delta}(\gamma;x)$ is elliptic in $\xi$ and
checking that residues vanish. Alternatively, this follows from Lemma~3.2 in~\cite{Rui02}. The function at issue there
deviates from $V_{b,\delta}(\gamma;x)$ by a~constant, but this constant is $\xi$-independent. Indeed, in~\cite{Rui02}
(and also in~\cite{RI09}) the factor ${\cal E}_{t,\delta}(\xi;\omega_{t,\delta})$ is replaced by ${\cal
E}_{t,\delta}(\xi;z_t)$, with $z_0=z_2=\pi/2r$ and $z_1=z_3=0$. Now $z_t$ does not depend on~$\xi$,
so~$\xi$-independence follows.

Our dif\/ferent choice of additive constants also ensures that the kernel identity~\eqref{kid} has no constant, as opposed
to equation~(3.36) of~\cite{RI09}. In point of fact, the present choice already arose in equation~(3.53) of {\it loc.~cit.}, and
it follows from the reasoning leading from this equation to equation~(3.58) that for this choice the kernel
identity~\eqref{kid} holds true.

From the~$G$-A$\Delta$Es~\eqref{Gades} it readily follows that the shift coef\/f\/icients~\eqref{Ve} and the Harish-Chandra
function~\eqref{c} are related via
\begin{gather}
\label{Vec}
V_{\delta}(x)=c(x)/c(x-ia_{-\delta}),\qquad \delta=+,-.
\end{gather}
In view of~\eqref{w}, this entails the identities
\begin{gather}
\label{Vew}
V_{\delta}(x+ia_{-\delta})w(x+ia_{-\delta})=V_{\delta}(-x)w(x),\qquad \delta=+,-,
\end{gather}
which we invoke in Appendixes~\ref{appendixB} and~\ref{appendixC}.

Next, we deduce from~\eqref{Vec} that the shift coef\/f\/icient~$V_{a,\delta}$ in the A$\Delta$O ${\cal
A}_{\delta}(\gamma;x)$~\eqref{cApmform} satisf\/ies
\begin{gather}
\label{Varel}
V_{a,\delta}(x)=V_{\delta}(-x)V_{\delta}(x+ia_{-\delta})=\frac{c(-x)}{c(x)}\cdot
\frac{c(x+ia_{-\delta})}{c(-x-ia_{-\delta})}.
\end{gather}
Thus it can be rewritten explicitly as
\begin{gather}
\label{Va}
V_{a,\delta}(\gamma;x)\equiv \frac{\prod\limits_{\mu=0}^7R_{\delta}(x+ia_{-\delta}/2\pm
i\gamma_{\mu})}{R_{\delta}(2x-ia_{\delta}/2)R_{\delta}(2x+2ia_{-\delta}+ia_{\delta}/2)R_{\delta}(2x+ia_{-\delta}\pm
ia_{\delta}/2)}.
\end{gather}

The simplest special cases of these formulas are obtained for the `free' Lam\'e cases $\gamma^f$~\eqref{gfree0} and
$\gamma_{\rm L}(a_{\pm})$~\eqref{gLame}. Specif\/ically, we obtain
\begin{gather}
c\big(\gamma^f;x\big)=V_{\pm}\big(\gamma^f;x\big)=V_{a,\pm}\big(\gamma^f;x\big)=1,\qquad V_{b,\pm}\big(\gamma^f;x\big)=0,
\end{gather}
and for instance (cf.~\eqref{cfree})
\begin{gather}
c(\gamma_{\rm L}(a_-);x)=1/R_+(2x+ia_+/2),\qquad V_{b,\pm}(\gamma_{\rm L}(a_-);x)=0,
\end{gather}
together with
\begin{gather}
\label{freesG2}
V_+ (\gamma_{\rm L}(a_-);x)=\frac{R_+(2x-2ia_-+ia_+/2)}{R_+(2x+ia_+/2)},\qquad V_{a,+}(\gamma_{\rm  L}(a_-);x)=\exp(-4ra_-),
\end{gather}
and
\begin{gather}
\label{freesG3}
V_- (\gamma_{\rm L}(a_-);x)= \exp(8irx)\exp(2ra_+),\qquad V_{a,-}(\gamma_{\rm L}(a_-);x)=\exp(-4ra_+).
\end{gather}

For the A$\Delta$Os~\eqref{AHpm} we get
\begin{gather}
A_{\delta}^H(\gamma;x)=
V_{\delta}^H(\gamma;x)\exp(-ia_{-\delta}d/dx)+V_{\delta}^H(\gamma;-x)\exp(ia_{-\delta}d/dx)+V_{b,\delta}(\gamma;x),
\end{gather}
with
\begin{gather}
\label{VH}
V_{\delta}^H(\gamma;x)=V_{\delta}(\gamma;x)P(\gamma;x)/P(\gamma;x-ia_{-\delta}).
\end{gather}
Using~\eqref{Vec},~\eqref{c},~\eqref{GE} and~\eqref{Pgam}, these coef\/f\/icients can be rewritten as
\begin{gather}
V_{\delta}^H(\gamma;x)=\frac{G(2x-2ia_{-\delta}+ia)}{G(2x+ia)}\prod\limits_{\mu=0}^7\frac{E(x\pm
i\gamma_{\mu})}{E(x-ia_{-\delta}\pm i\gamma_{\mu})},
\end{gather}
and when we now use the~$G$- and~$E$-A$\Delta$Es~\eqref{Gades} and~\eqref{Eades} we obtain
\begin{gather}
\label{VHfin}
V_{\delta}^H(\gamma;x)=\frac{1}{R_{\delta}(2x+ia_{\delta}/2)R_{\delta}(2x-ia_{-\delta}+ia_{\delta}/2)}\prod\limits_{\mu=0}^7\frac{1}{G_t(a_{\delta};-x+ia_{-\delta}/2\pm
i\gamma_{\mu})}.
\end{gather}
Notice that these functions are neither even nor quasi-periodic in~$x$.

As announced in the introduction, for the f\/irst~$\gamma$-regime, where~$\gamma$ is real, $D_8$-invariance of the
A$\Delta$Os ${\cal A}_{\pm}(\gamma;x)$ can now be read of\/f from~\eqref{Va} and~\eqref{Vbe}--\eqref{cEt};
likewise,~\eqref{VHfin} reveals $D_8$-invariance of $A_{\pm}^H(\gamma;x)$. We point out that the only term
in~$V_{b,\delta}(\gamma;x)$ that is not $B_8$-symmetric (i.e., invariant under arbitrary sign f\/lips) is the
product~$p_{2,\delta}(\gamma)$. Indeed, in view of~\eqref{Rade}, it changes sign under an odd number of sign f\/lips.

For the second regime we require the ordering~\eqref{imgam}, so we get $S_4\times S_4$-invariance. Furthermore, ${\cal
A}_{\pm}(\gamma;x)$ and~$A_{\pm}^H(\gamma;x)$ are still invariant under f\/lipping an even number of~$\gamma_{\mu}$ to~$-\gamma_{\mu}$. We also point out that~$V_{a,\delta}(\gamma;x)$ and $V_{\delta}^H(\gamma;x)$ are invariant under any
sign f\/lip of~$\operatorname{Im} \gamma_{\mu}$, $\mu=4,5,6,7$, whereas~$V_{b,\delta}(\gamma;x)$ is only invariant under an even
number of such sign f\/lips. (Now the product~$p_{3,\delta}(\gamma)$ changes sign under an odd number,
whereas~$p_{0,\delta}(\gamma)$, $p_{1,\delta}(\gamma)$ and~$p_{2,\delta}(\gamma)$ are invariant.)

Next, we note that~\eqref{cEt} implies
\begin{gather}
{\cal E}_{t,\delta}(\xi;\omega_{t,\delta})=\left(\frac{R_{\delta}(\xi -ia)}{R_{\delta}(ia)}\right)^2.
\end{gather}
Therefore, the additive constant diverges when $R_{\delta}(ia)$ vanishes. For the A$\Delta$O with the smallest shift
parameter this cannot happen, but for the A$\Delta$O with the largest one we must exclude that~$a_l$ is equal to an even
multiple of~$a_s$.

In fact, we impose the stronger requirements~\eqref{aa1}/\eqref{aa2} for the A$\Delta$O with the smal\-lest/largest shift
parameter, respectively. This not only ensures
\begin{gather}
R_{\pm}(ia)\ne 0,\qquad a=(a_++a_-)/2,
\end{gather}
but also implies that the poles of~$V_{\delta}(\gamma;x)$ and~$V_{b,\delta}(\gamma;x)$ are (at most) simple, and that
$V_{b,\delta}(\gamma;x)$ has no poles for real~$x$.

Choosing $g=\operatorname{Re} \gamma$ in the parameter space~$\tilde{\Pi}$~\eqref{Pit} for the remainder of this section, we
proceed to study whether the A$\Delta$Os~${\cal A}_{\pm}(\gamma;x)$ can be promoted to symmetric operators on suitable
dense subspaces of the Hilbert space~${\cal H}$~\eqref{cH}. These subspaces involve the auxiliary Harish-Chandra
function~\eqref{cPol}, and the locations of poles and zeros of~$c_P(\gamma;x)$ are crucial in the sequel. In view of the
zeros~\eqref{ze} of the~$E$-function, the pole locations are given~by
\begin{gather}
\label{cPpo1}
x\equiv 0 \pmod {\pi/2r},
\\
\label{cPpo2}
x\equiv i(ka_++(l+1)a_-)/2 \pmod {\pi/2r},
\\
\label{cPpo3}
x\equiv i((k+1)a_++la_-)/2 \pmod {\pi/2r},
\end{gather}
and the zero locations~by
\begin{gather}
\label{cPze}
x\equiv \pm i\gamma_{\mu}+ia+ika_++la_- \pmod {\pi/r},
\end{gather}
where
\begin{gather}
k,l\in{\mathbb N}\equiv \{0,1,2,\ldots \}.
\end{gather}

It is not immediate, but true that the shift coef\/f\/icients~$V_{a,\delta}$ admit an alternative representation
\begin{gather}
\label{Vaalt}
V_{a,\delta}(x)=\frac{c_P(-x)}{c_P(x)}\cdot \frac{c_P(x+ia_{-\delta})}{c_P(-x-ia_{-\delta})},\qquad \delta=+,-.
\end{gather}
This representation is of pivotal importance for what follows. It can be readily checked (by using
the~$E$-A$\Delta$Es~\eqref{Eades} and the relation~\eqref{GtR} between the trigonometric gamma function
and~$R_{\delta}$) that the right-hand side of~\eqref{Vaalt} indeed equals the right-hand side of~\eqref{Va}.
Alternatively,~\eqref{Vaalt} follows by verifying the identity
\begin{gather}
\label{ccP}
\frac{c(x)}{c(-x)}=\frac{c_P(x)}{c_P(-x)},
\end{gather}
and invoking~\eqref{Varel}.

Next, we note that the restrictions of the functions in the vector space~$S_t$~\eqref{St} to the interval $[0,\pi/2r]$
yield a~subspace of ${\cal H}$ (again denoted~$S_t$). This subspace is dense in ${\cal H}$, since it contains the
Chebyshev polynomials
\begin{gather}
p_n(\cos 2rx)\equiv \cos 2nrx,\qquad n\in{\mathbb N}.
\end{gather}
It readily follows that the subspace
\begin{gather}
\label{Dtgam}
D_t(\gamma)\equiv \frac{1}{c_P(\gamma;x)}S_t,\qquad t>0,\qquad g\in\tilde{\Pi},
\end{gather}
is also a~dense ${\cal H}$-subspace. We point out that this subspace depends on~$\gamma$, but is $B_8$-invariant for the
f\/irst regime and invariant under $S_4\times S_4$ and arbitrary~$\gamma_{\mu}$ sign f\/lips for the second one.
Furthermore, the~$\gamma$-restriction ensures that the poles~\eqref{cPze} of the prefactor are in the (open) upper half
plane (UHP). Hence we can view~$D_t(\gamma)$ as a~space of functions that are holomorphic for~$\operatorname{Im} x\in(-t,0]$
and meromorphic for~$\operatorname{Im} x\in(0,t)$. In particular, the action of the shift summand~$\exp(- ia_{\delta}d/dx)$
in~${\cal A}_{-\delta}(\gamma;x)$ on $f\in D_t(\gamma)$ yields a~function~$f_{\delta}(x)$ that is holomorphic for~$\operatorname{Im}  x\in (-t+ a_{\delta},a_{\delta}]$. Thus we need only choose
\begin{gather}
\label{tres}
t>a_{\delta},
\end{gather}
to ensure that this function is real-analytic on ${\mathbb R}$. Hence the restriction of~$f_{\delta}(x)$ to $[0,\pi/2r]$
is well def\/ined and belongs to~${\cal H}$.

Consider next the action of the summand~$V_{a,-\delta}(\gamma;x)\exp(ia_{\delta}d/dx)$ in~${\cal A}_{-\delta}(\gamma;x)$
on $f\in D_t(\gamma)$. Writing
\begin{gather}
f(x)=\frac{1}{c_P(x)}F(x),\qquad F\in S_t,
\end{gather}
we can use the representation~\eqref{Vaalt} to obtain
\begin{gather}
(V_{a,-\delta}(\gamma;x)\exp(ia_{\delta}d/dx)f)(x)=\frac{c_P(-x)}{c_P(x)}\cdot
\frac{1}{c_P(-x-ia_{\delta})}F(x+ia_{\delta}).
\end{gather}
Once more, the~$t$-requirement~\eqref{tres} ensures that the function~$F(x+ia_{\delta})$ is real-analytic on ${\mathbb
R}$, yielding a~vector in~${\cal H}$ upon restriction to~$[0,\pi/2r]$.

The key point is now that the three $c_P$-factors on the right-hand side yield a~meromorphic function that is {\it also}
real-analytic on~${\mathbb R}$. Indeed, since we require that~$\gamma$ belong to~$\tilde{\Pi}$, the poles of the three
factors are in the UHP, except those coming from the term $(1-\exp (4irx))^{-1}$ in~$c_P(-x)$. The latter poles,
however, are cancelled by zeros coming from the term~$(1-\exp (-4irx))$ in~$1/c_P(x)$. Thus no real poles remain, even
though $V_{a,\delta}(\gamma;x)$ itself does have real poles for generic~$\gamma$'s. (The latter can even be double
poles, namely when $a_s/a_l=2/3,2/5,2/7,\dots$, cf.~\eqref{Va}.)

Finally, the parameter restrictions imply that the function $V_{b,-\delta}(\gamma;x)$ is not only real-analytic
on~${\mathbb R}$, but also real-valued. Moreover, by evenness and $\pi/r$-periodicity its restriction to~${\mathbb R}$
is bounded below~by
\begin{gather}
\label{Vbmin}
M_{-\delta}(\gamma)\equiv \min_{x\in [0,\pi/2r]}V_{b,-\delta}(\gamma;x).
\end{gather}

We now focus on the A$\Delta$O~${\cal A}_s(\gamma;x)$ with the smallest shift parameter $a_s$, denoting its coef\/f\/icients
by $V_{a,s}(\gamma;x)$ and~$V_{b,s}(\gamma;x)$. (Thus these coef\/f\/icients can be expressed in terms of~$R_l(z)$ given
by~\eqref{Rsl}.) As a~consequence of the above analysis, we conclude that for $t>a_s$ we can def\/ine a~linear
operator~$\hat{{\cal A}}_{s}(t,\gamma)$ on the dense domain~$D_t(\gamma)\subset{\cal H}$ (given by~\eqref{Dtgam}), by
the composition of the action of ${\cal A}_s(\gamma;x)$ on the meromorphic extensions of functions in~$D_t(\gamma)$,
followed by restriction to~$[0,\pi/2r]$. We are now prepared for the f\/irst result of this section.

\begin{Theorem}
\label{theorem3.1}
With the assumption~\eqref{aa1} on $a_{\pm}$ in effect and $g\in\tilde{\Pi}$, the operator $\hat{{\cal A}}_s(t,\gamma)$
is well defined and symmetric on $D_t(\gamma)$ for any $t>a_s$. Furthermore, for all nonzero~$f\in D_t(\gamma)$ we have
a~lower bound
\begin{gather}
\label{lowb}
\big(f,\hat{{\cal A}}_s(t,\gamma)f\big)> M_s(\gamma)(f,f),\qquad M_s(\gamma)\equiv \min_{x\in [0,\pi/2r]}V_{b,s}(\gamma;x).
\end{gather}
\end{Theorem}
\begin{proof}
We have already shown that the parameter restriction~\eqref{aa1} and requirement~$g\in\tilde{\Pi}$ entail that the
operators $\hat{{\cal A}}_s(t,\gamma)$ are well def\/ined on the pertinent domains. Since the additive
coef\/f\/icient~$V_{b,s}(\gamma;x)$ is real-analytic and real-valued on ${\mathbb R}$, it gives rise to a~bounded and
self-adjoint multiplication operator on ${\cal H}$. Consequently, this summand of~$\hat{{\cal A}}_{s}(t,\gamma)$ plays
no role in domain and symmetry issues.

Fixing~$f_j(x)=F_j(x)/c_P(x)\in D_t(\gamma)$, $j=1,2$, with $t>a_s$, our next goal is to prove equality of
\begin{gather}
\label{IL}
I_L\equiv \int_{0}^{\pi/2r} \left(\frac{F_1^{*}(x+ia_s)}{c_P(-x-ia_s)}+V_{a,s}(-x)
\frac{F_1^{*}(x-ia_s)}{c_P(-x+ia_s)}\right) \frac{F_2(x)}{c_P(x)}dx,
\end{gather}
and
\begin{gather}
\label{IR}
I_R\equiv \int_{0}^{\pi/2r} \frac{F_1^{*}(x)}{c_P(-x)}\left(\frac{F_2(x-ia_s)}{c_P(x-ia_s)}+V_{a,s}(x)
\frac{F_2(x+ia_s)}{c_P(x+ia_s)}\right) dx.
\end{gather}
Here, the $*$ denotes the conjugate function (cf.~\eqref{Mconj}), and we used
\begin{gather}
V_{a,s}^{*}(\gamma;x)=V_{a,s}(\gamma;-x),\qquad c_P^{*}(\gamma;x)=c_P(\gamma;-x),
\end{gather}
cf.~\eqref{cons1},~\eqref{cons2}. In order to show this equality, we f\/irst substitute the representation~\eqref{Vaalt}
of~$V_{a,s}(x)$. If we then introduce
\begin{gather}
I_s(x)\equiv
\frac{F_1^{*}(x+ia_s/2)F_2(x-ia_s/2)}{c_P(\gamma;-x-ia_s/2)c_P(\gamma;x-ia_s/2)}=f_1^*(x+ia_s/2)f_2(x-ia_s/2),
\end{gather}
we can use evenness of~$F_1^{*}(x)$ and $F_2(x)$ to deduce that $I_L-I_R$ equals
\begin{gather}
\int_{0}^{\pi/2r} [I_s(x+ia_s/2)+I_s(-x+ia_s/2)-I_s(x-ia_s/2)-I_s(-x-ia_s/2)]dx.
\end{gather}
This integral can be rewritten as
\begin{gather}
\label{dint}
\int_{-\pi/2r}^{\pi/2r} [I_s(x+ia_s/2)-I_s(x-ia_s/2)]dx.
\end{gather}
Now $I_s(x)$ is not only $\pi/r$-periodic, but also holomorphic for $|\operatorname{Im} x|\le a_s/2$ (recall~\eqref{cPze}
and~$|\gamma_{\mu}|<a$). Hence the integral~\eqref{dint} vanishes by Cauchy's theorem and the symmetry assertion
results.

Finally, the estimate~\eqref{lowb} will follow once we prove
\begin{gather}
\label{auxb}
(f,(\hat{{\cal A}}_s(\gamma)-V_{b,s}(\gamma;\cdot))f)> 0,\qquad 0\ne f\in D_t(\gamma).
\end{gather}
Now we have already seen above that we have
\begin{gather}
I_R=\int_{-\pi/2r}^{\pi/2r} I_s(x-ia_s/2)dx=\int_{-\pi/2r}^{\pi/2r} I_s(x)dx.
\end{gather}
Choosing $f_1=f_2=f$, this implies that the left-hand side of~\eqref{auxb} is equal to the integral
\begin{gather}
\int_{-\pi/2r}^{\pi/2r}f^{*}(x+ia_s/2)f(x-ia_s/2) dx = \int_{-\pi/2r}^{\pi/2r}|f(x-ia_s/2)|^2 dx.
\end{gather}
This renders its positivity manifest, so that~\eqref{auxb} follows.
\end{proof}

The bound~\eqref{lowb} entails that the symmetric operator~$\hat{{\cal A}}_{s}(t,\gamma)$ admits self-adjoint
extensions. Moreover, whenever~$D_t(\gamma)$ is a~core (domain of essential self-adjointness~\cite{RS72})
for~$\hat{{\cal A}}_{s}(t,\gamma)$, it follows that the closure is a~self-adjoint operator that is greater
than~$M_s(\gamma)$; this operator is~$D_8$-invariant for the f\/irst~$\gamma$-regime, and invariant under $S_4\times S_4$
and an even number of~$\gamma_{\mu}$ sign f\/lips for the second one.

We are unable to prove that the core property holds true for arbitrary~$g=\operatorname{Re} \gamma\in\tilde{\Pi}$ and $t>a_s$.
In Section~\ref{section4}, however, we show its validity for all $g\in\Pi_r$ and $t\in(a_s,a]$. The~$g$-restriction is needed, since
the proof makes essential use of the integral operators in the previous section. (As already pointed out in the
introduction, the latter are not invariant under sign f\/lips.) On the other hand, since $D_8 $-invariance is built into
the def\/inition of~$\hat{{\cal A}}_{s}(t,\gamma)$ for the f\/irst regime, it follows that the core property holds for the
union of the $D_8$-transforms of~$\Pi_r$. This union is still a~proper subset of $\tilde{\Pi}$~\eqref{Pit}, however. In
particular, it does not contain the vector~$\gamma =0$. A~similar state of af\/fairs pertains to the second regime.

We conclude this section by reconsidering the A$\Delta$O~${\cal A}_l(\gamma;x)$ with the largest shift parameter $a_l$,
now requiring~\eqref{aa2}. Below~\eqref{Vbmin}, we restricted attention to~${\cal A}_s(\gamma;x)$, but upon scrutiny of
our line of reasoning, the reader will have no dif\/f\/iculty to see that it applies with obvious changes to~${\cal
A}_l(\gamma;x)$. However, the symmetric operator~$\hat{{\cal A}}_l^w(t,\gamma)$ on ${\cal H}$ thus obtained is `wrong'
in the following sense: For generic $g\in\Pi_r$, no self-adjoint extension of~$\hat{{\cal A}}_l^w(t,\gamma)$ has all of
the ${\cal T}(\gamma)$-eigenfunctions~$f_n(\gamma;x)$ in its domain and acts on them as the A$\Delta$O~${\cal
A}_l(\gamma;x)$. We detail this in Section~\ref{section5}, after presenting the `correct' Hilbert space version of this A$\Delta$O,
from whose def\/inition it shall follow that the functions~$f_n(\gamma;x)$ are joint eigenfunctions of~${\cal
A}_{\pm}(\gamma;x)$.

\section[Joint eigenvectors for~${\cal T} (\gamma)$ and $\hat{{\cal A}}_s(\gamma)$, $g\in\Pi_r$]{Joint eigenvectors for~$\boldsymbol{{\cal T} (\gamma)}$ and $\boldsymbol{\hat{{\cal A}}_s(\gamma)}$, $\boldsymbol{g\in\Pi_r}$}\label{section4}

Assuming that the parameters $a_+$, $a_-$ satisfy~\eqref{aa1}, we have shown in Section~\ref{section3} that the A$\Delta$O~${\cal
A}_s(\gamma;x)$ gives rise to symmetric operators~$\hat{{\cal A}}_{s}(t,\gamma)$ on~${\cal H}$ for all~$g\in\tilde{\Pi}$
and $t>a_s$. In this section, however, we restrict~$g$ to $\Pi_r$, so that the results of Section~\ref{section2} can be invoked. Our
main aim is to show that the~${\cal T}(\gamma)$-eigenfunctions~$f_n(\gamma;x)$~\eqref{fn} can be redef\/ined in such a~way
that they are also eigenfunctions of~$\hat{{\cal A}}_{s}(t,\gamma)$, provided $t\in(a_s,a]$. (Since we cannot rule out
degeneracies in the ${\cal T}(\gamma)$-spectrum, our original base choice for a~degenerate eigenspace need not yield
eigenvectors of~$\hat{{\cal A}}_{s}(t,\gamma)$.) This entails in particular that we obtain a~unique self-adjoint
closure~$\hat{{\cal A}}_s(\gamma)$, which is essentially self-adjoint on the span of the vectors~$f_n(\gamma;\cdot)$,
$n\in{\mathbb N}$, and which has solely discrete spectrum.

For our f\/irst result we need the non-obvious identity
\begin{gather}
\label{ccid}
\frac{c_P(\gamma;x)}{c(\gamma;x)}=\frac{\prod\limits_{\mu=0}^7E(\pm x +i\gamma_\mu)}{E(\pm 2x-ia)}.
\end{gather}
Using the def\/initions~\eqref{c},~\eqref{cPol}, and the relation~\eqref{GE} between the~$G$- and~$E$-functions, we see
that~\eqref{ccid} amounts to the identity
\begin{gather}
\label{Eid}
\frac{(1-\exp(-4irx))E(2x\pm i(a_+-a_-)/2)}{E(2x\pm ia)}=1.
\end{gather}
The latter can be verif\/ied directly by using~\eqref{E}, but it is more illuminating to invoke the
A$\Delta$Es~\eqref{Eades} and~\eqref{Gtade} to check~\eqref{Eid}.

\begin{Lemma}
\label{lemma4.1}
Assuming~$(a_+,a_-,g)\in(0,\infty)^2\times \Pi_r$, we have
\begin{gather}
\label{fnDt}
f_n(\gamma;x)\in D_t(\gamma),\qquad \forall\, t\in(a_s,a],\qquad \forall\, n\in{\mathbb N}.
\end{gather}
\end{Lemma}

\begin{proof}
In view of the def\/inition~\eqref{Dtgam}, it suf\/f\/ices to prove
\begin{gather}
c_P(\gamma;x)f_n(\gamma;x)\in S_a.
\end{gather}
Recalling~\eqref{fH}, we see that this amounts to
\begin{gather}
\frac{c_P(\gamma;x)}{c(\gamma;x)}\frac{1}{\prod\limits_{\mu}E(\pm x+i\gamma_{\mu})}H_n(\gamma;x)\in S_a.
\end{gather}
Thanks to the identity~\eqref{ccid} this can be rewritten as
\begin{gather}
\frac{H_n(\gamma;x)}{E(\pm 2x-ia)}\in S_a.
\end{gather}
Now from Theorem~\ref{theorem2.9} we have~$H_n\in S_{\sigma+a_l}$. Since the multiplier of~$H_n$ belongs to~$S_a$, the
lemma follows.
\end{proof}

The alert reader will have noted that we have used Theorem~\ref{theorem2.9} at the end of this proof, whereas we
announced in the introduction that we need only invoke Lemma~\ref{lemma2.4} for the purpose of this section (as opposed
to the next one). The point is that for our further conclusions it would suf\/f\/ice to use~\eqref{fnDt} for $t\in(a_s,\min
(a,\sigma +a_s)]$, and this weaker version does follow from Lemma~\ref{lemma2.4}.

For the remainder of this section we assume that the parameters~$a_{\pm}$ satisfy~\eqref{aa1}. We now def\/ine the space
of linear combinations of the functions $f_n(\gamma;x)$,
\begin{gather}
\label{cC}
{\cal C}(\gamma)\equiv {\rm LH}(f_n(\gamma;x))_{n=0}^{\infty},
\end{gather}
where LH stands for linear hull. Upon restricting~$x$ to~$[0,\pi/2r]$, this space yields a~dense subspace of~${\cal
H}$~\eqref{cH}. It follows from~Lemma~\ref{lemma4.1} and~Theorem~\ref{theorem3.1} that the operators~$\hat{{\cal
A}}_s(t,\gamma)$ are well def\/ined and symmetric on ${\cal C}(\gamma)$ for all $t\in(a_s,a]$. From now on we denote their
unique restriction to ${\cal C}(\gamma)$ by~$\hat{{\cal A}}_s(\gamma)$.

Our next goal is to show that~${\cal C}(\gamma)$ is a~core. To prove this, we need a~few auxiliary observations
involving the integral operator~${\cal I}(\gamma)$. To start with, we note that we have
\begin{gather}
({\cal I}(\gamma')f_n(\gamma;\cdot))(x)=\frac{1}{c(\gamma';x)}\int_{0}^{\pi/2r} {\cal
S}(\sigma;x,y)w(\gamma;y)F_n(\gamma;y)dy,
\end{gather}
cf.~\eqref{cIg} and~\eqref{fn}. Recalling the paragraph containing~\eqref{sing}, we readily deduce
\begin{gather}
\label{cIcC}
{\cal I}(\gamma'){\cal C}(\gamma)={\cal C}(\gamma').
\end{gather}
We are now prepared for the following lemma. Its proof is relegated to Appendix~\ref{appendixB}. Note that in view of~\eqref{cIcC},
both sides of~\eqref{cAcI} are well def\/ined.

\begin{Lemma}
\label{lemma4.2}
Let~$g=\operatorname{Re} \gamma\in\Pi_r$ and let $a_{\pm}$ satisfy~\eqref{aa1}. Then we have
\begin{gather}
\label{cAcI}
\hat{{\cal A}}_s(\gamma'){\cal I}(\gamma')f={\cal I}(\gamma') \hat{{\cal A}}_s(\gamma)f,\qquad \forall\, f\in {\cal
C}(\gamma).
\end{gather}
\end{Lemma}

\begin{Theorem}
\label{theorem4.3}
With the assumptions of Lemma~{\rm \ref{lemma4.2}}, the operator $\hat{{\cal A}}_s(\gamma)$ is essentially self-adjoint on
${\cal C}(\gamma)$. Its self-adjoint closure $($denoted by the same symbol$)$ has solely discrete spectrum, and it admits
eigenvectors that belong to the finite-dimensional eigenspaces of the positive trace class operator~${\cal T}(\gamma)$.
\end{Theorem}

\begin{proof}
Acting with~${\cal I}(\gamma)$ on~\eqref{cAcI} and using again~\eqref{cIcC} and~\eqref{cAcI}, we obtain
\begin{gather}
\hat{{\cal A}}_s(\gamma){\cal T}(\gamma)f={\cal T}(\gamma) \hat{{\cal A}}_s(\gamma)f,\qquad \forall\, f\in {\cal
C}(\gamma).
\end{gather}
Choosing~$f=f_n(\gamma;\cdot)$, this implies
\begin{gather}
{\cal T}(\gamma) \hat{{\cal A}}_s(\gamma)f_n(\gamma;\cdot)=\lambda_n(\gamma)^2\hat{{\cal
A}}_s(\gamma)f_n(\gamma;\cdot),\qquad \forall\, n\in{\mathbb N}.
\end{gather}
Therefore, the vector~$\hat{{\cal A}}_s(\gamma)f_n(\gamma;\cdot)$ belongs to the f\/inite-dimensional eigenspace~${\cal
E}_n(\gamma)$ of~${\cal T}(\gamma)$ corresponding to the eigenvalue~$\lambda_n(\gamma)^2$. As a~consequence, the
operator $\hat{{\cal A}}_s(\gamma)$ leaves~${\cal E}_n(\gamma)$ invariant, and since it is symmetric on~${\cal
C}(\gamma)$, it is self-adjoint on~${\cal E}_n(\gamma)$. From this the assertions are clear.
\end{proof}

In view of this theorem, we can choose a~new ONB for~${\cal E}_n(\gamma)$ consisting of joint eigenvectors of~${\cal
T}(\gamma)$ and~$\hat{{\cal A}}_s(\gamma)$. Recalling
\begin{gather}
f_n(\gamma;x)=m(\gamma;x)e_n(\gamma;x),
\end{gather}
(cf.~\eqref{fn},~\eqref{m} and~\eqref{Fn}), we see that this amounts to choosing a~new ONB~$e_n(\gamma;\cdot)$
for \linebreak $M(\gamma)^{-1}{\cal E}_n(\gamma)$, consisting of joint eigenvectors of~$T(\gamma)$ and the operator
\begin{gather}
\hat{H}_s(\gamma)\equiv M(\gamma)^{-1}\hat{{\cal A}}_s(\gamma)M(\gamma),\qquad g\in\Pi_r.
\end{gather}
It is readily checked that the latter commutes with complex conjugation, so we may and shall retain the real-valuedness
and sign f\/ixing~\eqref{signen} of~$e_n(\gamma;x)$ for the new base choice. (Of course, we need not change anything in
case~$\dim {\cal E}_n(\gamma)=1$, but we only know this for~$n=0$.)

The upshot is that we now have an ONB for ${\cal H}$ that consists of common eigenvectors~$f_n(\gamma;\cdot)$ of the
operators~$\hat{{\cal A}}_s(\gamma)$ and~${\cal T}(\gamma)$. Unfortunately we are unable to derive much information on
the eigenvalues of the former. Def\/ining these~by
\begin{gather}
\label{Enp}
\hat{{\cal A}}_s(\gamma)f_n(\gamma;\cdot)=:E_{n,s}(\gamma)f_n(\gamma;\cdot),
\end{gather}
we do get from~\eqref{lowb} a~f\/inite lower bound
\begin{gather}
E_{n,s}(\gamma)> M_s(\gamma),\qquad \forall\, n\in{\mathbb N},
\end{gather}
but at this stage we cannot even rule out that there exists a~f\/inite upper bound on the eigenvalues. (Equivalently, we
cannot rule out that~$\hat{{\cal A}}_s(\gamma)$ is bounded.) Now unboundedness will be shown in Section~\ref{section7}, but we have
to leave other natural questions open. In particular, we are unable to prove that~$E_{0,s}(\gamma)$ is the smallest
eigenvalue and that it is nondegenerate for all~$g\in\Pi_r$ and $a_{\pm}$ satisfying~\eqref{aa1}, even though these
features are highly plausible.

We continue to exploit~\eqref{Enp}, rewritten in the equivalent form
\begin{gather}
\label{Fnade}
V_s(-x)F_n(x\!+\!ia_s)+V_s(x)F_n(x\!-\!ia_s)+V_{b,s}(x)F_n(x)=E_{n,s}F_n(x),\!\!\qquad x\in(0,\pi/2r),  \!\!\!
\end{gather}
to clarify the analyticity properties of the functions $F_n(x)$ and $H_n(x)$. (Henceforth we often
suppress~$\gamma$-dependence, since from now on this usually does not cause ambiguities.) From Lemma~\ref{lemma2.4} we
see~$H_n(x)$ is holomorphic for $x\in{\mathbb C}_{\sigma+a_s}$, so that~$F_n(x)$ is meromorphic in~${\mathbb
C}_{\sigma+a_s}$. Therefore,~\eqref{Fnade} has a~meromorphic continuation to~$x\in{\mathbb C}_\sigma$. A~moment's
thought reveals that this state of af\/fairs implies that $F_n(x)$ has a~meromorphic continuation to all of~${\mathbb C}$.
Indeed, when we divide~\eqref{Fnade} by $V_s(-x)$, then we can continue~$F_n$ in steps of size~$a_s$ to the UHP.
Likewise, upon division by~$V_s(x)$ we can continue $F_n$ to the LHP.

Next, we aim to prove that~$H_n(x)$ is actually entire, so that the meromorphic function~$F_n(x)$ can only have poles at
the zeros of~$P(x)$, cf.~\eqref{Hn},~\eqref{Pgam}. We proceed with a~preview of the proof. A~key f\/irst step is to
replace~\eqref{Fnade} by the eigenvalue A$\Delta$E for the A$\Delta$O $A^H_s(\gamma;x)$~\eqref{AHpm}, i.e.,
\begin{gather}
\label{Hnade}
V_s^H(-x) H_n(x+ia_s)+V_s^H(x) H_n(x-ia_s)+(V_{b,s}(x)-E_{n,s})H_n(x)=0,
\end{gather}
and study the pole and zero features of the coef\/f\/icients of~$H_n(x\pm ia_s)$ and $H_n(x)$. In particular, it is crucial
that our assumption~\eqref{aa1} ensures that the three coef\/f\/icients have at most \emph{simple} poles.

Now Lemma~\ref{lemma2.4} yields holomorphy of $H_n(x)$ for $|\operatorname{Im} x|<\sigma+a_s$, and since $H_n(x)$ is also
$\pi/r$-periodic and even, we need only let $\operatorname{Im} x$ increase and analyze eventual poles of $H_n(x)$ for $\operatorname{Re} x\in\{0,\pi/2r\}$ that might arise from coef\/f\/icient poles in~\eqref{Hnade}. This analysis leads to a~sequence of
residue cancellation identities connecting $H_n(x)$-values. In view of the relation~\eqref{VH} between~$V_s^H$ and
$V_s$, we need~$P$-limits similar to those listed in Lemma~\ref{lemma2.7}. As another important ingredient for the
identities, we need the relations between the residues at the (at most) simple poles of the elliptic function
$V_{b,s}(x)$ and those of the $\pi/r$-periodic coef\/f\/icient functions~$V_s(\pm x)$, cf.~\eqref{Ve}--\eqref{p3}. Thanks to
the quasi-periodicity relation~\eqref{Vper}, we can obtain all of these relations from the following ones (which are
readily verif\/ied):
\begin{gather}
\label{resr1}
\big(\operatorname{Res}  \big(V_{b,s}(x)\big)+\operatorname{Res}  \big(V_s(x)\big)\big)\big|_{x=ia_s/2+k\pi/2r}=0,\qquad k\in{\mathbb Z},
\\
\label{resr2}
\big(\operatorname{Res}  \big(V_{b,s}(x)\big)+e^{4r(a-\sigma)}\operatorname{Res}  \big(V_s(x)\big)\big)\big|_{x=ia +k\pi/2r}=0,\qquad
k\in{\mathbb Z}.
\end{gather}
We are now prepared for the following theorem.

\begin{Theorem}
\label{theorem4.4}
Let $(n,a_{+},a_{-},g)\in {\mathbb N}\times (0,\infty)^2\times \Pi_r$, with~$a_+$, $a_-$ satisfying~\eqref{aa1}. Then the
function~$H_n(\gamma;x)$ is holomorphic in~${\mathbb C}$. Moreover, it obeys the identities
\begin{gather}
H_n(\gamma;\tilde{x}_{\tau}+ik a_l/2)=\exp(4k r(\sigma -a)) H_n(\gamma;\tilde{x}_{\tau}-ik a_l/2)
\nonumber\\
\qquad{}\times \prod\limits_{\mu=0}^7\prod\limits_{m=1}^k \big(1-(-)^{\tau}\exp[2r(\gamma_{\mu}+(k +1-2m)a_l/2)]\big),\qquad k
\in{\mathbb N}^*,\quad \tau=0,1,  \label{Hnident2}
\end{gather}
where
\begin{gather}
\label{tilx}
\tilde{x}_0:=ia_s/2,\qquad \tilde{x}_1:=\pi/2r+ia_s/2,
\end{gather}
and
\begin{gather}
H_n(\gamma;x_{\tau,k}+ia_s)=\exp(4k r(\sigma -a)) H_n(\gamma;x_{\tau,k}-ia_s)\label{Hnident3}
\\
\qquad{}\times \prod\limits_{\mu=0}^7\prod\limits_{m=1}^k \big(1-(-)^{\tau}\exp[2r(\gamma_{\mu}\pm a_s/2+(k +1-2m)a_l/2)]\big),\qquad
k \in{\mathbb N}^*,\quad \tau=0,1,\nonumber
\end{gather}
where
\begin{gather}
x_{0,k}:=ika_l/2,\qquad x_{1,k}:=\pi/2r+ika_l/2,\qquad k\in{\mathbb N}^*.
\end{gather}
\end{Theorem}

\begin{proof}
Recalling~\eqref{VHfin}, we see that~$V_s^H(-x)$ has no zeros in the UHP. Moreover, its poles in the UHP are simple and
located at
\begin{gather}
\label{up1}
x\equiv -ia_s/2+ika_l/2 \pmod{\pi/2r},\qquad k\in{\mathbb N}^*,
\end{gather}
and
\begin{gather}
\label{up2}
x\equiv ika_l/2 \pmod{\pi/2r},\qquad k\in{\mathbb N}^*.
\end{gather}

Likewise, $V_s^H(x)$ can only have at most simple poles in the UHP at
\begin{gather}
\label{dp1}
x\equiv ia_s/2+ika_l/2 \pmod{\pi/2r},\qquad k\in{\mathbb N},
\end{gather}
and
\begin{gather}
\label{dp2}
x\equiv ika_l/2 \pmod{\pi/2r},\qquad k\in{\mathbb N}^*.
\end{gather}
(The $G_t$-product yields zeros in the UHP at
\begin{gather}
x\equiv ia\pm i\gamma_{\mu}+i\ell a_l\pmod{\pi/r},\qquad \ell\in{\mathbb N},
\end{gather}
which may cancel some of the poles~\eqref{dp1},~\eqref{dp2}.)

Finally, the coef\/f\/icient of~$H_n(x)$ can only have at most simple UHP-poles at the locations~\eqref{up1}
and~\eqref{dp1}.

As announced above, we need only study the consequences of~\eqref{Hnade} for~$x$ equal to $it$ and~$it+\pi/2r$, with
$t>0$. As~$t$ increases, the f\/irst coef\/f\/icient poles that are met are those for $t=a_s/2$, cf.~\eqref{dp1}, or for
$t=-a_s/2+a_l/2$, cf.~\eqref{up1}. Since $H_n\in S_{\sigma+a_s}$ by Lemma~\ref{lemma2.4}, we see (using also~\eqref{VH}
and~\eqref{resr1}) that the residues at the former poles cancel. Now $V_s^H(-x)$ is regular and nonzero at these
locations, and so $H_n(x+ia_s)$ must be regular as well. Thus $H_n(x)$ cannot have poles for $\operatorname{Im} x=3a_s/2$.

Since the poles for $t=-a_s/2+a_l/2$ arise from the coef\/f\/icient of~$H_n(x+ia_s)$, and the second and third term
in~\eqref{Hnade} have at most simple poles, it follows that~$H_n(x+ia_s)$ must be regular for~$\operatorname{Im}
x=-a_s/2+a_l/2$, lest non-simple poles arise for the f\/irst term, which are not present for the other terms. In turn,
this implies that the values $H_n(\tilde{x}_{\tau}+ia_l/2)$ and $H_n(-\tilde{x}_{\tau}+ia_l/2)$ are related in such
a~way that the residues at the corresponding poles of $V_s^H(-x)$ and $V_{b,s}(x)$ cancel.

In order to obtain these relations and subsequent ones, we invoke the limits
\begin{gather}
\lim_{x\to \tilde{x}_{\tau}}\frac{P(\gamma;\tilde{x}_{\tau}+ika_l/2)}{P(\gamma;\tilde{x}_{\tau}-ika_l/2)}\nonumber\\
\qquad{} =
\prod\limits_{\mu=0}^7\prod\limits_{m=1}^k \big(1-(-)^{\tau}\exp[2r(\gamma_{\mu}+(k +1-2m)a_l/2)]\big),\qquad  k\in{\mathbb
N}^*,\quad \tau=0,1,\label{Pql}
\end{gather}
whose proof is easily adapted from the proof of~\eqref{Pid}. Using~\eqref{Pql} with $k=1$, combined with~\eqref{resr2},
we obtain the relations~\eqref{Hnident2} with~$k=1$. It is readily verif\/ied that these relations also ensure that the
residues at the poles~\eqref{dp1} with $k=1$ cancel, so that $H_n(x)$ is regular for $\operatorname{Im} x= 3a_s/2+a_l/2$ as
well.

The next pole of the coef\/f\/icient~$V_s^H(-x)$ occurs for $t=a_l/2$, cf.~\eqref{up2}. Once more, $H_n(x+ia_s)$ must be
regular to prevent the occurrence of non-simple poles. Again, this entails a~relation between the values
$H_n(\tilde{x}_{\tau}+ia_s)$ and $H_n(\tilde{x}_{\tau}-ia_s)$ encoding residue cancellation, which we shall return to
shortly. Letting~$t$ increase further, we get $V_s^H(-x)$-poles for $t=-a_s/2+a_l$, cf.~\eqref{up1}, at which
$H_n(x+ia_s)$ must be regular. This gives rise to the residue cancellation identity~\eqref{Hnident2} with $k=2$,
obtained via~\eqref{Pql} with $k=2$ and~\eqref{resr1}. This identity also ensures residue cancellation at the
poles~\eqref{dp1} with $k=2$, implying regularity of $H_n(x)$ for $\operatorname{Im} x= 3a_s/2+a_l$.

Clearly, this line of reasoning inductively yields entireness of~$H_n(x)$, with the identities~\eqref{Hnident2}
guaranteeing that the residues at the poles~\eqref{up1} and~\eqref{dp1} cancel.

The identities~\eqref{Hnident3} ensure that the residues at the $V_s^H(\pm x)$-locations~\eqref{up2} and~\eqref{dp2}
cancel. In more detail, we can invoke the easily verif\/ied limit
\begin{gather}
\lim_{x\to x_{\tau,k}}\frac{V_s(x)}{V_s(-x)}=-\exp(8kr(\sigma-a)), \qquad k\in{\mathbb N}^*,\qquad \tau =0,1,
\end{gather}
and f\/irst write
\begin{gather}
\frac{P(x+ia_s)}{P(x-ia_s)}=\prod\limits_{\mu=0}^7\frac{G_t(a_l;x-i\gamma_{\mu}\pm ia_s/2)}{G_t(a_l;-x-i\gamma_{\mu}\pm
ia_s/2)}.
\end{gather}
If we then let~$x$ converge to $x_{\tau,k}$, we arrive at the product in~\eqref{Hnident3}.
\end{proof}

We conjecture that $H_n(\gamma;x)$ is entire for the two excluded ratios $a_s/a_l=1,1/2$ as well. In this connection we
add that the identities~\eqref{HHsp} for $a_s=a_l$ and~\eqref{Hnident} with $\ell=2$ for $a_s=a_l/2$ can be used to show
that the functions~$A_s(\gamma;x)F_n(\gamma;x)$, $g\in\Pi_r$, are regular on ${\mathbb R}$ (even though the A$\Delta$O
coef\/f\/icients have second-order real poles), and hence belong to ${\cal H}_w$~\eqref{cHw}. This strongly suggests that
for the excluded ratios the functions~$F_n(\gamma;x)$ from Section~\ref{section2} are once more eigenfunctions of~$A_s(\gamma;x)$.

We point out that the identities~\eqref{Hnident2} correspond to~\eqref{Hnident} with~$a_s$ and~$a_l$ interchanged.
Moreover, for $k=1$ the identities~\eqref{Hnident2} and~\eqref{Hnident3} \emph{coincide} with~\eqref{Hnident} for
$\ell=1$ and~$\ell=2$, respectively. Note that this implies that the latter identity also holds for $a_s\in (a_l/2,a_l)$
(entailing $L=1$).

We conclude this section by deriving some insight into the degeneracy issue. We shall improve on this later on, so the
following lemma has an auxiliary character. We stress that the issue of degeneracy of the $E_{n,s}$-eigenspace refers to
the Hilbert space operator~$\hat{{\cal A}}_s(\gamma)$, and not to the A$\Delta$O~${\cal A}_s(\gamma;x)$. For the latter,
we know that its $E_{n,s}$-eigenspace contains the meromorphic function~$f_n(\gamma;x)$. Therefore, it also contains all
functions of the form~$\zeta(x)f_n(\gamma;x)$ with~$\zeta\in{\cal P}(a_s)$, cf.~\eqref{cP}. However, since the
coef\/f\/icients of the A$\Delta$O belong to~${\cal P}(a_l)$, all functions~$f_n(\gamma;x+ika_l)$ with~$k\in{\mathbb Z}$ are
also in the eigenspace, and so are their linear combinations with coef\/f\/icients from~${\cal P}(a_s)$. The lemma shows
that the $E_{n,s}$-eigenspace of~$\hat{{\cal A}}_s(\gamma)$ is far more restricted.

\begin{Lemma}
\label{lemma4.5}
Let $a_+$, $a_-$ satisfy~\eqref{aa1}. Fixing $n_0\in{\mathbb N}$, there are at most finitely many distinct~$n_1,\ldots,n_M$
for which
\begin{gather}
E_{n_0,s}(\gamma)=E_{n_j,s}(\gamma), \qquad j=1,\ldots,M.
\end{gather}
Moreover, for any such $n_j$, there exists $\zeta_j\in{\cal P}(a_s)$ such that
\begin{gather}
f_{n_j}(\gamma;x)=\zeta_j(x)f_{n_0}(\gamma;x).
\end{gather}
\end{Lemma}
\begin{proof}
Assuming $E_{n_0,s}$ is degenerate, consider the Casorati determinant of $f_{n_0}(x)$ and~$f_{n_1}(x)$. It is given~by
the meromorphic function
\begin{gather}
{\cal D}(x):=f_{n_0}(x+ia_s/2)f_{n_1}(x-ia_s/2)-f_{n_0}(x-ia_s/2)f_{n_1}(x+ia_s/2),
\end{gather}
and this function satisf\/ies the f\/irst order A$\Delta$E
\begin{gather}
\frac{{\cal D}(x+ia_s/2)}{{\cal D}(x-ia_s/2)}=\frac{1}{V_{a,s}(x)}=\frac{c(x)c(-x-ia_s)}{c(-x)c(x+ia_s)},
\end{gather}
cf.~\eqref{cApmform} and~\eqref{Varel}. Therefore, it is of the form
\begin{gather}
\label{cDf}
{\cal D}(x)=\zeta(x)c(-x-ia_s/2)/c(x+ia_s/2),\qquad \zeta\in{\cal P}(a_s).
\end{gather}

Now we have
\begin{gather}
{\cal D}_H(x):=P(x\pm ia_s/2)c(x\pm ia_s/2){\cal D}(x)
\nonumber\\
\hphantom{{\cal D}_H(x)~}{}
=H_{n_0}(x+ia_s/2)H_{n_1}(x-ia_s/2)-H_{n_0}(x-ia_s/2)H_{n_1}(x+ia_s/2),
\end{gather}
so by Theorem~\ref{theorem4.3} the function~${\cal D}_H(x)$ is holomorphic. In view of~\eqref{cDf}, it can be written as
\begin{gather}
{\cal D}_H(x)=\zeta(x)P(x\pm ia_s/2)c(\pm x- ia_s/2)
\nonumber\\
\hphantom{{\cal D}_H(x)}{}
=\zeta(x)\frac{R_l(2x+ia_l/2)}{R_s(2x+ia_s/2)}\prod\limits_{\mu=0}^7E(\pm x -ia_s/2\pm i\gamma_{\mu}).\label{cDH}
\end{gather}
(We used~\eqref{c},~\eqref{GE} and~\eqref{Gades} in the second step.)

We claim that $\zeta(x)$ must be constant. Indeed, assuming it is not, it has a~sequence of poles~$p_0+ika_s$,
$k\in{\mathbb Z}$. Consider now those zeros of the four factors of the~$E$-product involving~$\gamma_{\mu}$ that can be
at minimal distance to the real line. They are located at
\begin{gather}
\pm x\equiv ia+ia_s/2\pm i\gamma_{\mu} \pmod{\pi/r}.
\end{gather}
It easily follows from this that the distance of the UHP-zeros to the LHP-zeros of the~$E$-product is greater than~$a_s$
(recall $|g_{\mu}|<a$). The same is true for the non-real zeros of the factor $R_l(2x-ia_l/2)$. The poles of~$\zeta(x)$
can therefore not be matched by zeros, contradicting holomorphy of~${\cal D}_H(x)$. Hence $\zeta(x)$ must be constant.

Next, we note that the denominator function has a~zero for~$x=ia_s/2$ that is not matched by a~zero of the numerator,
unless $\zeta(x)=0$. As a~result, the function~${\cal D}_H(x)$ vanishes identically. From this we deduce
\begin{gather}
H_{n_1}(x)=\zeta_1(x)H_{n_0}(x),\qquad \zeta_1(x)\in{\cal P}(a_s).
\end{gather}
Finally, we note that by holomorphy $H_{n_0}(x)$ can only have f\/initely many zeros in the rectangle with
corners~$0,\pi/r,\pi/r+ia_s,ia_s$. Thus the vector space of multipliers in~${\cal P}(a_s)$ such that their product
with~$H_{n_0}(x)$ is holomorphic is f\/inite-dimensional. From this the lemma readily follows.
\end{proof}

\section[The operator $\hat{{\cal A}}_l(\gamma)$, $g\in \Pi_r$]{The operator $\boldsymbol{\hat{{\cal A}}_l(\gamma)}$, $\boldsymbol{g\in \Pi_r}$}\label{section5}

In this section we focus on the Hilbert space features of the A$\Delta$O~${\cal A}_l(\gamma;x)$, assuming~$g\in\Pi_r$
and~\eqref{aa2} throughout. In Section~\ref{section3} we have seen that the action of~${\cal A}_l(\gamma;x)$ on functions
in~$D_t(\gamma)$ yields functions whose restrictions to~$[0,\pi/2r]$ are in~${\cal H}$, provided~$t>a_l$. Moreover, the
Hilbert space operator thus obtained is symmetric and bounded below.

For generic $g\in\Pi_r$ and $t>a$, however, the functions $f_n(\gamma;x)$ do \emph{not} belong to~$D_t(\gamma)$ for all
$n\in{\mathbb N}$. This will become clear shortly, but for now it may be illuminating to see why this is plausible~by
recalling Lemma~\ref{lemma4.1} and its proof. The point is that in general $H_n(\gamma;x)$ has no reason to vanish for
$x=ia$, so that~$t$ in~\eqref{fnDt} cannot be chosen larger than~$a$.

We therefore proceed in a~dif\/ferent way. Consider the action of the A$\Delta$O $A^H_l(\gamma;x)$~\eqref{AHpm} on the
holomorphic functions $H_n(\gamma;x)$. It is given~by
\begin{gather}
A^H_l(x)H_n(x)=V_l^H(-x) H_n(x+ia_l)+V_l^H(x) H_n(x-ia_l)+V_{b,l}(x)H_n(x).
\end{gather}
For generic~$g$ the coef\/f\/icients~$V_l^H(\pm x)$ have simple poles for $x\equiv 0\pmod{\pi/2r}$, but since $H_n(x)$ is
even and $\pi/r$-periodic, the (a priori meromorphic) function on the right-hand side has no real poles. Therefore it
belongs to~${\cal H}_{w_H}$.

As a~consequence, the A$\Delta$O~${\cal A}_l(\gamma;x)$ maps functions in the span~${\cal C}(\gamma)$ of the ONB
functions~$f_n(\gamma;x)$ to functions whose restrictions to~$[0,\pi/2r]$ belong to~${\cal H}$. (In particular, the
image functions have no real poles, whereas $V_{a,l}(\gamma;x)$ does have real poles.) We denote the operator thus
obtained by~$\hat{{\cal A}}_l(\gamma)$. We are now in the position to state and prove the f\/irst result of this section.

\begin{Lemma}
\label{lemma5.1}
Assuming~\eqref{aa2} and $g\in\Pi_r$, the operator $\hat{{\cal A}}_l(\gamma)$ is symmetric on ${\cal C}(\gamma)$.
\end{Lemma}
\begin{proof}
We reason in the same way as in the proof of Theorem~\ref{theorem3.1}, but now choose~$f_j(x)\in{\cal C}(\gamma)$,
$j=1,2$. Then we set
\begin{gather}
\phi_j(x)\equiv c(\gamma;x) f_j(x),\qquad j=1,2,
\end{gather}
yielding even functions that belong to the span of the ${\cal H}_{w}$-ONB given by~$\{F_n(\gamma;x)\}_{n=0}^{\infty}$.
Now we get as the analogs of~\eqref{IL},~\eqref{IR}
\begin{gather}
\label{ILl}
I_L\equiv \int_{0}^{\pi/2r} \big(V_l(\gamma;-x)\phi_1^{*}(x+ia_l) +V_l(\gamma;x) \phi_1^{*}(x-ia_l) \big)
\phi_2(x)w(\gamma;x)dx,
\end{gather}
and
\begin{gather}
\label{IRl}
I_R\equiv \int_{0}^{\pi/2r} \phi_1^{*}(x)\big(V_l(\gamma;x)\phi_2(x-ia_l)+V_l(\gamma;-x) \phi_2(x+ia_s)\big)
w(\gamma;x)dx.
\end{gather}
Next, we introduce
\begin{gather}
I_l(x)   \equiv   V_l(\gamma;-x+ia_l/2)\phi_1^*(x+ia_l/2)\phi_2(x-ia_l/2)w(\gamma;x-ia_l/2)
\nonumber
\\
\hphantom{I_l(x)}{}  =   \frac{\phi_1^*(x+ia_l/2)\phi_2(x-ia_l/2)}{c(\gamma;\pm x-ia_l/2)}=f_1^*(x+ia_l/2)f_2(x-ia_l/2),
\end{gather}
where we used~\eqref{Vec} and~\eqref{w} in the second step. Then we obtain, using evenness and $\pi/r$-periodicity,
\begin{gather}
\label{alint}
I_L-I_R= \int_{-\pi/4r}^{3\pi/4r} [I_l(x+ia_l/2)-I_l(x-ia_l/2)]dx.
\end{gather}

In order to show that the integral~\eqref{alint} vanishes, we rewrite $I_l(x)$ in terms of the functions
\begin{gather}
h_j(x)\equiv P(\gamma;x)\phi_j(x),\qquad j=1,2.
\end{gather}
More specif\/ically, we set
\begin{gather}
I_l(x)=h_1^*(x+ia_l/2)h_2(x-ia_l/2)/Q(x),
\end{gather}
so that
\begin{gather}
Q(x)=P(x\pm ia_l/2)c(\pm x-ia_l/2) = \frac{R_s(2x+ia_s/2)}{R_l(2x+ia_l/2)}\prod\limits_{\mu=0}^7E(\pm x-ia_l/2\pm
i\gamma_{\mu}),
\end{gather}
cf.~\eqref{cDH}.

The function $1/Q(x)$ is even and its only poles in the strip~$|\operatorname{Im} x|\le a_l/2$ are simple and located at
\begin{gather}
x\equiv \pm i\ell a_s/2,\qquad \ell=1,\ldots,L \pmod{\pi/2r}.
\end{gather}
By evenness and $\pi/r$-periodicity, its residue sums at $x=\pm i\ell a_s/2$ and at $x=\pi/2r \pm i\ell a_s/2$ vanish.
Using Cauchy's theorem and evenness of $h_2(x)$, we then deduce
\begin{gather}
I_L-I_R=\sum\limits_{\tau=0,1}\sum\limits_{\ell=1}^L\rho_{\ell,\tau}(\gamma)\big[h_1^*(x_{\tau}-i\ell
a_s/2)h_2(x_{\tau}+i\ell a_s/2)
\nonumber\\
\hphantom{I_L-I_R=}{} - h_1^*(x_{\tau}+i\ell a_s/2)h_2(x_{\tau}-i\ell a_s/2)\big].
\end{gather}
Here we have
\begin{gather}
\rho_{\ell,\tau}(\gamma):=i\pi\frac{R_l(x_{\tau}+i\ell a_s)}{R_s'(\tilde{x}_{\tau}+i\ell a_s)}
\prod\limits_{\mu=0}^7\frac{1}{E(-x_{\tau}\pm i\ell a_s/2\pm i\gamma_{\mu})},
\end{gather}
and $x_{\tau}$/$\tilde{x}_{\tau}$ are given by~\eqref{xtau}/\eqref{tilx}.

Now the holomorphic functions~$H_n(\gamma;x)$ satisfy
\begin{gather}
H_n^*(\gamma;x)=H_n(\gamma;x),\qquad n\in{\mathbb N}, \qquad g\in\Pi_r,\qquad  x\in{\mathbb C},
\end{gather}
since they are real-valued for real~$x$. Since $h_1$ is a~linear combination of the $H_n$'s, it follows
from~\eqref{Hnident} that it satisf\/ies
\begin{gather}
\label{h1ident}
h_1^*(x_{\tau}+i\ell a_s/2)=\pi_{\ell,\tau}(\gamma) h_1^*(x_{\tau}-i\ell a_s/2).
\end{gather}
Likewise, these identities hold when~$h_1^*$ is replaced by~$h_2$. Therefore, the dif\/ference terms in square brackets
all vanish. As a~result we get $I_L=I_R$, and so it follows that~$\hat{{\cal A}}_l(\gamma)$ is symmetric on~${\cal
C}(\gamma)$.
\end{proof}

This proof reveals that the analog of the lower bound~\eqref{lowb} does not follow in the same way as in the proof of
Theorem~\ref{theorem3.1}. Indeed, we now get for $0\ne f\in{\cal C}(\gamma)$ an identity
\begin{gather}
(f,(\hat{{\cal A}}_l(\gamma)-V_{b,l}(\gamma;\cdot))f)= \int_{-\pi/2r}^{\pi/2r}|f(x-ia_l/2)|^2 dx\nonumber\\
\hphantom{(f,(\hat{{\cal A}}_l(\gamma)-V_{b,l}(\gamma;\cdot))f)=}{}
+\sum\limits_{\tau=0,1}\sum\limits_{\ell=1}^L\rho_{\ell,\tau}(\gamma)\pi_{\ell,\tau}(\gamma) |h(x_{\tau}-i\ell
a_s/2)|^2,
\end{gather}
with $h(x)=P(\gamma;x)c(\gamma;x)f(x)$. Since we have~$(-)^{\ell}\rho_{\ell,\tau}(\gamma)<0$ and the sign of the numbers
$\pi_{\ell,\tau}(\gamma)$ depends on~$\gamma$, the residue sum is not positive in general.

At this stage we can also explain why the symmetric operator~$\hat{{\cal A}}_l^w(t,\gamma)$ from Section~\ref{section3} has the
suf\/f\/ix `w' (for `wrong'). To this end we need only make some quite weak assumptions on~$g\in\Pi_r$. For example, it
suf\/f\/ices to assume
\begin{gather}
\sigma>a_l/2-La_s/2,
\end{gather}
together with
\begin{gather}
\label{a2}
\pi_{L,0}(\gamma)\ne 0.
\end{gather}
The f\/irst assumption entails that we have
\begin{gather}
\lambda_nH_n(\gamma;\frac{i}{2}(a_l-La_s))=P\left(\gamma;\frac{i}{2}(a_l-La_s)\right)\nonumber\\
\hphantom{\lambda_nH_n(\gamma;\frac{i}{2}(a_l-La_s))=}{}
\times \int_0^{\pi/2r}{\cal
S}(\sigma;\frac{i}{2}(a_l-La_s)),y)w(\gamma';y)F_n(\gamma';y)dy,
\end{gather}
cf.~\eqref{FnF}. The functions~$P(\gamma;x)$ and~${\cal S}(\sigma;x,y)$, $y\in{\mathbb R}$, are both positive for
$x=i(a_l-La_s)/2$, so by completeness the left-hand side cannot vanish for all $n\in{\mathbb N}$; in particular, it is
positive for~$n=0$. Using next the second assumption~\eqref{a2}, we deduce that~$H_n(\gamma;i(a_l+L a_s)/2)$ does not
vanish for all $n\in{\mathbb N}$. But~$E(2x-ia)$ does vanish for $x=i(a_l+La_s)/2$, and since we have (cf.~the proof of
Lemma~\ref{lemma4.1})
\begin{gather}
c_P(\gamma;x)f_n(\gamma;x)=H_n(\gamma;x)/E(\pm 2x-ia),
\end{gather}
it follows that there exist $f_n $ for which $c_P(x)f_n(x)$ has a~pole at $i(a_l+La_s)/2$. Because $ (a_l+La_s)/2 <a_l$
by~\eqref{Ldef}, we conclude that for these~$n$-values (including $n=0$) the critical functions~$f_n(\gamma;x)$ are not
in the def\/inition domain~$D_t(\gamma)$ of~$\hat{{\cal A}}_l^w(t,\gamma)$. (Recall we need to impose $t>a_l$ to ensure
symmetry of~$\hat{{\cal A}}_l^w(t,\gamma)$ on~$D_t(\gamma)$.)

A priori, however, these functions could still belong to the domain of a~self-adjoint extension of~$\hat{{\cal
A}}_l^w(t,\gamma)$. (Recall that at least one self-adjoint extension exists, since $\hat{{\cal A}}_l^w(t,\gamma)$ is
bounded below.) Now our third assumption is that the action of this extension on the $f_n(\gamma;x)$ in question is
given by the action of the A$\Delta$O~${\cal A}_l(\gamma;x)$. (Obviously, this is the case when the self-adjoint
extension equals~$\hat{{\cal A}}_l(\gamma)$ on~${\cal C}(\gamma)$.)

To arrive at a~contradiction, we f\/ix an exceptional~$n$; we may as well take $n=0$, since it is always exceptional. Then
our third assumption implies that $I_L$~\eqref{ILl} and~$I_R$~\eqref{IRl} are equal for the choices
\begin{gather}
\phi_2(x)=c(\gamma;x)f_0(\gamma;x),\qquad \phi_1(x)=c(\gamma;x)f(x)/c_P(\gamma;x),\qquad f\in S_t,\qquad t>a_l.
\end{gather}
On the other hand, when we follow the same path as before, we arrive at
\begin{gather}
I_L-I_R=\sum\limits_{\tau=0,1}\sum\limits_{\ell=1}^L\rho_{\ell,\tau}(\gamma)H_0(\gamma;x_{\tau}-i\ell
a_s/2)\nonumber\\
\hphantom{I_L-I_R=}{}
\times \big[h_1^*(x_{\tau}-i\ell a_s/2)\pi_{\ell,\tau}(\gamma) - h_1^*(x_{\tau}+i\ell a_s/2) \big].\label{ressumsp}
\end{gather}
(Indeed, the function $h_1(x)$ is holomorphic for $|\operatorname{Im} x|<t$, since it equals~$E(\pm x-ia)f(x)$, cf.~the proof
of Lemma~\ref{lemma4.1}.) This equality holds true for arbitrary $f\in S_t$, so we can ensure that~$h_1^*(x_0-iLa_s/2)$
is nonzero, whereas all other values~$h_1^*(x_{\tau}\pm i\ell a_s/2)$ vanish. In view of~\eqref{a2} this implies that
the right-hand side is nonzero, yielding the desired contradiction.

More generally, just as the proof of Lemma~\ref{lemma5.1}, this reasoning shows the crucial role of the
identities~\eqref{Hnident}. It should be stressed that there appears to be no way to prove them without the detailed
study undertaken in Subsections~\ref{section2.3} and~\ref{section2.4}. To be sure, they could be imposed on the initial domain
of~$\hat{A}^H_l(\gamma)$ from the outset, but then it would be fully unclear that dense subspaces of~${\cal H}_{w_H}$
with these features \emph{exist}. In that connection we point out that for the desired residue cancellation to take
place, it already suf\/f\/ices that the numbers~$\pi_{\ell,\tau}(\gamma)$ in~\eqref{h1ident} are real. (That is, their
precise values are irrelevant.)

Of course, since ${\cal A}_s(\gamma;x)$ and~${\cal A}_l(\gamma;x)$ commute, the functions~${\cal
A}_l(\gamma;x)f_n(\gamma;x)$ are eigenfunctions of ${\cal A}_s(\gamma;x)$ with eigenvalue $E_{n,s}$. However, we cannot
conclude that they (or rather their restrictions to~$[0,\pi/2r]$) belong to the $E_{n,s}$-eigenspace of~$\hat{{\cal
A}}_s(\gamma)$, unless they belong to~${\cal C}(\gamma)$. From the following counterparts of Lemma~\ref{lemma4.2} and
Theorem~\ref{theorem4.3} it follows in particular that this is indeed the case. The proof of the lemma is relegated to
Appendix~\ref{appendixC}.

\begin{Lemma}
\label{lemma5.2}
Let~$g=\operatorname{Re} \gamma\in\Pi_r$ and let $a_{\pm}$ satisfy~\eqref{aa2}. Then we have
\begin{gather}
\hat{{\cal A}}_l(\gamma'){\cal I}(\gamma')f={\cal I}(\gamma') \hat{{\cal A}}_l(\gamma)f,\qquad \forall\, f\in {\cal
C}(\gamma).
\end{gather}
\end{Lemma}

\begin{Theorem}
\label{theorem5.3}
With the assumptions of Lemma~{\rm \ref{lemma5.2}}, the operator $\hat{{\cal A}}_l(\gamma)$ is essentially self-adjoint on
${\cal C}(\gamma)$. Its self-adjoint closure $($denoted by the same symbol$)$ has solely discrete spectrum and it admits
eigenvectors that belong to the finite-dimensional eigenspaces of the positive trace class operator~${\cal T}(\gamma)$.
These eigenvectors can be chosen to be eigenvectors of~$\hat{{\cal A}}_s(\gamma)$ as well.
\end{Theorem}

\begin{proof}
The f\/irst two assertions follow by adapting the proof of Theorem~\ref{theorem4.3} in an obvious way. Also, the action of
the operators~$\hat{{\cal A}}_{\pm}(\gamma)$ on the f\/inite-dimensional eigenspace~${\cal E}_n(\gamma)$ of~${\cal
T}(\gamma)$ with eigenvalue~$\lambda_n(\gamma)^2$ is given by the A$\Delta$O-action. Since ${\cal A}_s(\gamma;x)$ and
${\cal A}_l(\gamma;x)$ commute, we can choose a~new ONB that consists of joint eigenvectors of~$\hat{{\cal
A}}_{\pm}(\gamma)$.
\end{proof}

In Section~\ref{section4} we switched to an ONB of joint eigenvectors of~${\cal T}(\gamma)$ and~$\hat{{\cal A}}_s(\gamma)$, but of
course a~new choice for $f_n(\gamma;x)$ is not needed when the eigenvalue~$\lambda_n(\gamma)^2$ of~${\cal T}(\gamma)$ is
nondegene\-ra\-te. Likewise, we need not change the ONB from Section~\ref{section4} whenever all joint eigenvalue
pairs~$(\lambda_n(\gamma)^2,E_{n,s}(\gamma))$ are nondegenerate. In fact, it may well be that already~${\cal T}(\gamma)$
has nondegenerate spectrum, but a~further analysis of this issue seems elusive.

We shall return to the joint degeneracy issue shortly, but f\/irst we obtain the $H_n$-identities that express residue
cancellation at the poles occurring in the eigenvalue A$\Delta$E following from Theorem~\ref{theorem5.3}:
\begin{gather}
V_l^H(-x) H_n(x+ia_l)+V_l^H(x) H_n(x-ia_l)+(V_{b,l}(x)-E_{n,l})H_n(x)=0.
\end{gather}
(Recall entireness of~$H_n(x)$ has already been proved in Theorem~\ref{theorem4.4}.)

\begin{Corollary}
\label{corollary5.4}
Let $(n,a_{+},a_{-},g)\in {\mathbb N}\times (0,\infty)^2\times \Pi_r$ and let~$a_+$, $a_-$ satisfy~\eqref{aa2}. With~$L$
defined by~\eqref{Ldef}, $x_{\tau}$ by~\eqref{xtau}, and $\pi_{\ell,\tau}(\gamma)$ by~\eqref{pijtau}, $H_n(\gamma;x)$
obeys
\begin{gather}
\label{Hnident4}
H_n(\gamma;x_{\tau}+i\ell a_s/2)= \pi_{\ell,\tau}(\gamma) H_n(\gamma;x_{\tau}-i\ell a_s/2),\qquad \ell >L,\qquad \tau=0,1.
\end{gather}
Also, setting
\begin{gather}
\tilde{x}_{0,k}:=ika_s/2,\qquad  \tilde{x}_{1,k}:=\pi/2r+ika_s/2,\qquad k\in{\mathbb N}^*,
\end{gather}
we have
\begin{gather}
H_n(\gamma;\tilde{x}_{\tau,k}+ia_l)=\exp(4k r(\sigma -a)) H_n(\gamma;\tilde{x}_{\tau,k}-ia_l) \label{Hnident5}
\\
\qquad{}\times \prod\limits_{\mu=0}^7\prod\limits_{m=1}^k \big(1-(-)^{\tau}\exp[2r(\gamma_{\mu}\pm a_l/2+(k +1-2m)a_s/2)]\big),\qquad
k \in{\mathbb N}^*,\quad \tau=0,1.\nonumber
\end{gather}
\end{Corollary}

\begin{proof}
From~\eqref{VHfin} we see that~$V_l^H(-x)$ has no zeros in the UHP, while its poles in the UHP are simple and located at
\begin{gather}
\label{up1l}
x\equiv -ia_l/2+i\ell a_s/2 \pmod{\pi/2r},\qquad \ell\ge L+1,
\end{gather}
and
\begin{gather}
\label{up2l}
x\equiv ika_s/2 \pmod{\pi/2r},\qquad k\in{\mathbb N}^*.
\end{gather}
Moreover, $V_l^H(x)$ can only have at most simple poles in the UHP at
\begin{gather}
\label{dp1l}
x\equiv ia_l/2+i\ell a_s/2 \pmod{\pi/2r},\qquad \ell\ge -L,
\end{gather}
and
\begin{gather}
\label{dp2l}
x\equiv ika_s/2 \pmod{\pi/2r},\qquad k\in{\mathbb N}^*.
\end{gather}
(The $G_t$-product may cancel some of the poles~\eqref{dp1l},~\eqref{dp2l}.) Finally, the coef\/f\/icient of~$H_n(x)$ can
only have at most simple UHP-poles at the locations~\eqref{up1l} and~\eqref{dp1l}.

The identities~\eqref{Hnident4}/\eqref{Hnident5} now encode residue cancellation at the poles~\eqref{up1l}/\eqref{up2l}
of the coef\/f\/icient~$V_l^H(-x)$ of~$H_n(x+ia_l)$. This follows in the same way as in the proof of
Theorem~\ref{theorem4.4}, noting that the ratio restriction~\eqref{aa2} entails that the residues at the simple poles
of~$V_l(x)$ and~$V_{b,l}(x)$ are related by~\eqref{resr1},~\eqref{resr2} with $s\to l$. We point out that residue
cancellation at the poles~\eqref{dp1l} for $|\ell|\le L$ is ensured by~\eqref{Hnident}.
\end{proof}

As already mentioned, we believe $H_n(\gamma;x)$ is entire without restrictions on the ratio $a_s/a_l$. Likewise, we
surmise that the $H_n$-identities in Theorem~\ref{theorem4.4} and Corollary~\ref{corollary5.4} hold true for arbitrary
ratios.

We continue with the counterpart of Lemma~\ref{lemma4.5}.

\begin{Lemma}
\label{lemma5.5}
Let $a_+$, $a_-$ satisfy~\eqref{aa2}. Fixing $n_0\in{\mathbb N}$, there are at most finitely many distinct~$n_1,\ldots,n_M$
for which
\begin{gather}
E_{n_0,l}(\gamma)=E_{n_j,l}(\gamma), \qquad j=1,\ldots,M.
\end{gather}
Moreover, for any such $n_j$, there exists $\zeta_j\in{\cal P}(a_l)$ such that
\begin{gather}
f_{n_j}(\gamma;x)=\zeta_j(x)f_{n_0}(\gamma;x).
\end{gather}
\end{Lemma}

\begin{proof}
The proof of Lemma~\ref{lemma4.5} can be easily adapted.
\end{proof}

We are now in the position to obtain a~remarkable nondegeneracy result.

\begin{Theorem}
\label{theorem5.6}
Assume $\gamma\in\Pi_r$ and $a_+/a_-$ is irrational. Then all joint eigenspaces of the operators~$\hat{{\cal
A}}_{\pm}(\gamma)$ are one-dimensional.
\end{Theorem}
\begin{proof}
We assume
\begin{gather}
(E_{m,+}(\gamma),E_{m,-}(\gamma))=(E_{n,+}(\gamma),E_{n,-}(\gamma)),\qquad m\ne n,
\end{gather}
so as to derive a~contradiction. From Lemma~\ref{lemma4.5} and $E_{m,s}(\gamma)=E_{n,s}(\gamma)$ we infer
\begin{gather}
H_n(\gamma;x)=\zeta_s(x)H_m(\gamma;x),\qquad \zeta_s\in{\cal P}(a_s).
\end{gather}
Similarly, from Lemma~\ref{lemma5.5} and $E_{m,l}(\gamma)=E_{n,l}(\gamma)$ we obtain
\begin{gather}
H_n(\gamma;x)=\zeta_l(x)H_m(\gamma;x),\qquad \zeta_l\in{\cal P}(a_l).
\end{gather}
Thus we have $\zeta_s=\zeta_l=:\zeta$ with~$\zeta$ in ${\cal P}(a_s)\cap{\cal P}(a_l)$. Since the latter intersection
consists of the constants when $a_s/a_l\notin{\mathbb Q}$, we arrive at the announced contradiction.
\end{proof}

\section{Weyl group spectral symmetry}\label{section6}

The principal aim of this section is to obtain more information on the eigenvalues~$E_{n,\pm}(\gamma)$ of the
self-adjoint Hilbert space operators $\hat{{\cal A}}_{\pm}(\gamma)$. Along the way, however, we obtain a~commutativity
result about the trace class operators ${\cal T}(\gamma)$ that is of considerable interest in itself. We recall that for
the latter operators we restrict attention to~$\gamma$'s that obey~$g=\operatorname{Re} \gamma \in\Pi_r$ and that are either
real or satisfy~\eqref{imgam},~\eqref{sumres}. By contrast, for $\hat{{\cal A}}_{\pm}(\gamma)$ we can allow more
general~$\gamma$'s by exploiting the $D_8$-invariance of the A$\Delta$Os~${\cal A}_{\pm}(\gamma;x)$. We begin with
a~lemma that is an easy corollary of our previous results.

\begin{Lemma}
\label{lemma6.1}
Let $g\in\Pi_r$ and let $a_{\pm}$ satisfy~\eqref{aa1}/\eqref{aa2} for~$\hat{{\cal A}}_{s}(\gamma)$/$\hat{{\cal
A}}_{l}(\gamma)$, respectively. Then we have
\begin{gather}
\label{Ens}
E_{n,s}(\gamma)=E_{n,s}(\gamma'),
\\
\label{Enl}
E_{n,l}(\gamma)=E_{n,l}(\gamma').
\end{gather}
\end{Lemma}

\begin{proof}
This follows from Lemma~\ref{lemma4.2} and Theorem~\ref{theorem4.3} for the case~\eqref{Ens}, and from
Lemma~\ref{lemma5.2} and Theorem~\ref{theorem5.3} for~\eqref{Enl}. Indeed, ${\cal I}(\gamma')$ maps the eigenvector
$f_n(\gamma;\cdot)$ to a~positive multiple of the eigenvector $f_n(\gamma';\cdot)$.
\end{proof}

We recall from Subsection~\ref{section2.1} that~$\gamma$ is allowed to vary over two distinct regimes. In both cases, we introduced
subsets $\Pi_r^s(j)\subset\Pi_r$, $j=1,2$, for which the self-duality relation~\eqref{selfd} holds true, by requiring
the relevant permutation symmetry. Since $V_{\delta}$ and $V_{a,\delta}$ are determined by the~$c$-function
(cf.~\eqref{Vec} and~\eqref{Va}), and since the product functions $p_{t,\delta}$ def\/ining $V_{b,\delta}$
(cf.~\eqref{Vbe}--\eqref{cEt}) are invariant under these permutations, we deduce
\begin{gather}
\hat{{\cal A}}_{\pm}(\gamma)=\hat{{\cal A}}_{\pm}(\gamma'),\qquad g=\operatorname{Re} \gamma\in \Pi_r^s(j),\qquad j=1,2.
\end{gather}
For these self-dual~$\gamma$-choices, therefore, the equations~\eqref{Ens},~\eqref{Enl} reduce to tautologies.

On the other hand, even sign f\/lips need not leave the self-dual subsets invariant. In fact, f\/lipping all of the eight
signs entails $\sigma\to -\sigma$, so that this transformation maps \emph{all} $g\in\Pi_r$ out of~$\Pi_r$. The latter
map is well def\/ined for both~$\gamma$-regimes, but we recall that for the second regime we only allow~$\phi$ in the
group~$\Phi$ of even sign f\/lips for which the $\phi(\gamma)$-component sum is real, cf.~\eqref{imgam},~\eqref{sumres}.

To ease the exposition we now specialize to the f\/irst regime until further notice. We recall that for $\gamma\in\Pi_r$
the Hilbert space action of $\hat{{\cal A}}_{\pm}(\gamma)$ on the core ${\cal C}(\gamma)$ is given by the action of the
A$\Delta$Os ${\cal A}_{\pm}(\gamma;x)$. The latter are manifestly $D_8$-invariant for the f\/irst regime, so at face value
it would seem that operators $\hat{{\cal A}}_{\pm}(\gamma)$ whose~$\gamma$'s belong to $\Pi_r$ and are related via even
sign f\/lips coincide. In fact, however, a~further analysis reveals that this is not at all obvious.

To f\/ix the thoughts, let us choose $\gamma_0$ close to $-a$ and~$\gamma_1,\ldots,\gamma_7$ close to $0$ and rationally
independent. Then we can freely f\/lip an even number of signs among $\gamma_1,\ldots,\gamma_7$, yet stay in~$\Pi_r$. The
$\sigma(\gamma)$'s obtained from these f\/lips are distinct, yielding 64 distinct positive trace class opera\-tors~${\cal
T}(\gamma)$. A~priori, therefore, we obtain 64 distinct ONBs of ${\cal T}(\gamma)$-eigenvectors. The dif\/f\/iculty is now
that we cannot show that these ONBs coincide when we only use the $D_8$-invariance of the A$\Delta$Os; indeed, this
$D_8$-symmetry is even compatible with getting 64 disjoint spectra for~$\hat{{\cal A}}_s(\gamma)$ and~$\hat{{\cal
A}}_l(\gamma)$.

It follows in particular from our next result that all of these spectra (and indeed the operators themselves) are
actually equal. Even more remarkably, it also implies that the 64 distinct trace class operators form a~commutative
family. As the proof reveals, these two facts are intimately related. To state the general assertions, we need the new
notion of a~$\gamma$-cluster.

For a~given $\dot{\gamma}\in\Pi_r$, we def\/ine this cluster $K(\dot{\gamma})$ as the set of~$\gamma$'s in $\Pi_r$ that
are related by an even sign f\/lip to~$\dot{\gamma}$. We also denote the number of distinct~$\gamma$'s in the cluster by
$N(\dot{\gamma})$. Thus we have
\begin{gather}
N(\dot{\gamma})=|K(\dot{\gamma})|, \qquad 1\le N(\dot{\gamma})\le 64,\qquad (\mathrm{f\/irst\ regime}).
\end{gather}
Due to the $\Pi_r$-restriction, the cluster depends on $(a_+,a_-)$, but just as with other objects, we do not indicate
this dependence explicitly. Notice also that~$\gamma$'s in the same cluster may be related by a~permutation, and that
the case $N(\dot{\gamma})=1$ occurs when all components of $\dot{\gamma}$ but one vanish.

Turning to the second~$\gamma$-regime, we recall that we only allow even sign f\/lips that leave~\eqref{sumres} invariant.
This yields a~restricted f\/lip group~$\Phi_r$ with 48 elements, as is readily verif\/ied. Fi\-xing~$\dot{\gamma}$
satisfying~\eqref{imgam},~\eqref{sumres} and~$\dot{g}\in\Pi_r$, we def\/ine~$K(\dot{\gamma})$ as the set of~$\gamma$'s
that are related to~$\dot{\gamma}$ by a~f\/lip from~$\Phi_r$ and obey~$g\in\Pi_r$. Due to the latter restriction, its
cardinality is at most 24:
\begin{gather}
N(\dot{\gamma})=|K(\dot{\gamma})|, \qquad 4\le N(\dot{\gamma})\le 24,\qquad (\mathrm{second\ regime}).
\end{gather}
We are now prepared for the following theorem.

\begin{Theorem}
\label{theorem6.2}
Fix $\dot{\gamma}$ with $\dot{g}= \operatorname{Re} \dot{\gamma} \in\Pi_r$. Letting $(a_+,a_-)\in(0,\infty)^2$,
the~$N(\dot{\gamma})$ positive trace class operators ${\cal T}(\gamma)$ with~$\gamma$ in the cluster $K(\dot{\gamma})$
form a~commutative family. Moreover, requiring~\eqref{aa1}/\eqref{aa2}, the operators $\hat{{\cal
A}}_s(\gamma)$/$\hat{{\cal A}}_l(\gamma)$ for $\gamma\in K(\dot{\gamma})$ are equal.
\end{Theorem}

\begin{proof}

We recall f\/irst that $\hat{{\cal A}}_l(\gamma)$ is def\/ined via the A$\Delta$O-action on ${\cal C}(\gamma)$, which in
turn is def\/ined as the span of the ${\cal T}(\gamma)$-eigenvectors. Since the ${\cal T}(\gamma)$ for $\gamma\in
K(\dot{\gamma})$ are generically distinct, the resulting spaces ${\cal C}(\gamma)$ and operators $\hat{{\cal
A}}_l(\gamma)$ are, a~priori, distinct as well. For $\hat{{\cal A}}_s(\gamma)$, however, the def\/inition
domain~$D_t(\gamma)$ in Theorem~\ref{theorem3.1} does not involve the HS operators at all. Whenever~$D_t(\gamma)$ is
a~core, therefore, we obtain a~self-adjoint closure that is invariant under the pertinent permutations and even sign
f\/lips. (At this point it may not be superf\/luous to stress again the distinction between the Hilbert space
operators~$\hat{{\cal A}}_{\pm}(\gamma)$ and the A$\Delta$Os~${\cal A}_{\pm}(\gamma;x)$: The \emph{latter} are
manifestly the same for all $\gamma\in K(\dot{\gamma})$.)

The crux is now that Lemma~\ref{lemma4.1} reveals that each of the spaces ${\cal C}(\gamma)$ with $\gamma\in
K(\dot{\gamma})$ is a~subspace of the same space $D_t(\gamma)=D_t(\dot{\gamma})$, provided~$t$ belongs to $(a_s,a]$.
Since~$\hat{{\cal A}}_s(\gamma)$ is symmetric on~$D_t(\gamma)$ for $t>a_s$ (by Theorem~\ref{theorem3.1}), and each of
the subspaces ${\cal C}(\gamma)$ is a~core for~$\hat{{\cal A}}_s(\gamma)$ (by Theorem~\ref{theorem4.3}), we deduce that
$D_t(\dot{\gamma})$ is a~core for all $t\in (a_s,a]$. As a~result, the operators $\hat{{\cal A}}_s(\gamma)$ and the
corresponding orthogonal decomposition of ${\cal H}$ in $\hat{{\cal A}}_s(\gamma)$-eigenspaces are the same for all of
the~$\gamma$'s in the cluster.

Now by Lemma~\ref{lemma4.2} each of the ${\cal T}(\gamma)$'s leaves the latter eigenspaces invariant. Therefore, all of
these operators commute whenever all eigenspaces are one-dimensional, or equivalently, whenever the spectrum of
$\hat{{\cal A}}_s(\gamma)$ is nondegenerate. We believe that this is always true, but unfortunately we can only prove
that all eigenspaces are f\/inite-dimensional, cf.~Lemma~\ref{lemma4.5}. Thus we cannot deduce commutativity of the ${\cal
T}(\gamma)$'s when we only take the properties of~$\hat{{\cal A}}_s(\gamma)$ into account.

Requiring that the parameters $a_{\pm}$ satisfy~\eqref{aa2}, however, we can invoke the operators~$\hat{{\cal
A}}_l(\gamma)$ as well. As we have proved in Section~\ref{section5}, they leave the eigenspaces of~$\hat{{\cal A}}_s(\gamma)$
invariant. Moreover, we have just shown that these eigenspaces coincide for all~$\gamma$'s in the cluster, and the
action of~$\hat{{\cal A}}_l(\gamma)$ is that of the A$\Delta$O ${\cal A}_l(\gamma;x)$. Therefore the
operators~$\hat{{\cal A}}_l(\gamma)$ in the cluster coincide, and we can choose an ONB of~$\hat{{\cal
A}}_l(\gamma)$-eigenvectors in each $\hat{{\cal A}}_s(\gamma)$-eigenspace.

When we now f\/irst assume that $a_+/a_-$ is irrational, then the joint eigenspaces of~$\hat{{\cal A}}_{\pm}(\gamma)$ are
one-dimensional by Theorem~\ref{theorem5.6}. Since they are left invariant by the ${\cal T}(\gamma)$'s, it follows that
the ${\cal T}(\gamma)$'s commute. By Lemma~\ref{lemma2.1} we have HS-continuity of ${\cal T}(\gamma)$ in $a_+$ and
$a_-$, so we are now in the position to deduce that all of the cluster ${\cal T}(\gamma)$'s commute for \emph{all}
$(a_+,a_-)\in(0,\infty)^2$.
\end{proof}

Even though it is now clear that the eigenvectors~$f_0(\dot{\gamma}),f_1(\dot{\gamma}),\ldots$ of ${\cal
T}(\dot{\gamma})$ can be chosen to be eigenvectors of ${\cal T}(\gamma)$ for all $\gamma\in K(\dot{\gamma})$, it does
not follow that $f_n(\gamma)$ \emph{equals} $f_n(\dot{\gamma})$. This is because we are unable to prove that the
ordering of the eigenvalues of~${\cal T}(\gamma)$ on $f_0(\dot{\gamma}),f_1(\dot{\gamma}),\ldots$ coincides with that of
the eigenvalues $\lambda_0(\gamma),\lambda_1(\gamma),\ldots$. (We do have equality for $n=0$, though,
cf.~\eqref{e0pos}.)

At this point it is illuminating to illustrate the above developments with explicit examples. Letting $a_s=a_+<a_l=a_-$
from now on, we choose
\begin{gather}
\dot{\gamma}:=\gamma_{\rm L}(a_s)=\big({-}a_l/2,-a_l/2+a_s/2,0,a_s/2,
\nonumber\\
\hphantom{\dot{\gamma}:=\gamma_{\rm L}(a_s)=}{}
{-}a_l/2+i\pi/2r,-a_l/2+a_s/2+i\pi/2r,-i\pi/2r,a_s/2-i\pi/2r\big),\label{gamdot}
\end{gather}
cf.~\eqref{gfree0},~\eqref{gLame}. Combining~\eqref{Fnfree} and~\eqref{cfree} with~\eqref{sR}, we obtain
\begin{gather}
\label{efnfree}
e_n(\dot{\gamma};x)=\sqrt{\frac{4r}{\pi}}\sin2(n+1)rx,\qquad
f_n(\dot{\gamma};x)=\sqrt{\frac{r}{\pi}}\big(e^{2nirx}-e^{-(2n+4)irx}\big).
\end{gather}
It is also easily checked that we have (cf.~\eqref{freesG2},~\eqref{freesG3})
\begin{gather}
H_{\delta}(\dot{\gamma};x)=\exp(-2ra_{-\delta})\big(\exp(-ia_{-\delta}d/dx) + \exp(ia_{-\delta}d/dx)\big),
\\
{\cal A}_{\delta}(\dot{\gamma};x)=\exp(-ia_{-\delta}d/dx) +\exp(-4ra_{-\delta}) \exp(ia_{-\delta}d/dx),
\end{gather}
so that the functions~\eqref{efnfree} yield eigenvalues
\begin{gather}
\label{Enf}
E_{n,\delta}(\dot{\gamma})=2\exp(-2ra_{-\delta})\cosh(2(n+1) ra_{-\delta}), \qquad n\in{\mathbb N},\qquad \delta=+,-.
\end{gather}

We now consider~$\gamma$'s related to $\dot{\gamma}$ by a~sign f\/lip in $\Phi_r$. Since we have $\dot{\gamma}_2=0$ and
$\dot{\gamma}_6=-i\pi/2r$, we can freely f\/lip signs for the remaining 6 components. Now $\sigma(\dot{\gamma})$ equals
$a_l/2-a_s/2$, so when we f\/lip signs of $\dot{\gamma}_3, \dot{\gamma}_6$ and $\dot{\gamma}_7$ and denote the result~by
$\gamma^{(l)}$, we readily obtain
\begin{gather}
\sigma(\gamma^{(l)})=a_l/2, \qquad w(\gamma^{(l)};x)=p_l^4s_l(x)^2s_l(x+\pi/2r)^2.
\end{gather}
Moreover, this is a~self-dual~$\gamma$-choice, so $\gamma^{(l)}$ belongs to the cluster $K(\dot{\gamma})$ and
$I(\gamma^{(l)})$ is a~positive HS operator with kernel
\begin{gather}
p_l^4s_l(x)s_l(\pi/2r-x)s_l(y)s_l(\pi/2r-y)/R_l(x+y)R_l(x-y).
\end{gather}

We have met this integral operator in our study of the Heun case, cf.~the kernel below equation~(4.1) in~\cite{RIII09}. From
this previous work we conclude that $I(\gamma^{(l)})$ has eigenvectors $e_n(\dot{\gamma})$ with corresponding ordered
eigenvalues
\begin{gather}
\label{laml}
\lambda_n(\gamma^{(l)})=\frac{\pi e^{ra_l}}{p_l\cosh(n+1)ra_l},\qquad n\in{\mathbb N}.
\end{gather}
In particular, in agreement with Theorem~\ref{theorem6.2}, $I(\gamma^{(l)})$ commutes with $I(\dot{\gamma})$. (Recall
that the ordered eigenvalues of the latter are given by~\eqref{lamfree}.)

Next, we also f\/lip the signs of $\dot{\gamma}_1$, $\dot{\gamma}_5$ (and f\/lip back to $\dot{\gamma}_6$). This yields
a~vector~$\gamma^{(s)}$ that is obtained from~$\gamma^{(l)}$ via $a_s\leftrightarrow a_l$ and a~permutation. Thus we
need only take the suf\/f\/ix~$l$ to~$s$ in the above quantities, showing $\gamma^{(s)}\in K(\dot{\gamma})$ and yielding
$I(\gamma^{(s)})$-eigenvalues
\begin{gather}
\label{lams}
\lambda_n(\gamma^{(s)})=\frac{\pi e^{ra_s}}{p_s\cosh(n+1)ra_s},\qquad n\in{\mathbb N},
\end{gather}
on the ONB-vectors $e_n(\dot{\gamma})$.

Consider next the vector $\gamma^{(0)}$ obtained from $\dot{\gamma}$ by a~f\/lip of $\dot{\gamma}_0$. This yields
$\sigma(\gamma^{(0)})=a_l/4-a_s/2$, so it is necessary to require $a_l>2a_s$ for this vector to belong to
$K(\dot{\gamma})$. But this is not suf\/f\/icient: We also need $(\gamma^{(0)'})_0>-a$, entailing $a_l<4a_s$. Assuming
$a_l\in(2a_s,4a_s)$ until further notice, we do obtain $\gamma^{(0)}\in K(\dot{\gamma})$, but it is plain
that~$\gamma^{(0)}$ is not self-dual, so that $I(\gamma^{(0)})$ is not self-adjoint.

From Theorem~\ref{theorem6.2} we can still deduce that the vectors $e_n(\dot{\gamma})$ are
$T(\gamma^{(0)})$-eigenvectors, but now we have no information on the eigenvalues, except that they are all positive and
that the largest one $\lambda_0(\gamma^{(0)})^2$ corresponds to $e_0(\dot{\gamma})$. To be quite precise about the
ambiguity, it may be in order to add that we do not know whether $e_n(\dot{\gamma})$ is equal to $e_n(\gamma^{(0)})$ for
$n>0$. Indeed, we have agreed to order the eigenvectors in accordance with the ordering of the singular values, so there
may be a~nontrivial permutation of the ONB-vectors.

In sharp contrast to our ignorance concerning the explicit values and ordering of the singular values
$\lambda_0(\gamma^{(0)})$ of the HS operators $I(\gamma^{(0)})$ and $I(\gamma^{(0)'})$, we can conclude from
Lemma~\ref{lemma6.1} that the spectra of the operators $\hat{{\cal A}}_{\pm}(\gamma^{(0)})$ and $\hat{{\cal
A}}_{\pm}(\gamma^{(0)'})$ are explicitly given by the numbers in~\eqref{Enf}. Furthermore, the corresponding $\hat{{\cal
A}}_{\pm}(\gamma^{(0)'})$-eigenvectors are the multiples of ${\cal I}(\gamma^{(0)'})f_n(\dot{\gamma})$. We stress that
the operators $\hat{{\cal A}}_{\pm}(\gamma^{(0)'})$ are not free (i.e., the A$\Delta$Os ${\cal
A}_{\pm}(\gamma^{(0)'};x)$ have~$x$-dependent coef\/f\/icients), yet they have the free spectra~\eqref{Enf}~by
Lemma~\ref{lemma6.1}.

As a~fourth f\/lip choice, consider the vector $\gamma^{(3)}$ obtained from $\dot{\gamma}$ by f\/lipping $\dot{\gamma}_3$.
This gives $\sigma(\gamma^{(3)})=a_l/2-a_s/4$, so we get $\gamma^{(3)}\in K(\dot{\gamma})$ without restricting
$a_s/a_l$. Like in the previous case, $\gamma^{(3)}$ is not self-dual, and we need only change the suf\/f\/ix $(0)$ to $(3)$
in the above to reach the same conclusions.

Finally, consider the vector~$\gamma^{(p)}$ obtained from~$\dot{\gamma}$ by f\/lipping $\dot{\gamma}_1,\dot{\gamma}_5$ and
$\dot{\gamma}_6$. This yields $\sigma(\gamma^{(p)})=0$, so~$\gamma^{(p)}$ does not belong to the cluster
$K(\dot{\gamma})$ and the integral kernel ${\cal K}(\gamma^{(p)};x,y)$~\eqref{cK} has poles for real $x,y$.

After this detailed analysis, the reader will have no dif\/f\/iculty to establish what happens for any remaining
$\dot{\gamma}$-f\/lip. Returning to the f\/irst~$\gamma$-regime, we introduce the parameter space
\begin{gather}
\label{Pi1}
\Pi(1):= W(D_8)\Pi_r.
\end{gather}
Here and from now on, $W(\mathfrak{g})$ denotes the (obvious action of the) Weyl group of the (complex, simple) Lie
algebra~$\mathfrak{g}$. Clearly, $\Pi(1)$ is a~subset of $\tilde{\Pi}$~\eqref{Pit}. As we have seen above, with the
parameter restriction~\eqref{aa1} in force and $\gamma\in\Pi(1)$, the operator~$\hat{{\cal A}}_s(\gamma)$ is
a~well-def\/ined self-adjoint operator on ${\cal H}$~\eqref{cH}, which does not change as~$\gamma$ varies over
a~$W(D_8)$-orbit. By contrast, requiring~\eqref{aa2} and f\/ixing $\dot{\gamma}\in\Pi_r$, the operator $\hat{{\cal
A}}_l(\gamma)$ is thus far not def\/ined for all $\gamma\in W(D_8)\dot{\gamma}$. Indeed, it is only def\/ined and equal to
$\hat{{\cal A}}_l(\dot{\gamma})$ for all~$\gamma$ in the cluster $ K(\dot{\gamma})\subset \Pi_r$. However, we may and
shall def\/ine it for all of $\gamma\in\Pi(1)$ by requiring that it be invariant under $D_8$-transformations. Note that
this def\/inition is the obvious one in view of the $D_8$-invariance of the A$\Delta$O ${\cal A}_l(\gamma;x)$.

With~\eqref{aa2} in ef\/fect, spectral invariance of the operators $\hat{{\cal A}}_{\pm}(\gamma)$ for all~$\gamma$ on
a~$W(D_8)$-orbit in $\Pi(1)$ is now plain. Indeed, as announced in the introduction, the operators themselves are
$D_8$-invariant. (On the other hand, it should be noted that the A$\Delta$Os $A_{\pm}(\gamma;x)$ are generically
distinct under the 128 even sign f\/lips, since their coef\/f\/icients $V_{\pm}(\gamma;x)$ are generically distinct,
cf.~\eqref{Ve}.)

Lemma~\ref{lemma6.1}, however, enables us to deduce spectral equality for operators $\hat{{\cal A}}_{\pm}(\gamma)$ and
$\hat{{\cal A}}_{\pm}(\gamma')$ that are in general vastly dif\/ferent. Now the $E_8$-ref\/lection~$J$ given by~\eqref{J}
commutes with permutations, but it commutes with only one sign f\/lip, namely, the map
\begin{gather}
\label{rho}
\rho(\gamma)\equiv -\gamma.
\end{gather}
Denoting one of the remaining 126 f\/lips by~$\phi$, we get $(\phi(\gamma))'\in\Pi_r $ whenever $\phi(\gamma)\in\Pi_r$ (by
virtue of the def\/inition of $\Pi_r$, cf.~\eqref{Pir}), but when $\phi(\gamma)$ does not belong to $\Pi_r$, it need not
even be true that $(\phi(\gamma))'$ belongs to $\tilde{\Pi}$~\eqref{Pit}. This can be seen for example by letting
$a_l>4a_s$ and choosing~$\gamma$ equal to the real part of $\dot{\gamma}$~\eqref{gamdot}; then the sign f\/lip of
$\dot{\gamma}_0$ yields a~vector that is mapped out of $\tilde{\Pi}$ by $-J$. Consequently, $J\Pi(1)$ is not a~subset
of~$\tilde{\Pi}$.

Put in a~more conceptual way: even though $-J$ leaves $\Pi_r$ invariant (by def\/inition), it does not leave $\Pi(1)$
invariant. The example just given reveals what is happening from a~geometric point of view: Letting $\gamma\in
\Pi(1)\setminus\Pi_r$, we clearly have $\| \gamma\|_{\infty}<a$, whereas~$J$ can increase the modulus of
a~$\gamma$-component so that it becomes larger than~$a$.

We can ensure $\| J\gamma\|_{\infty}<a$ by restricting attention to~$\gamma$ obeying $\| \gamma\|_2<a$. (Indeed,
since~$J$ is a~ref\/lection, we have $\| J\gamma\|_{\infty}\le\| J\gamma\|_2=\| \gamma\|_2$.) Accordingly, we introduce
the parameter spaces
\begin{gather}
\label{PiE8}
\Pi(E_8)\equiv \{\gamma\in{\mathbb R}^8\,|\, \| \gamma\|_2 <a\},\qquad \Pi(E_8)^*\equiv \Pi(E_8)\setminus \{0\}.
\end{gather}
Since any $w\in W(E_8)$ is an isometry, these spaces are invariant under~$W(E_8)$, so that $\Pi(E_8)$ is a~subset
of~$\Pi$~\eqref{Pi}. We now claim that we have
\begin{gather}
\label{Pisub}
\Pi(E_8)^*\subset \Pi(1).
\end{gather}
To prove this, we f\/irst note that the Schwarz inequality implies $\| \gamma\|_1 <\sqrt{8}a $, so that
\begin{gather}
|\sigma(\gamma)|\le \| \gamma\|_1/4< a/\sqrt{2}<a,\qquad \forall\, \gamma\in\Pi(E_8).
\end{gather}
For a~given $\gamma\in\Pi(E_8)^*$, it is therefore enough to f\/ind an even sign f\/lip~$\phi$ such that
\begin{gather}
\langle \zeta, \phi(\gamma)\rangle <0,
\end{gather}
and a~moment's thought suf\/f\/ices to see that this can always be done.

We are now prepared to state and prove one of the principal results of this paper.

\begin{Theorem}
\label{theorem6.3}
Let~$\gamma$ vary over $\Pi(E_8)^*$. Assuming $a_{\pm}$ obey~\eqref{aa1}, the spectrum of the opera\-tor~$\hat{{\cal
A}}_s(\gamma)$ is $E_8$-symmetric. Assuming next $a_{\pm}$ obey~\eqref{aa2}, the spectrum of the operator $\hat{{\cal
A}}_l(\gamma)$ is $E_8$-symmetric, too.
\end{Theorem}

\begin{proof}
We f\/irst note that by virtue of~\eqref{Pisub} the operators at issue are well-def\/ined self-adjoint operators. Now the
orbit of~$\gamma$ under $W(E_8)$ is generated by the ref\/lection~$J$ combined with the action of~$W(D_8)$. Under the
latter action we obtain the same operators, hence isospectrality. It is therefore enough to show that we retain
isospectrality under the action of~$J$.

To this end, let us f\/ix $\gamma\in\Pi(E_8)^*$. Now there are three cases. First, let $\langle \zeta, \gamma\rangle <0$.
Then we can appeal to Lemma~\ref{lemma6.1}, combined with spectral invariance under~$\rho$~\eqref{rho}. Second, let
$\langle \zeta, \gamma\rangle >0$. Then we have $J\gamma\in\Pi_r$, so by Lemma~\ref{lemma6.1} $J\gamma$ yields the same
spectrum as $(-J)J\gamma=-\gamma$. But $-\gamma$ gives rise to the same operators as~$\gamma$, so that we again retain
spectral invariance. Finally, in the third case $\langle \zeta, \gamma\rangle =0$ we see from~\eqref{J} that
$J\gamma=\gamma$.
\end{proof}

Evidently, the zero vector is f\/ixed under $W(E_8)$, but we have far less information in this special case. More
specif\/ically, we do not know whether $\hat{{\cal A}}_s(t,0)$ is essentially self-adjoint on its def\/inition domain
$D_t(0)$ for $t\in (a_s,a]$, and we cannot use~${\cal I}(0)$ to associate a~well-def\/ined symmetric operator to the
A$\Delta$O ${\cal A}_l(0;x)$. Also, there is no guarantee that continuity arguments can be used to handle this special
case. Indeed, even for the simpler setting of ref\/lectionless second order A$\Delta$Os it can happen that
self-adjointness breaks down for a~single point along an ${\mathbb R}$-orbit, cf.~the paragraph containing equation~(6.7)
in~\cite{Rui05FC}.

In order to gain more perspective on Theorem~\ref{theorem6.3} and its Lie-algebraic aspects, we f\/irst recall some
information that can be found, e.g.,~in~\cite{Kac90}, slightly changing conventions to suit our context. Denoting the
canonical base vectors of ${\mathbb R}^8$ by $e_0,\ldots,e_7$, the $D_8$ roots are given by
\begin{gather}
\delta e_j+\delta'e_k,\qquad 0\le j<k\le 7,\qquad \delta,\delta'\in \{-1,1\},
\end{gather}
and the $E_8$ roots are the union of the $D_8$ roots and the additional roots
\begin{gather}
\frac12\sum\limits_{j=0}^7\delta_j e_j,\qquad \sum\limits_{j=0}^7\delta_j\in 2{\mathbb Z},\qquad \delta_j\in \{-1,1\}.
\end{gather}
The simple $E_8$ roots can be chosen as
\begin{gather}
\alpha_0=\frac12 (-e_0-e_1-e_2+e_3+e_4+e_5+e_6-e_7),\nonumber\\
 \alpha_j=e_{j-1}-e_j,\qquad j=1,\ldots,6,\qquad \alpha_7=e_6+e_7.  
\end{gather}
Choosing~$v$ in the Weyl chamber
\begin{gather}
C(E_8)\equiv \big\{v\in{\mathbb R}^8\,|\, \langle \alpha_j,v\rangle \ge 0,\ j=0,\ldots,7\big\},
\end{gather}
we see that $v_7$ can vary over ${\mathbb R}$, whereas we have
\begin{gather}
\label{uineq}
|v_7|\le v_6\le v_5\le \cdots \le v_0.
\end{gather}
Note, however, that when $v\in{\mathbb R}^8$ satisf\/ies these inequalities, it need not belong to~$C(E_8)$.

To connect this to our parameter spaces, we need the highest root
\begin{gather}
\theta \equiv 3\alpha_0+2\alpha_1+4\alpha_2+6\alpha_3+5\alpha_4+4\alpha_5+3\alpha_6+2\alpha_7,
\end{gather}
and the Weyl alcove
\begin{gather}
A(E_8)\equiv \{v\in C(E_8)\,|\, \langle \theta,v\rangle \le 1\}.
\end{gather}
The crux is that~$\theta$ equals $\zeta/2$, as is easily checked. Let us now ask: For what $c>0$ is $-cA(E_8)$ in the
closure of the parameter space $\Pi_r$? For the f\/irst condition $\sigma (-cv)\le a$ we need $c\le 2a$, while for the
second condition $\| cv\|_{\infty}\le a$ we must require $cv_0\le a$, cf.~\eqref{uineq}. It is not dif\/f\/icult to see that
the maximum of $v_0$ on $A(E_8)$ equals $3/4$ and is taken on in the unique boundary point $(3,1,\ldots,1,-1)/4$, whose
$\ell^2$-norm equals~1. Thus we can take $c=4a/3$, and on this `optimal' alcove multiple $-(4a/3)A(E_8)$ the maximal
$\ell^{\infty}$-norm equals~$a$ and is taken on in the unique boundary point
\begin{gather}
\gamma_b:=a(-1,-1/3,\ldots,-1/3,1/3),\qquad \sigma(\gamma_b)=3a/4,\qquad \|\gamma_b\|_2=4a/3.
\end{gather}

When we now delete the boundary point~$\gamma_b$ and the zero vector, we obtain a~set $G(E_8)$ that belongs to $\Pi_r$.
The orbit set
\begin{gather}
\label{Gorbit}
{\cal O}(E_8)\equiv W(E_8)G(E_8),
\end{gather}
contains~$\gamma$'s satisfying $\|\gamma\|_2\in [a,4a/3)$, so it is not a~subset of $\Pi(E_8)$. Whenever the set $d
{\cal O}(E_8)$, $d>0$, is a~subset of $\tilde{\Pi}$~\eqref{Pit}, it follows as in the proof of Theorem~\ref{theorem6.3}
that we get spectral $E_8$-invariance on it. Clearly, this is true for $d\le 3/4$, since then we obtain a~subset of
$\Pi(E_8)$. We leave the question open whether~3/4 is the largest~$d$-value for which
\begin{gather}
d{\cal O}(E_8)\subset \tilde{\Pi}.
\end{gather}

Finally, let us recall the Weyl group orders
\begin{gather}
|W(E_8)|=4!\cdot 6!\cdot 8!,\qquad |W(D_8)|=2^7\cdot 8!.
\end{gather}
For generic $\gamma\in\Pi(E_8)$, therefore, Theorem~\ref{theorem6.3} entails that we obtain 135 distinct isospectral
operators on the orbit $W(E_8)\gamma$.

For the second~$\gamma$-regime the role of $W(D_8)$ is played by the subgroup $D(2)$ that is the semidirect product of
$S_4\times S_4$ and the restricted f\/lip group $\Phi_r$. The analog of $\Pi(1)$~\eqref{Pi1} is the parameter space
\begin{gather}
\Pi(2):= D(2) \{\gamma \in{\mathbb C}^8\,|\, \operatorname{Re} \gamma\in\Pi_r,\ \operatorname{Im} \gamma \
\mathrm{satisf\/ies~\eqref{imgam},~\eqref{sumres}}\}.
\end{gather}
Following the analysis for the f\/irst regime, we see that we can choose as the counterpart of $\Pi(E_8)$~\eqref{PiE8} the
space
\begin{gather}
\Pi(E(2)):= \{\gamma \in{\mathbb C}^8\,|\, \|\operatorname{Re} \gamma\|_2<a,\ \operatorname{Im} \gamma\
\mathrm{satisf\/ies~\eqref{imgam},~\eqref{sumres}}\}.
\end{gather}
Here $E(2)$ denotes the subgroup of $W(E_8)$ that is obtained upon combining $D(2)$ with the ref\/lection~$J$.

We are now in the position to obtain the analog of Theorem~\ref{theorem6.3}.

\begin{Theorem}
\label{theorem6.4}
Let~$\gamma$ vary over
\begin{gather}
\Pi(E(2))^*:= \{\gamma\in\Pi(E(2))\,|\, \operatorname{Re} \gamma \ne 0\}.
\end{gather}
Assuming $a_{\pm}$ obey~\eqref{aa1}, the spectrum of the operator $\hat{{\cal A}}_s(\gamma)$ is $E(2)$-symmetric.
Assuming next $a_{\pm}$ obey~\eqref{aa2}, the spectrum of the operator $\hat{{\cal A}}_l(\gamma)$ is $E(2)$-symmetric,
too.
\end{Theorem}
\begin{proof}
The proof of Theorem~\ref{theorem6.3} applies with obvious changes.
\end{proof}

\section{Eigenvalue and eigenvector asymptotics}\label{section7}

In this f\/inal section we focus on issues that have an asymptotic character. As a~pivotal auxiliary tool, we need the
large-$n$ asymptotic behavior of the orthonormal polynomials
\begin{gather}
\label{pn}
p_n(\gamma;\cos 2rx)=:P_n(\gamma;x),\qquad g=\operatorname{Re} \gamma\in\tilde{\Pi},\qquad n\in{\mathbb N},
\end{gather}
with respect to the weight function $w_P(\gamma;x)$~\eqref{wP} on $[0,\pi/2r]$. More specif\/ically, we introduce
\begin{gather}
D_n(\gamma;x):= c_P(\gamma;x)e^{2inrx}+c_P(\gamma;-x)e^{-2inrx},
\end{gather}
(with $c_P$ given by~\eqref{cPol}), and note that $c_P(\gamma;x/2r)$ belongs to the space ${\cal C}_{1,1}$ def\/ined by
equation~(2.14) in~\cite{Rui05MR}. From Theorem~2.4 in~\cite{Rui05MR} we then deduce a~decay bound
\begin{gather}
\label{PnDn}
\| P_n-D_n\|_P=O(\exp(-2nrd)),\qquad d=a_s-\epsilon,\qquad \epsilon>0,\qquad n\to\infty,
\end{gather}
where $\|\cdot\|_P$ denotes the norm on the weighted $L^2$-space ${\cal H}_P$~\eqref{cHP}. (The decay rate~$d<a_s$ is
determined by the convergence radius of the factor $1/ E(2x\pm i(a_+-a_-)/2)$ in $c_P$~\eqref{cPol}, viewed as a~power
series in $\exp(-2irx)$, cf.~\eqref{E}.)

In the present context it is more convenient to work with
\begin{gather}
\label{psin}
\psi_n(\gamma;x):= \sqrt{\frac{r}{\pi}}P_n(\gamma;x)/c_P(\gamma;x),\qquad g=\operatorname{Re} \gamma\in\tilde{\Pi},\qquad
n\in{\mathbb N},
\\
\label{an}
a_n(\gamma;x):= \sqrt{\frac{r}{\pi}} D_n(\gamma;x)/c_P(\gamma;x)= \sqrt{\frac{r}{\pi}}
\Big(e^{2inrx}-\exp(-4irx)u(\gamma;-x)e^{-2inrx}\Big),
\end{gather}
where we have introduced the `$S$-matrix'
\begin{gather}
\label{u}
u(\gamma;x):= -\exp(-4irx)c(\gamma;x)/c(\gamma;-x),
\end{gather}
and used the identity~\eqref{ccP}. The factor $ (r/\pi)^{1/2}/c_P(\gamma;x)$ gives rise to a~unitary similarity between
${\cal H}_P$ and ${\cal H}$~\eqref{cH}, so the vectors $\{\psi_n(\gamma;\cdot)\}_{n\in{\mathbb N}}$ yield an ONB for
${\cal H}$. Moreover,~\eqref{PnDn} implies
\begin{gather}
\label{psia}
\| \psi_n-a_n\|=O(\exp(-2nrd)),\qquad d=a_s-\epsilon,\qquad \epsilon>0,\qquad n\to\infty,
\end{gather}
with $\|\cdot\|$ denoting the ${\cal H}$-norm.

For a~better understanding of these formulas (and for later purposes), we mention that in the special case when~$\gamma$
equals $\dot{\gamma}$~\eqref{gamdot}, we get (using the doubling formula~\eqref{Edup} for the~$E$-function)
\begin{gather}
c_P(\dot{\gamma};x)=1/(1-\exp(-4irx)),\qquad u(\dot{\gamma};x)=1,
\end{gather}
so that
\begin{gather}
\label{Tch2}
P_n(\dot{\gamma};x) =\sin 2(n+1)rx/\sin 2rx,\qquad  \psi_n(\dot{\gamma};x)=a_n(\dot{\gamma};x)=f_n(\dot{\gamma};x),
\end{gather}
cf.~\eqref{efnfree}.

Next, we restrict attention to~$\gamma$ satisfying $g\in\Pi_r$. Then the integral operator ${\cal I}(\gamma)$~\eqref{cI}
is a~complete HS operator. We proceed to study the functions
\begin{gather}
b_n(\gamma';x):= ({\cal I}(\gamma')a_n(\gamma;\cdot))(x)
\nonumber\\
\hphantom{b_n(\gamma';x)~}{}
= \frac{1}{c(\gamma';x)}\int_0^{\pi/2r}{\cal S}(\sigma;x,y) \sqrt{\frac{r}{\pi}}
\left(\frac{e^{2inry}}{c(\gamma;-y)}+\frac{e^{-2inry}}{c(\gamma;y)}\right)dy,
\end{gather}
where we used~\eqref{an},~\eqref{u} in the second step. Since ${\cal S}$ is even in~$y$, we can rewrite this as
\begin{gather}
b_n(\gamma';x) =\sqrt{\frac{r}{\pi}} \frac{1}{c(\gamma';x)}\int_{-\pi/2r}^{\pi/2r}I(\gamma;x,y)dy,
\\
I(\gamma;x,y):={\cal S}(\sigma;x,y) \exp(2inry)G(-2y+ia)\prod\limits_{\mu=0}^7G(y+i\gamma_{\mu}),\qquad g\in\Pi_r,
\end{gather}
where we also used~\eqref{c}. The integrand is $\pi/r$-periodic in~$y$, and for $x\in{\mathbb R}$ its poles in the UHP
nearest to ${\mathbb R}$ occur at $y\equiv i\sigma\pm x \pmod{\pi/2r}$, cf.~\eqref{cS}. Fixing $x\in(0,\pi/2r)$, we now
shift up the contour by a~distance
\begin{gather}
\label{eta}
\xi\in (\sigma,\eta),\qquad \eta:=\min(a,\sigma+a_s),
\end{gather}
so that only the two poles at $y=i\sigma \pm x$ are met. Hence we need the residues
\begin{gather}
\operatorname{Res}  I(\gamma;x,y)|_{y=i\sigma +\delta x}=-r_0\frac{G(2i\sigma-ia)G(2i\sigma +2\delta x-ia)}{G(2\delta
x+ia)}\exp(2inr(i\sigma +\delta x))
\nonumber\\
\hphantom{\operatorname{Res}  I(\gamma;x,y)|_{y=i\sigma +\delta x}=}{}
{}\times G(-2i\sigma -2\delta x+ia)\prod\limits_{\mu} G(i\sigma +\delta x +i\gamma_{\mu})
\\
\hphantom{\operatorname{Res}  I(\gamma;x,y)|_{y=i\sigma +\delta x}}{}
=-r_0 \exp(-2nr\sigma)G(2i\sigma -ia)c(\gamma';\delta x)\exp(2i\delta nrx),\qquad \delta=+,-,\nonumber
\end{gather}
where we used $\sigma+\gamma_{\mu}=-\gamma_{\mu}'$. Using now Cauchy's theorem we deduce
\begin{gather}
b_n(\gamma';x)=\kappa_n(\sigma)a_n(\gamma';x)+ \sqrt{\frac{r}{\pi}} \frac{1}{c(\gamma';x)}\int_{{\cal C}}
I(\gamma;x,y)dy,
\end{gather}
where (using~\eqref{r0})
\begin{gather}
\kappa_n(\sigma)  :=   -2\pi i r_0\exp(-2nr\sigma)G(2i\sigma -ia)
\nonumber\\
\hphantom{\kappa_n(\sigma)~}{} =   \frac{\pi G(2i\sigma -ia)}{r\prod\limits_{k=1}^{\infty}(1-\exp(-2kra_+))(1-\exp(-2kra_-))}\cdot \exp(-2nr\sigma),\label{kapn}
\end{gather}
and where ${\cal C}$ is the contour $\operatorname{Re} y\in [-\pi/2r,\pi/2r]$, $\operatorname{Im} y=\xi $. From this we readily obtain an
estimate
\begin{gather}
|b_n(\gamma';x)-\kappa_n(\sigma)a_n(\gamma';x)|=O(\exp(-2nr (\eta-\varepsilon))),\qquad n\to\infty,
\end{gather}
where the implied constant depends on $\varepsilon>0$, but not on $x\in (0,\pi/2r)$. Hence we get
\begin{gather}
\|b_n(\gamma';\cdot)-\kappa_n(\sigma)a_n(\gamma';\cdot)\|=O(\exp(-2nr (\eta-\varepsilon))),\qquad n\to\infty.
\end{gather}
Comparing this bound to~\eqref{psia} and noting $a_s\le \eta$ (cf.~\eqref{eta}), we can telescope to deduce the
following lemma, which summarizes the key consequence of the above developments.

\begin{Lemma}
\label{lemma7.1}
Let $a_+,a_-\in(0,\infty)$ and $g=\operatorname{Re} \gamma\in\Pi_r$. The functions $\psi_n(\gamma;x)$, $x\in[0,\pi/2r]$,
defined by~\eqref{psin}, yield an ONB $\{\psi_n(\gamma)\}_{n\in{\mathbb N}}$ for ${\cal H}$~\eqref{cH} and obey
\begin{gather}
\label{psib}
\|{\cal I}(\gamma')\psi_n(\gamma)-\kappa_n(\sigma)\psi_n(\gamma')\|=O(\exp(-2nr d)),\\
 d=a_s-\epsilon,\qquad
\epsilon>0,\qquad n\to\infty,  \nonumber
\end{gather}
where $\kappa_n(\sigma)$ is given by~\eqref{kapn}.
\end{Lemma}

A moment's thought now leads to the conjecture
\begin{gather}
\label{conj}
\lambda_n(\gamma)\sim \kappa_n(\sigma(\gamma)),\qquad n\to \infty.\quad (?)
\end{gather}
In words, it seems plausible that the asymptotic behavior of the singular values $\lambda_n(\gamma)$ of the HS operators
${\cal I}(\gamma)$ and ${\cal I}(\gamma')$ is given by the numbers $\kappa_n(\sigma)$. Moreover, we expect that the
polynomial-type ONB in the lemma approximates the ${\cal T}(\gamma)$-eigenvector ONB $f_n(\gamma)$, in the sense that
\begin{gather}
\label{psifnconj}
\|\psi_n(\gamma)-f_n(\gamma) \|\to 0,\qquad n\to\infty. \quad (?)
\end{gather}
We continue by supplying evidence for these conjectures. First, we can test them for the three~$\gamma$-choices for
which we explicitly know the singular values, namely~$\gamma$ equal to $\dot{\gamma},\gamma^{(l)}$ and $\gamma^{(s)}$,
cf.~\eqref{lamfree},~\eqref{laml} and~\eqref{lams}, resp. The relevant~$\sigma$-values are $(a_l-a_s)/2$, $a_l/2$ and
$a_s/2$, and this already suf\/f\/ices to check that the decay rates of the three sets of $\lambda_n$'s and $\kappa_n$'s do
match. To compare the proportionality constants, we need the values of $G(z)$ for~$z$ equal to
$(ia_l-3ia_s)/2,(ia_l-ia_s)/2$, and $(ia_s-ia_l)/2$. They follow from~\eqref{Gspec}, using also
the~$G$-A$\Delta$Es~\eqref{Gades} and~\eqref{sR},~\eqref{pde} in the f\/irst case. This yields equality of the constants
in all three cases. Moreover, in view of~\eqref{Tch2} and sign f\/lip invariance, the six ONB's in question coincide.

Next, we are heading for a~proof of the conjecture concerning $\lambda_n(\gamma)$ for the case that the decay rate
parameter~$\sigma$ of $\kappa_n(\sigma)$ is smaller than the parameter $a_s$ on the right-hand side of~\eqref{psib}. In
fact, we shall obtain a~far more precise asymptotic behavior. In the following lemma we isolate the general result from
which this can be deduced.

\begin{Lemma}
\label{lemma7.2}
Assume~$T$ is a~compact and complete operator on ${\cal H}$ with singular value decomposition
\begin{gather}
T=\sum\limits_{n=0}^{\infty}\nu_n (f_n,\cdot) f_n',\qquad \nu_0\ge \nu_1\ge \cdots >0,
\end{gather}
and associated ONBs $\{f_n^{(')}\}_{n\in{\mathbb N}}$. Assume $\{p_n^{(')}\}_{n\in{\mathbb N}}$ are ONBs such that
\begin{gather}
\label{Test}
\|Tp_n-e^{-ns}p_n'\|\le C e^{-na}, \qquad \forall\, n>N,
\end{gather}
where
\begin{gather}
\label{sa}
0<s<a.
\end{gather}
Then there exists $K\ge N$ such that the intervals
\begin{gather}
\label{In}
I_n:= \big[e^{-ns}-Ce^{-na},e^{-ns}+Ce^{-na}\big]
\end{gather}
belong to $(0,\infty)$ and are separated by a~distance
\begin{gather}
\label{Ind}
d(I_{n+1},I_n)> e^{-ns}(1-e^{-s})/2,\qquad \forall\, n>K.
\end{gather}
Moreover, there exists $M\ge K$ such that
\begin{gather}
\label{nuIn}
\nu_n\in I_n,\qquad \forall\, n>M,
\end{gather}
and~$T$ is trace class.
\end{Lemma}

\begin{proof}
We begin by noting that by~\eqref{Test} we have an estimate
\begin{gather}
|\| Tp_n\|-e^{-ns}|=|\| Tp_n\|-\|e^{-ns}p_n'\||\le \| Tp_n-e^{-ns}p_n'\|\le Ce^{-na}.
\end{gather}
Clearly this implies
\begin{gather}
\label{Tpn}
\| Tp_n\| \in I_n,\qquad \forall\, n>N.
\end{gather}
Using~\eqref{sa}, it is also easy to see that we can achieve positivity of the intervals~$I_n$ and the
separation~\eqref{Ind} between them by letting $n>K$, with $K\ge N$ chosen suf\/f\/iciently large.

Next, we introduce projections
\begin{gather}
P_n:=\sum\limits_{m=0}^n(p_m,\cdot)p_m,\qquad n\ge K,
\end{gather}
and set
\begin{gather}
\mu_K:= \min (\|Tp\|: \|p\|=1, p\in P_K{\cal H}).
\end{gather}
Since~$T$ has a~trivial null space, we have $\mu_K>0$. Thus there exists $M\ge K$ such that
\begin{gather}
\label{mKb}
e^{-Ms}+Ce^{-Ma}<\mu_K.
\end{gather}

Taking henceforth $n> M$, we now analyze the singular values $\nu_n$ via the max/min and min/max principles. Denoting
subspaces of ${\cal H}$ with dimension~$n$ by ${\cal H}_n$, we f\/irst use
\begin{gather}
\nu_n= \max (\min (\| Tf\|: \|f\|=1, f\in {\cal H}_{n+1})).
\end{gather}
Choosing ${\cal H}_{n+1}=P_n{\cal H}$, it follows from~\eqref{mKb} and~\eqref{Tpn} that
\begin{gather}
\label{nulb}
\nu_n\ge e^{-ns}-Ce^{-na}.
\end{gather}
Next, we use
\begin{gather}
\nu_n= \min \big(\max \big(\| Tf\|: \|f\|=1, f\in {\cal H}_n^{\perp}\big)\big).
\end{gather}
Choosing ${\cal H}_n=P_{n-1}{\cal H}$, we deduce from~\eqref{Tpn} that we also have
\begin{gather}
\label{nuub}
\nu_n\le e^{-ns}+Ce^{-na}.
\end{gather}
Combining the lower and upper bounds~\eqref{nulb} and~\eqref{nuub}, we obtain~\eqref{nuIn}.

Finally, from~\eqref{In}--\eqref{nuIn} it is clear that the singular value sequence is in $\ell^1({\mathbb N})$, which
implies~$T$ is trace class.
\end{proof}

We proceed to apply this lemma to our special context, as encoded in Lemma~\ref{lemma7.1}. To this end we set
\begin{gather}
T=\frac{iG(ia-2i\sigma)}{2\pi r_0}{\cal I}(\gamma'),\qquad \nu_n=\frac{iG(ia-2i\sigma)}{2\pi r_0} \lambda_n(\gamma),  \nonumber
\\
f_n^{(')}=f_n(\gamma^{(')}), \qquad p_n^{(')}=\psi_n(\gamma^{(')}), \qquad s=2r\sigma,\qquad a=2r(a_s-\epsilon). \label{Tsubst}
\end{gather}
Then we see that the assumptions of Lemma~\ref{lemma7.2} are satisf\/ied, provided $\sigma<a_s$. (Indeed, the
bound~\eqref{psib} amounts to~\eqref{Test}.) As a~result, we obtain the following theorem.

\begin{Theorem}
\label{theorem7.3}
Let $a_+,a_-\in(0,\infty)$, $g\in\Pi_r$ and $\sigma<a_s$. Fixing $\epsilon\in(0,a_s-\sigma)$, there exist $M\in{\mathbb
N}$ and $C>0$ such that for all $n>M$ the positive numbers~$\lambda_n(\gamma)$ are distinct and satisfy
\begin{gather}
|\lambda_n(\gamma)-\kappa_n(\sigma)|<C\exp(-2nr(a_s-\epsilon)),
\end{gather}
where $\kappa_n(\sigma)$ is defined by~\eqref{kapn}.
\end{Theorem}

For the case $\sigma<a_s$, this theorem implies in particular the validity of the conjecture~\eqref{conj}, but it yields
more detailed information. We continue to partially prove the conjecture~\eqref{psifnconj}. We distinguish two cases
with dif\/ferent assumptions. The f\/irst one applies to the 4-dimensional $\Pi_r$-subsets $\Pi_r^s(j)$, $j=1,2$, on which
${\cal I}(\gamma)$ is positive, cf.~\eqref{selfd}--\eqref{g2}. The pertinent general result now follows.

\begin{Lemma}
\label{lemma7.4}
Assume~$T$ satisfies the assumptions of Lemma~{\rm \ref{lemma7.2}} and in addition
\begin{gather}
\label{extra}
f_n'=f_n,\qquad p_n'=p_n,\qquad \forall\, n\in{\mathbb N}.
\end{gather}
Then we have
\begin{gather}
\label{fpb}
|(f_n,p_n)| =1- O(\exp(-2n(a-s))),\qquad n\to\infty.
\end{gather}
\end{Lemma}

\begin{proof}
On the one hand we have an upper bound
\begin{gather}
\| (T-\nu_n)p_n\|\le \| Tp_n-e^{-ns}p_n\| +|e^{-ns}-\nu_n| =O\big(e^{-na}\big),
\end{gather}
where we used~\eqref{Test},~\eqref{extra} and~\eqref{nuIn}. On the other hand, from~\eqref{extra}
and~\eqref{Ind},~\eqref{nuIn} we get a~lower bound
\begin{gather}
\| (T-\nu_n)p_n\|^2=\| \sum\limits_{m=0}^{\infty}(\nu_m-\nu_n) (f_m,p_n)f_m
\|^2=\sum\limits_{m=0}^{\infty}(\nu_m-\nu_n)^2 |(f_m,p_n)|^2
\\ 
\hphantom{\| (T-\nu_n)p_n\|^2}{}
\ge \min_{m\ne n} (\nu_m-\nu_n)^2 \sum\limits_{m\ne n}|(f_m,p_n)|^2\ge Ce^{-2ns}\big(1-|(f_n,p_n)|^2\big),\qquad \forall\,
n>M.\nonumber
\end{gather}
Combining these bounds, we readily deduce~\eqref{fpb}.
\end{proof}

Substituting once more~\eqref{Tsubst}, we can use this lemma to prove the following theorem.

\begin{Theorem}
\label{theorem7.5}
Let $a_+,a_-\in(0,\infty)$, $g\in\Pi_r^s(j)$, $j=1,2$, and $\sigma<a_s$. Fixing $\rho\in (0,a_s-\sigma)$, there exists
a~sequence of signs $s_n\in \{+,-\}$ such that
\begin{gather}
\label{fpsidif}
\| f_n(\gamma)-s_n\psi_n(\gamma)\| =O(\exp (-2nr\rho)),\qquad n\to\infty.
\end{gather}
\end{Theorem}

\begin{proof}
Since the extra~$g$-assumption implies that~$\gamma$ and $\gamma'$ are related by a~permutation, we deduce not only
$f_n(\gamma')=f_n(\gamma)$, but also $\psi_n(\gamma')=\psi_n(\gamma)$, by symmetry of $w_P(\gamma;x)$~\eqref{wP}.
Therefore, with~\eqref{Tsubst} in ef\/fect, Lemma~\ref{lemma7.4} applies, yielding
\begin{gather}
\label{fpsisp}
|(f_n(\gamma),\psi_n(\gamma))|=1-O(\exp(-4nr\rho)),\qquad n\to\infty.
\end{gather}

We now claim that the inner products $(f_n(\gamma),\psi_n(\gamma))$ are real-valued. Taking this for granted, we can use
\begin{gather}
(f_n(\gamma)-s\psi_n(\gamma),f_n(\gamma)-s\psi_n(\gamma))=2- 2s(f_n(\gamma),\psi_n(\gamma)),\qquad s=+,-,
\end{gather}
and~\eqref{fpsisp} to infer~\eqref{fpsidif}. Thus it remains to prove the reality claim.

To this end we recall that we have
\begin{gather}
f_n(\gamma;x)=e_n(\gamma;x)/c(\gamma;x)w(\gamma;x)^{1/2},\qquad x\in(0,\pi/2r),
\end{gather}
cf.~\eqref{Fn}. Now we have normalized the phase of $e_n(\gamma;x)$ by requiring that it be real-valued and (to f\/ix the
remaining sign) that it be positive for~$x$ near 0, cf.~\eqref{signen}. Since $w(\gamma;x)$ and the quotient
$c(\gamma;x)/c_P(\gamma;x)$ are positive on $(0,\pi/2r)$ (in view of~\eqref{ccid}), and the polyno\-mials~$P_n(\gamma;x)$
are real-valued on this interval, we need only recall the def\/inition~\eqref{psin} to deduce that the inner products are
real.
\end{proof}

Quite likely, we need $s_n=+$ in~\eqref{fpsidif}, but we are not able to prove this beyond doubt.

Our f\/inal theorem concerns again the general case $g\in\Pi_r$, but now we need to require $\sigma <a_s/2$, since we only
have the following lemma available to handle the case where~$T$ is not self-adjoint.

\begin{Lemma}
\label{lemma7.6}
Assume~$T$ satisfies the assumptions of Lemma~{\rm \ref{lemma7.2}} and in addition
\begin{gather}
2s<a,
\end{gather}
and
\begin{gather}
\label{Tstar}
\|T^*p_n'-e^{-ns}p_n\|\le C e^{-na}, \qquad \forall\, n>N.
\end{gather}
Then we have
\begin{gather}
\label{fpb2}
|(f_n,p_n)| =1- O(\exp(-2n(a-2s))),\qquad n\to\infty.
\end{gather}
\end{Lemma}

\begin{proof}
Using the bounds~\eqref{Test} and~\eqref{Tstar} we have
\begin{gather}
\| T^*Tp_n-e^{-2ns}p_n\|\le \| T^*(Tp_n-e^{-ns}p_n')\| + \| e^{-ns}(T^*p_n'-e^{-ns}p_n)\|
\nonumber\\
\hphantom{\| T^*Tp_n-e^{-2ns}p_n\|}{}
=O\big(e^{-na}\big)+O\big(e^{-n(s+a)}\big)=O\big(e^{-na}\big).
\end{gather}
We can now invoke Lemma~\ref{lemma7.4} with $T^*T$ and $2s$ in the role of~$T$ and~$s$ to deduce~\eqref{fpb2}
from~\eqref{fpb}.
\end{proof}

From this lemma the following theorem will be clear by now.

\begin{Theorem}
\label{theorem7.7}
Let $a_+,a_-\in(0,\infty)$, $g\in\Pi_r$ and $2\sigma<a_s$. Fixing $\rho\in (0,a_s-2\sigma)$, there exists a~sequence of
signs $s_n\in \{+,-\}$ such that
\begin{gather}
\| f_n(\gamma)-s_n\psi_n(\gamma)\| =O(\exp (-2nr\rho)),\qquad n\to\infty.
\end{gather}
\end{Theorem}

We now turn to the conjectures~\eqref{conjs},~\eqref{conjl}. To begin, we can test them for the `free'
$\dot{\gamma}$-cluster, and this f\/irst test is clearly passed, cf.~\eqref{Enf}.

Next, consider the functions $a_n(\gamma;x)$. Viewed as vectors $a_n(\gamma)\in{\cal H}$, they approximate the unit
vectors $\psi_n(\gamma)$ for $n\to\infty$, cf.~\eqref{psia}. (In particular, this entails $\|a_n\|\to 1$ for
$n\to\infty$.) In turn, the vectors $\psi_n(\gamma)$ approximate the ${\cal T}(\gamma)$-eigenvectors $f_n(\gamma)$ for
$n\to\infty$. More precisely, we have proved this under further restrictions on~$\sigma$ in Theorems~\ref{theorem7.5} and~\ref{theorem7.7}.

Now for generic~$\gamma$ the vectors $a_n(\gamma)$ belong to the domain $D_t(\gamma)$~\eqref{Dtgam} for $t\le a_s/2$,
but not for $t>a_s/2$, cf.~\eqref{cPpo2},~\eqref{cPpo3}. Therefore, they are not likely to belong to the domains of the
self-adjoint Hilbert space operators $\hat{{\cal A}}_{\pm}(\gamma)$. Even so, we may study the action of the
\emph{analytic difference operators} ${\cal A}_{\pm}(\gamma;x)$ on the functions~$a_n(\gamma;x)$. Writing the latter as
\begin{gather}
a_n(\gamma;x)= \sqrt{\frac{r}{\pi}} \left(e^{2inrx}+\frac{c_P(\gamma;-x)}{c_P(\gamma;x)}e^{-2inrx}\right),
\end{gather}
and the former as (cf.~\eqref{cApmform} and~\eqref{Vaalt})
\begin{gather}
{\cal A}_{\delta}(\gamma;x)=\exp(-ia_{-\delta}\partial_x)\nonumber\\
\hphantom{{\cal A}_{\delta}(\gamma;x)=}{}+
\frac{c_P(\gamma;-x)}{c_P(\gamma;x)}\exp(ia_{-\delta}\partial_x)\frac{c_P(\gamma;x)}{c_P(\gamma;-x)}+V_{b,\delta}(\gamma;x),\qquad \delta=+,-,
\end{gather}
we readily calculate
\begin{gather}
{\cal A}_{\delta}(\gamma;x)a_n(\gamma;x)=\big(\exp(2nra_{-\delta}) +V_{b,\delta}(\gamma;x)\big)a_n(\gamma;x)+
\exp(-2nra_{-\delta})\rho_n(\gamma;x),
\end{gather}
where the remainder function reads
\begin{gather}
\rho_n(\gamma;x):=
\sqrt{\frac{r}{\pi}}\frac{1}{c_P(\gamma;x)}\left(\frac{c_P(\gamma;-x)c_P(\gamma;x+ia_{-\delta})}{c_P(\gamma;-x-ia_{-\delta})}\exp(2inrx)
+(x\to -x)\right).
\end{gather}
For generic~$\gamma$, the even function in brackets has double poles for $x\equiv 0 \pmod{\pi/2r}$, whereas the factor
$c_P(\gamma;x)^{-1}$ yields simple zeros for $x\equiv 0 \pmod{\pi/2r}$, cf.~\eqref{cPpo1}--\eqref{cPpo3}. Thus
$\rho_n(\gamma;x)$ does not belong to ${\cal H}$. But obviously we do have
\begin{gather}
{\cal A}_{\delta}(\gamma;x)a_n(\gamma;x)= \exp(2nra_{-\delta}) a_n(\gamma;x)+O(1),\qquad n\to \infty, \qquad x\in(0,\pi/2r).
\end{gather}
This is the second piece of evidence we have on of\/fer.

The third test comes from the relativistic Lam\'e case~\eqref{gLame}. The connection to our previous work on this
context can be readily made by using~\eqref{clam}. The point is that the A$\Delta$Os ${\cal A}_{\pm}(\gamma_{\rm
L}(b);x)$ amount to the squares of the relativistic Lam\'e A$\Delta$Os occurring in~\cite{Rui03} (up to an additive
constant and slightly dif\/ferent conventions). This is nearly immediate for the free case $b=a_s$, and from this
perspective the f\/irst test is a~special case of the third one. More generally, the Hilbert space versions of the Lam\'e
A$\Delta$Os def\/ined in~\cite{Rui03} have large-eigenvalue asymptotics $\exp(nra_{-\delta})$, so it remains to show that
their squares do yield the operators $\hat{{\cal A}}_{\pm}(\gamma_{\rm L}(b))$ up to an additive constant. However, this
is beyond our present scope, as well as a~more precise description of the relation.

The fourth test pertains only to the conjecture~\eqref{conjs}. This is because it involves the extra requirement
$\sigma> a_s$, and we cannot require that~$\sigma$ be greater than $a_l$. (Indeed, this would violate our standing
assumption $\sigma<a$.) The test arises upon combining~\eqref{conjs} with our conjecture~\eqref{conj}, which we can only
prove for $\sigma<a_s$, cf.~Theorem~\ref{theorem7.3}. The point is that for $\sigma>a_s$ we deduce from~\eqref{conjs}
and~\eqref{conj} that the product $ E_{n,s}(\gamma)\lambda_n(\gamma)$ should vanish exponentially for $n\to\infty$, so
that the operator $\hat{{\cal A}}_s(\gamma){\cal I}(\gamma)$ should be HS. It is not hard to verify that this is indeed
the case, the crux being that we can take the shifts by $\pm ia_s$ under the integral for $a_s<\sigma$.

The reader who remains sceptic about the conjectures~\eqref{conjs},~\eqref{conjl} after looking at this evidence is in
the company of the author. In fact, thus far we have not even presented a~complete proof that the operators $\hat{{\cal
A}}_{\pm}(\gamma)$ are unbounded.

This f\/law, however, can be remedied for the case of $\hat{{\cal A}}_s(\gamma)$. Indeed, consider the functions
\begin{gather}
\label{tn}
t_n(\gamma;x):= \frac{1}{c_P(\gamma;x)}\big(e^{2inrx}+e^{-2inrx}\big),\qquad g=\operatorname{Re} \gamma\in\Pi_r.
\end{gather}
Clearly, they satisfy
\begin{gather}
\lim_{n\to \infty}(t_n(\gamma),t_n(\gamma))= 2\int_0^{\pi/2r}w_P(\gamma;x)dx.
\end{gather}
Moreover, they belong to $D_t(\gamma)$~\eqref{Dtgam} for all $t>0$. Thus they are in the def\/inition domain of~$\hat{{\cal A}}_s(\gamma)$. Now we calculate
\begin{gather}
{\cal A}_s(\gamma;x)t_n(\gamma;x)= e^{2nra_s}d_n(\gamma;x) +V_{b,s}(\gamma;x)t_n(\gamma;x) +e^{-2nra_s}r_n(\gamma;x),
\end{gather}
where $d_n$ is the dominant function
\begin{gather}
d_n(\gamma;x):=\frac{1}{c_P(\gamma;x)}\left(\frac{c_P(\gamma;x)}{c_P(\gamma;x-ia_s)}e^{2inrx}+(x\to -x)\right),
\end{gather}
and $r_n$ the remainder function
\begin{gather}
r_n(\gamma;x):=\frac{1}{c_P(\gamma;x)}\left(\frac{c_P(\gamma;-x)}{c_P(\gamma;-x-ia_s)}e^{2inrx}+(x\to -x)\right).
\end{gather}
As is easily checked, these functions satisfy
\begin{gather}
\lim_{n\to \infty}(q_n(\gamma),q_n(\gamma))=2\int_0^{\pi/2r}\frac{1}{c_P(\gamma;\pm x-ia_s)}dx,\qquad q=d,r.
\end{gather}
Putting the pieces together, it is clear that $\|\hat{{\cal A}}_s(\gamma)t_n(\gamma)\|$ diverges as $n\to\infty$,
whereas $\|t_n(\gamma)\|$ remains f\/inite. Thus $\hat{{\cal A}}_s(\gamma)$ is unbounded.

This argument applies with obvious changes to the `wrong' operator $\hat{{\cal A}}_l^w(\gamma)$. But it does \emph{not}
apply to~$\hat{{\cal A}}_l(\gamma)$. This is because for generic~$\gamma$ the vectors~$t_n(\gamma)$ are not in the
domain of~$\hat{{\cal A}}_l(\gamma)$. We proceed to demonstrate this by an argument that illuminates the importance of
the identities~\eqref{Hnident} for the residue cancellations in the proofs of Lemmas~\ref{lemma5.1} and~\ref{lemma5.2}.

We can transform the vectors $t_n(\gamma)$~\eqref{tn} in the Hilbert space ${\cal H}$~\eqref{cH} to vectors in the
Hilbert space ${\cal H}_{w_H}$~\eqref{cHwH} by using the identity
\begin{gather}
P(\gamma;x)c(\gamma;x)/c_P(\gamma;x)=E(\pm 2x -ia),
\end{gather}
cf.~\eqref{Pgam} and~\eqref{ccid}. This yields~$\gamma$-independent vectors $v_n\in{\cal H}_{w_H}$ given~by
\begin{gather}
\label{vn}
v_n(x)=E(\pm 2x -ia)\big(e^{2inrx}+e^{-2inrx}\big).
\end{gather}
Now consider~\eqref{ressumsp} with $v_n$ in the role of $h_1$. From~\eqref{vn} we read of\/f
\begin{gather}
v_n(x_{\tau}+i\ell a_s/2)=0,\qquad v_n(x_{\tau}-i\ell a_s/2)\ne 0,
\end{gather}
so for these functions we do not get residue cancellation in~\eqref{ressumsp} (assuming generic~$\gamma$, of course).
Consequently, $v_n$ is not in the domain of $\hat{A}_l^H(\gamma)$.

\appendix


\section[The functions~$G$, $R_{\pm}  $, $s_{\pm}$,~$E$ and $G_t$]{The functions~$\boldsymbol{G}$, $\boldsymbol{R_{\pm}}$, $\boldsymbol{s_{\pm}}$,~$\boldsymbol{E}$ and $\boldsymbol{G_t}$}\label{appendixA}

We begin this appendix by collecting properties of the elliptic gamma function. It can be def\/ined by the product
representation
\begin{gather}
\label{Gell}
G(r,a_+,a_-;z) \equiv \prod\limits_{m,n=0}^\infty \frac{1-\exp \big(-(2m+1)ra_+-(2n+1)ra_--2irz\big)}{1-\exp
\big(-(2m+1)ra_+-(2n+1)ra_-+2irz\big)}.
\end{gather}
Just as in the main text, we often suppress the dependence on the parameters and require that they be positive
(cf.~\eqref{parp}). Clearly, $G(z)$ is meromorphic for $z\in{\mathbb C}$, and holomorphic and nonzero for~$z$ in the
strip
\begin{gather}
\label{strip}
|\operatorname{Im} z|<a,\qquad a\equiv (a_{+}+a_{-})/2.
\end{gather}
Its poles and zeros are given~by
\begin{gather}
p_{klm}= -ia-ika_+-ila_-+m\pi/r,\qquad k,l\in{\mathbb N},\qquad m\in{\mathbb Z},
\end{gather}
\begin{gather}
\label{ze}
z_{klm}= ia+ika_++ila_-+m\pi/r,\qquad k,l\in{\mathbb N},\qquad m\in{\mathbb Z},
\end{gather}
where~${\mathbb N}$ denotes the nonnegative integers. For~$z$ in the strip~\eqref{strip} we have an alternative
representation
\begin{gather}
\label{Gg}
G(z)=\exp(ig(z)),\qquad |\operatorname{Im} z|<a,
\end{gather}
where
\begin{gather}
\label{g}
g(z)\equiv\sum\limits_{n=1}^{\infty}\frac{\sin(2nrz)}{2n\sinh(nra_{+})\sinh(nra_{-})},\qquad |\operatorname{Im} z|<a.
\end{gather}
From this (and also from~\eqref{Gell}), the following properties are immediate:
\begin{gather}
\label{refl}
G(-z) = 1/G(z),\qquad ({\rm ref\/lection\ equation}),
\\
G(z+\pi/r)=G(z),\qquad ({\rm periodicity}),
\\
\label{minv}
G(a_-,a_+;z) = G(a_+,a_-;z),\qquad ({\rm modular\ invariance}).
\end{gather}

The elliptic gamma function can be viewed as a~minimal solution to an analytic dif\/ference equation (A$\Delta$E) that
involves a~right-hand side function def\/ined~by
\begin{gather}
\label{defR}
R(r,\alpha;z) =\prod\limits_{k=1}^{\infty}[1-\exp(2irz-(2k-1)\alpha r)][1-\exp(-2irz-(2k-1)\alpha r)].
\end{gather}
Indeed, letting
\begin{gather}
\label{Rdel}
R_\delta(z)= R(r,a_\delta;z),\qquad \delta=+,-,
\end{gather}
it satisf\/ies the two A$\Delta$Es
\begin{gather}
\label{Gades}
\frac{G(z+ia_\delta/2)}{G(z-ia_\delta/2)} = R_{-\delta}(z),
\qquad
\delta=+,-,
\end{gather}
with the modular symmetry feature~\eqref{minv} entailing that only one of the two needs to be verif\/ied. It easily
follows from~\eqref{defR} and~\eqref{Rdel} that the functions $R_+$ and~$R_-$ are entire, even and $\pi/r$-periodic, and
$R_{\delta}(z)$ satisf\/ies
\begin{gather}
\label{Rade}
\frac{f(z+ia_\delta/2)}{f(z-ia_\delta/2)} = -\exp(-2irz).
\end{gather}
They have representations
\begin{gather}
\label{Rrep}
R_{\delta}(z)=\exp \left(-\sum\limits_{n=1}^{\infty}\frac{\cos 2nrz}{n\sinh nra_{\delta}}\right),\qquad |\operatorname{Im}
z|<a_{\delta}/2,\qquad \delta=+,-,
\end{gather}
as is readily checked from~\eqref{Gg},~\eqref{g} and~\eqref{Gades}. The latter~$G$-A$\Delta$Es also entail the
identities
\begin{gather}
\label{RR}
\frac{G(z+ia)}{G(z-ia)}=R_+(z+ia_+/2)R_-(z-ia_-/2)=R_+(z-ia_+/2)R_-(z+ia_-/2).
\end{gather}
Another identity we need is the~$G$-duplication formula
\begin{gather}
G(r,a_{+},a_{-};2z)   =   \prod\limits_{l,m =+,-}G(r,a_{+},a_{-};z-i(la_{+}+ma_{-})/4)
\nonumber
\\
\hphantom{G(r,a_{+},a_{-};2z)   =}{}    \times G(r,a_{+},a_{-};z-i(la_{+}+ma_{-})/4-\pi /2r).\label{Gdup}
\end{gather}
It implies the duplication formulas
\begin{gather}
R_{\delta}(2z)=\prod\limits_{\tau=+,-}R_{\delta}(z-\tau ia_{\delta}/4)R_{\delta}(z-\tau ia_{\delta}/4-\pi/2r),\qquad
\delta=+,-.
\end{gather}

Occasionally, it is more revealing to use the functions
\begin{gather}
\label{sR}
s_{\delta}(z)\equiv s(r,a_{\delta};z)\equiv -ie^{irz}R_{\delta}(z+ia_{\delta}/2)/p_{\delta},\qquad \delta=+,-,
\end{gather}
where
\begin{gather}
\label{pde}
p_{\delta}\equiv p(r,a_{\delta})\equiv 2r\prod\limits_{k=1}^{\infty}(1-e^{-2kra_{\delta}})^2.
\end{gather}
It is easily verif\/ied that the function~$s_{\delta}(z)$ is entire, odd and $\pi/r$-antiperiodic, and that it also
satisf\/ies the A$\Delta$E~\eqref{Rade}. Its relation to the Weierstrass~$\sigma$-function is given~by
\begin{gather}
s_{\delta}(z)= \exp\big({-}\eta(\pi/2r,ia_{\delta}/2)\, z^2r/\pi\big)\sigma(z;\pi/2r,ia_{\delta}/2),
\end{gather}
where we use the notation of Whittaker/Watson~\cite{WW73}.

In the main text we often work with alternative parameters
\begin{gather}
a_s\equiv \min (a_+,a_-),\qquad a_l\equiv \max(a_+,a_-),
\end{gather}
and corresponding functions
\begin{gather}
\label{Rsl}
R_s(z)\equiv R(r,a_s;z),\qquad R_l(z)\equiv R(r,a_l;z).
\end{gather}
For the case $a_s<a_l$ there is a~unique~$L\in{\mathbb N}^*$ such that
\begin{gather}
La_s<a_l,\qquad (L+1)a_s\ge a_l,\qquad L\in{\mathbb N}^*.
\end{gather}
Denoting the residues of $G(z)$ at the simple poles~$-ia-ika_s$, $k=0,\ldots, L$, by~$r_k$, we obtain ratios
\begin{gather}
\label{rqu}
\frac{r_0}{r_k}=\lim_{z\to 0}\frac{G(z-ia)}{G(z-ia-ika_s)}= \prod\limits_{j=1}^k R_l(ia_l/2+ija_s),\qquad k=1,\ldots,L,
\end{gather}
that are used in Section~\ref{section2}. Moreover, in Section~\ref{section7} we need the explicit formulas
\begin{gather}
\label{r0}
r_0=i\Big/2r\prod\limits_{k=1}^{\infty}(1-\exp(-2kra_+))(1-\exp(-2kra_-)),
\\
\label{Gspec}
G((ia_{\delta}-ia_{-\delta})/2)=\prod\limits_{k=1}^{\infty}\frac{1-\exp(-2kra_{-\delta})}{1-\exp(-2kra_{\delta})},\qquad
\delta=+,-.
\end{gather}

All of the above material can be found in Subsection~III~B of~\cite{Rui97}, and some of it also in quite a~few later
papers (mostly with dif\/ferent conventions). In the present work it will be crucial to make use of the function
\begin{gather}
\label{E}
E(r,a_+,a_-;z)\equiv	 \prod\limits_{m,n=0}^\infty \big(1-\exp \big({-}(2m+1)ra_+-(2n+1)ra_--2irz\big)\big).
\end{gather}
It seems it has not been studied or used before, but of course it is closely related to the elliptic gamma function,
since we have
\begin{gather}
\label{GE}
G(z)=E(z)/E(-z).
\end{gather}
It is obviously entire in~$z$ with zeros given by~\eqref{ze}. It is not dif\/f\/icult to see that it admits an alternative
representation
\begin{gather}
\label{Ee}
E(z)=\exp(e(z)),\qquad \operatorname{Im} z<a,
\end{gather}
where
\begin{gather}
\label{erep}
e(z)\equiv\sum\limits_{n=1}^{\infty}\frac{-\exp(-2inrz)}{4n\sinh(nra_{+})\sinh(nra_{-})},\qquad \operatorname{Im} z<a.
\end{gather}
Clearly, it is also $\pi/r$-periodic and modular invariant. Moreover, from~\eqref{Ee} and~\eqref{erep} we deduce the
duplication formula
\begin{gather}
E(r,a_{+},a_{-};2z)   =   \prod\limits_{l,m =+,-}E(r,a_{+},a_{-};z-i(la_{+}+ma_{-})/4)
\nonumber
\\
 \hphantom{E(r,a_{+},a_{-};2z)   =}{}   \times E(r,a_{+},a_{-};z-i(la_{+}+ma_{-})/4-\pi /2r).\label{Edup}
\end{gather}

Finally, the~$E$-function satisf\/ies two A$\Delta$Es that involve the trigonometric gamma function
\begin{gather}
\label{Gt}
G_t(r,\alpha;z)\equiv \prod\limits_{n=0}^{\infty} (1-q^{2n+1}\exp(2irz))^{-1},\qquad q\equiv \exp(-\alpha r),
\end{gather}
cf.~Subsection~III~C in~\cite{Rui97}. Indeed, it is a~minimal solution to both of the A$\Delta$Es
\begin{gather}
\label{Eades}
\frac{E(z+ia_\delta/2)}{E(z-ia_\delta/2)} =\frac{1}{G_t(r,a_{-\delta};-z)},
\qquad
\delta=+,-,
\end{gather}
which arises by iteration. Likewise, $G_t(r,\alpha;z)$ is the iteration solution to
\begin{gather}
\label{Gtade}
\frac{G_t(z+i\alpha/2)}{G_t(z-i\alpha/2)}=1-\exp(2irz).
\end{gather}
We also note the relations{\samepage
\begin{gather}
\label{GtR}
G_t(r,a_{\delta};z)G_t(r,a_{\delta};-z)=1/R_{\delta}(z),\qquad \delta=+,-,
\end{gather}
which can be used to derive the~$G$-A$\Delta$Es~\eqref{Gades} from the~$E$-A$\Delta$Es~\eqref{Eades}.}

We conclude this appendix by listing the conjugate functions of the above meromorphic functions. (The conjugate is
def\/ined~by
\begin{gather}
\label{Mconj}
M^{*}(w)\equiv \overline{M(\overline{w})},\qquad w\in{\mathbb C}^N,
\end{gather}
where $M(w)$ is meromorphic.) In view of our standing assumption~\eqref{parp} we clearly have
\begin{gather}
\label{cons1}
F^*(z)=F(-z),\qquad F=G,\ E,\ G_t,
\\
\label{cons2}
\theta^*(z)=\theta(z),\qquad \theta=R_{\pm},\ s_{\pm}.
\end{gather}



\section{Proof of Lemma~\ref{lemma4.2}}\label{appendixB}

By virtue of~\eqref{cC} it suf\/f\/ices to show that we have
\begin{gather}
\label{aim}
{\cal A}_{s}(\gamma';x)\int_{-\pi/2r}^{\pi/2r} {\cal K}(\gamma';x,y)f_n(\gamma;y)dy= \int_{-\pi/2r}^{\pi/2r} {\cal
K}(\gamma';x,y){\cal A}_{s}(\gamma;y)f_n(\gamma;y)dy,
\end{gather}
for $x\in(0,\pi/2r)$. In order to do so, we distinguish three cases, namely, (i) $a_s<\sigma$, (ii) $a_s>\sigma$, and
(iii) $a_s=\sigma$. In case (i), we are allowed to interchange the shifts~$x\to x\pm ia_s$ and the integration on the
left-hand side of~\eqref{aim}, since no poles are met. Then we can use the kernel identity~\eqref{cAid} to conclude
that~\eqref{aim} is equivalent to
\begin{gather}
\label{i}
\int_{-\pi/2r}^{\pi/2r} f_n(\gamma;y){\cal A}_s(\gamma;-y){\cal K}(\gamma';x,y)dy= \int_{-\pi/2r}^{\pi/2r} {\cal
K}(\gamma';x,y){\cal A}_{s}(\gamma;y)f_n(\gamma;y)dy.
\end{gather}
To prove this equality, we choose~$x$ equal to some positive number~$p\in(0,\pi/2r)$ and consider the conjugate function
\begin{gather}
\label{cKst}
{\cal K}^*(\gamma';p,y)=\frac{{\cal S}(\sigma;p,y)}{c(\gamma';-p)c(\gamma;y)},
\end{gather}
cf.~\eqref{Mconj} and~\eqref{cons1}. We assert that as a~function of~$y$, it belongs to~$D_{\sigma}(\gamma)$. Indeed,
this assertion amounts to the claim
\begin{gather}
\label{Ssig}
\frac{c_P(\gamma;y)}{c(\gamma;y)}{\cal S}(\sigma;p,y)\in S_{\sigma},
\end{gather}
cf.~\eqref{Dtgam}. To see that~\eqref{Ssig} holds true for~$a_s<\sigma$, we need only recall the identity~\eqref{ccid}.

The point is now that we can reinterpret~\eqref{i} as an equality of inner products, viz.,
\begin{gather}
\label{ipeq}
\big(\hat{{\cal A}_s}(\gamma){\cal K}^*(\gamma';p,\cdot),f_n\big)= \big({\cal K}^*(\gamma';p,\cdot),\hat{{\cal A}_s}(\gamma)f_n\big).
\end{gather}
Since~$\hat{{\cal A}_s}(\gamma)$ is symmetric on~$D_\sigma(\gamma)$ by Theorem~\ref{theorem3.1}, this equality does
hold.

Turning to case (ii), our previous reasoning fails on two distinct counts. First, for~$a_s>\sigma$ we cannot take the
shift~$x\to x+i\delta a_s$ under the integral sign, since the two poles of~${\cal S}(\sigma;x,y)$ at $y=\pm (x
-i\delta\sigma)$ pass the contour. Second, even after handling residue terms, we can no longer appeal to~\eqref{ipeq},
since its left-hand side is ill def\/ined for~$a_s>\sigma$. Indeed, the four~$x$-dependent poles of~${\cal S}(\sigma;x,y)$
at distance~$\sigma$ from the contour prevent the function~\eqref{cKst} from belonging to~$D_t(\gamma)$ for some~$t>
a_s$.

In order to bypass these snags, we invoke our previous results concerning analytic continuation of the function
\begin{gather}
g_n(\gamma;x)=\lambda_n F_n(\gamma';x),
\end{gather}
cf.~\eqref{gnF},~\eqref{FnF} and Lemma~\ref{lemma2.4}. Also, in order to appeal to evenness properties, it is expedient
to switch from~${\cal A}_s$ to~$A_s$, cf.~\eqref{cApm}. Upon multiplying~\eqref{cAcI} by~$c(\gamma';x)$, we see that we
should show that the function
\begin{gather}
\label{iiL}
2A_s(\gamma';x)g_n(\gamma;x)=A_s(\gamma';x)\int_{-\pi/2r}^{\pi/2r} {\cal S}(\sigma;x,y)w(\gamma;y)F_n(\gamma;y)dy,
\end{gather}
evaluated for $x=p\in(0,\pi/2r)$, equals
\begin{gather}
\label{iiR}
\int_{-\pi/2r}^{\pi/2r} {\cal S}(\sigma;p,y)w(\gamma;y)A_s(\gamma;y)F_n(\gamma;y)dy,\qquad p\in(0,\pi/2r).
\end{gather}

To this end we invoke the representation~\eqref{gns2} for the analytic continuation of~$g_n(\gamma;x)$ to~$\operatorname{Im}
x\in (\sigma, \sigma+a_s)$. A~priori, this only yields an expression for the value~$g_n(\gamma;p+ia_s)$, but the
value~$g_n(\gamma;p-ia_s)$ equals~$g_n(\gamma;-p+ia_s)$ by evenness. Hence, using also evennness of ${\cal
S}(\sigma;x,y)$ in~$x$, we obtain for~\eqref{iiL} evaluated at~$x=p$ the following result:
\begin{gather}
    R_n(p)+\sum\limits_{\tau=+,-}V_s(\gamma';\tau p) \int_{-\pi/2r}^{\pi/2r} {\cal S}(\sigma;p-i\tau
a_s,y)w(\gamma;y)F_n(\gamma;y)dy
\nonumber
\\
\hphantom{R_n(p)}{}    +V_{b,s}(\gamma';p)\int_{-\pi/2r}^{\pi/2r} {\cal S}(\sigma;p,y)w(\gamma;y)F_n(\gamma;y)dy.\label{RV}
\end{gather}
Here, $R_n(p)$ is the residue sum (cf.~\eqref{res1})
\begin{gather}
\label{Rn}
R_n(p)=4\sum\limits_{\tau=+,-}V_s(\gamma';\tau p)\rho_n^{(1)}(\gamma';-\tau p+ia_s).
\end{gather}

We are now in the position to invoke the kernel identity~\eqref{kid}. It enables us to rewrite~\eqref{RV} as
\begin{gather}
R_n(p)+\int_{-\pi/2r}^{\pi/2r} w(\gamma;y)F_n(\gamma;y)\left(\sum\limits_{\tau=+,-}V_s(\gamma;\tau y) {\cal
S}(\sigma;p,y-i\tau a_s) +V_{b,s}(\gamma;y){\cal S}(\sigma;p,y)\right)dy
\nonumber\\
\hphantom{R_n(p)}{}
=R_n(p)+\int_{-\pi/2r}^{\pi/2r} w(\gamma;y)F_n(\gamma;y)\big(2 V_s(\gamma;- y) {\cal S}(\sigma;p,y+i a_s)
\nonumber\\
\hphantom{R_n(p)}{}
+V_{b,s}(\gamma;y){\cal S}(\sigma;p,y)\big)dy,\label{RV2}
\end{gather}
where the second step follows by taking~$y\to -y$ in the $\tau=+$ term and using evenness of~$w$, $F_n$ and~${\cal S}$.
We proceed to study the term
\begin{gather}
\label{t1}
2\int_{-\pi/2r}^{\pi/2r} m_H(\gamma;y)H_n(\gamma;y) V_s(\gamma;-y) {\cal S}(\sigma;p,y+ia_s)dy.
\end{gather}
We used~\eqref{HF}, so as to invoke Lemma~\ref{lemma2.4}. The latter shows that~$H_n(\gamma;y)$ is holomorphic for~$\operatorname{Im}  y\in [-a_s,0]$. It is readily verif\/ied that the meromorphic function~$m_H(\gamma;y)V_s(\gamma;-y)$ is also
pole-free for~$\operatorname{Im} y\in [-a_s,0]$. Therefore, when we shift the contour down by~$a_s$, we only encounter the two
(simple) poles of~${\cal S}(\sigma;p,y+ia_s)$ at $y=\pm p -ia_s +i\sigma$. As a~result,~\eqref{t1} equals
\begin{gather}
\label{t1new}
4\sum\limits_{\tau=+,-}\rho_n^{(2)}(\gamma;\tau p) +2\int_{-\pi/2r}^{\pi/2r} w(\gamma;y-ia_s)F_n(\gamma;y-ia_s)
V_s(\gamma;-y+ia_s) {\cal S}(\sigma;p,y)dy,\!\!\!
\end{gather}
where we have introduced
\begin{gather}
\rho_n^{(2)}(\gamma;x):= i\pi r_0 G(2i\sigma -ia)G(2x-ia)G(-2x+2i\sigma-ia)
\nonumber\\
\hphantom{\rho_n^{(2)}(\gamma;x):=}{}
\times V_s(\gamma;x+ia_s-i\sigma)w(\gamma;x+ia_s-i\sigma)F_n(x+ia_s-i\sigma).\label{res2}
\end{gather}

We now use the identity~\eqref{Vew} to rewrite the product of the~$w$- and $V_s$-term in the integral occurring
in~\eqref{t1new}. The upshot is that~\eqref{RV2} equals
\begin{gather}
4\sum\limits_{\tau =+,-}\big(\rho_n^{(2)}(\gamma;\tau p)+V_s(\gamma';\tau p)\rho_n^{(1)}(\gamma';ia_s-\tau p)\big)
\nonumber\\
\qquad{}+
\int_{-\pi/2r}^{\pi/2r} {\cal S}(\sigma;p,y)w(\gamma;y)A_s(\gamma;y) F_n(\gamma;y)dy,
\end{gather}
where we used~\eqref{Rn}. Comparing this to~\eqref{iiR}, we conclude that the lemma holds true in case~(ii), provided
the residue sum vanishes. Now from~\eqref{res1} and~\eqref{res2} we see that this is the case when we have an identity
\begin{gather}
    G(2x-ia)G(-2x+2i\sigma-ia) V_s(\gamma;x+ia_s-i\sigma)
\nonumber
\\
\qquad{}
    =V_s(\gamma';-x)G(2(x+ia_s)-ia)G(2i\sigma-2(x+ia_s)-ia).\label{VGid}
\end{gather}

Finally, to check~\eqref{VGid}, we f\/irst use the def\/inition~\eqref{Ve} of~$V_s$ and the
relation~$\gamma_{\mu}'=-\gamma_\mu-\sigma$ to establish that the $\gamma_{\mu}$-dependent factors are equal. Thus it
remains to verify
\begin{gather}
    \frac{G(2x-ia)G(-2x+2i\sigma-ia)}{R_l(2x-2i\sigma+2ia_s+ia_l/2)R_l(2x-2i\sigma+ia_s+ia_l/2)}
\nonumber
\\
\qquad{}    =\frac{G(2x+2ia_s-ia)G(-2x-2ia_s+2i\sigma-ia)}{R_l(2x- ia_l/2)R_l(2x+ia_s-ia_l/2)}.
\end{gather}
Using the~$G$-A$\Delta$Es~\eqref{Gades} it is straightforward to do so. Hence the proof of the lemma for case~(ii) is
complete.

It remains to prove~\eqref{aim} for the special case~$a_s=\sigma$. In this case we show that~\eqref{iiL}, evaluated for
\begin{gather}
\label{qp}
x=q\equiv p+im(\gamma)/2,\qquad p\in (0,\pi/2r),
\end{gather}
equals
\begin{gather}
\label{iiiR}
\int_{-\pi/2r}^{\pi/2r} {\cal S}(a_s;q,y)w(\gamma;y)A_s(\gamma;y)F_n(\gamma;y)dy.
\end{gather}
Thanks to the positive imaginary part of~$q$ we can take the down shift under the integral. For the up shift we use once
more~\eqref{gns2}. Then we get as the counterpart of~\eqref{RV} and~\eqref{Rn}
\begin{gather}
    4V_s(\gamma';-q)\rho_n^{(1)}(\gamma';q+ia_s) +V_{b,s}(\gamma';q)\int_{-\pi/2r}^{\pi/2r} {\cal
S}(a_s;q,y)w(\gamma;y)F_n(\gamma;y)dy
\nonumber
\\
 \qquad{}   +\sum\limits_{\tau=+,-}V_s(\gamma';\tau q) \int_{-\pi/2r}^{\pi/2r} {\cal S}(a_s;q-i\tau
a_s,y)w(\gamma;y)F_n(\gamma;y)dy.
\end{gather}

As in case (ii), we can now use the kernel identity~\eqref{kid}. This yields in particular the term~\eqref{t1} with
$p\to q$ and $\sigma\to a_s$. When we shift the contour down by~$a_s$, we now meet only the pole of~${\cal
S}(a_s;q,y+ia_s)$ at~$y=-q$, so the analog of~\eqref{t1new} is
\begin{gather}
4\rho_n^{(2)}(\gamma;q) +2\int_{-\pi/2r}^{\pi/2r} w(\gamma;y-ia_s)F_n(\gamma;y-ia_s) V_s(\gamma;-y+ia_s) {\cal
S}(a_s;q,y)dy.
\end{gather}
Following the reasoning in case (ii), we can once more use the identity~\eqref{VGid} to verify that the residue terms in
the counterpart of~\eqref{RV2} cancel, and so equality to~\eqref{iiiR} follows. By analyticity, this yields equality
for~$x\in(0,\pi/2r)$, so the proof of Lemma~\ref{lemma4.2} is complete.



\section{Proof of Lemma~\ref{lemma5.2}}\label{appendixC}

The proof runs along the same lines as the proof of Lemma~\ref{lemma4.2} in Appendix~\ref{appendixB}. Of course, we always have
$a_l>\sigma$, so there is no analog of case~(i). However, a~greater ef\/fort is needed to verify residue cancellation. The
analog of the case~(ii) assumption~$a_s>\sigma$ amounts to assuming
\begin{gather}
\label{ii}
a_l-\sigma\in (\nu a_s,(\nu+1)a_s),\!\qquad \nu\in\{0,\ldots,L-1\}\!\qquad \mathrm{or}\!\qquad a_l\in (\sigma +La_s,(L+1)a_s], \!\!\!
\end{gather}
as opposed to
\begin{gather}
\label{iii}
a_l=\sigma+\nu a_s,\qquad \nu\in\{1,\ldots,L\},
\end{gather}
which is the counterpart of the case (iii) assumption $a_s=\sigma$.

We proceed to treat case~\eqref{ii}. To show equality of~\eqref{iiL} and~\eqref{iiR} with $A_s$ replaced by~$A_l$, we
can then appeal to~\eqref{HHnu}. We need to evaluate the even function~$F_n(\gamma';x)=H_n(\gamma';x)/P(\gamma';x)$ at
points $x=\pm p +ia_l$ where it is holomorphic, so the counterpart of~\eqref{RV} is
\begin{gather}
    Q_n(p)+\sum\limits_{\tau=+,-}V_l(\gamma';\tau p) \int_{-\pi/2r}^{\pi/2r} {\cal S}(\sigma;p-i\tau
a_l,y)w(\gamma;y)F_n(\gamma;y)dy
\nonumber
\\
\hphantom{Q_n(p)}{}    +V_{b,l}(\gamma';p)\int_{-\pi/2r}^{\pi/2r} {\cal S}(\sigma;p,y)w(\gamma;y)F_n(\gamma;y)dy,\label{RVl}
\end{gather}
with residue sum
\begin{gather}
Q_n(p)=\sum\limits_{\tau=+,-}\frac{V_l(\gamma';\tau p)}{P(\gamma';ia_l-\tau
p)}\sum\limits_{\ell=0}^{\nu}\mu_{\ell}(\gamma;ia_l-\tau p)H_n(\gamma;ia_l-\tau p-i\sigma-i\ell a_s)
\nonumber\\
\hphantom{Q_n(p)}{}
=-4i\pi\sum\limits_{\tau=+,-}V_l(\gamma';\tau p)\sum\limits_{\ell=0}^{\nu}w(\gamma;ia_l-\tau p -i\sigma-i\ell
a_s)F_n(\gamma;\tau p +i\sigma +i\ell a_s-ia_l)
\nonumber\\
\hphantom{Q_n(p)=}{}
\times r_{\ell}G(2i\sigma+i\ell a_s-ia)G(2(ia_l-\tau p)-i\ell a_s-ia)
\nonumber\\
\hphantom{Q_n(p)=}{}
\times
G(-2(ia_l-\tau p -i\sigma)+i\ell a_s -ia),\label{Qndef}
\end{gather}
where we used~\eqref{muell},~\eqref{xiell} in the second step.

Using the kernel identity~\eqref{kid}, we deduce~\eqref{RVl} equals
\begin{gather}
\label{Qnt}
Q_n(p)+\int_{-\pi/2r}^{\pi/2r} w(\gamma;y)F_n(\gamma;y)\big(2 V_l(\gamma;- y) {\cal S}(\sigma;p,y+i a_l)
+V_{b,l}(\gamma;y){\cal S}(\sigma;p,y)\big)dy,
\end{gather}
yielding the analog of~\eqref{RV2}.

Consider now the analog of~\eqref{t1}, i.e., the second term in~\eqref{Qnt}. When we shift the contour down by~$a_l$,
we see that the integral becomes equal to~\eqref{iiR} with~$A_s\to A_l$. Therefore, just as in Appendix~\ref{appendixB} it remains to
show residue cancellation. In this case, however, we meet not only the simple poles of~${\cal S}(\sigma;p,y+ia_l)$ at
\begin{gather}
\label{cSp}
y=\pm p -ia_l+i\sigma +i\ell a_s,\qquad \ell=0,\ldots,\nu,
\end{gather}
but also the poles of the remaining factors for $\operatorname{Im} y=0$ and $\operatorname{Im} y= \pm\pi/2r$. We can ensure that we do
not meet the poles at $\operatorname{Im} y= -\pi/2r$ by f\/irst shifting the contour to the right by some $\epsilon\in(0,\pi/2r
-p)$.

We show f\/irst that the residue sum at the~$p$-independent poles vanishes. To this end we rewrite the relevant term as
\begin{gather}
\label{rewt}
2\int_{-\pi/2r+\epsilon}^{\pi/2r+\epsilon} B(\gamma;y+ia_l/2)H_n(\gamma;y) {\cal S}(\sigma;p,y+ia_l)dy,
\end{gather}
where we have introduced a~new function
\begin{gather}
B(\gamma;y):=m_H(\gamma;y-ia_l/2)V_l(\gamma;-y+ia_l/2).
\end{gather}
We proceed to rewrite this function. First we use~\eqref{mH},~\eqref{w} and~\eqref{Vec} to get
\begin{gather}
B(\gamma;y)=1/P(\gamma;y-ia_l/2)c(\gamma;\pm y- ia_l/2).
\end{gather}
Next we use~\eqref{Pgam},~\eqref{c}, and then the~$G$- and~$E$-A$\Delta$Es~\eqref{Gades} and~\eqref{Eades} to deduce the
representation
\begin{gather}
B(\gamma;y)=\left(\frac{R_l(2y+ia_l/2)}{R_s(2y+ia_s/2)}\prod\limits_{\mu=0}^7\frac{1}{E(\pm
y-ia_l/2-i\gamma_{\mu})}\right)\Big/\prod\limits_{\mu=0}^7G_t(a_s;-y-i\gamma_{\mu}).
\end{gather}
This representation reveals the salient properties of~$B(\gamma;y)$: In the strip~$|\operatorname{Im} y|\le a_l/2$ it has (at
most) simple poles at
\begin{gather}
y\equiv \pm i\ell a_s/2,\qquad \ell=1,\ldots,L \pmod{\pi/2r}.
\end{gather}
Moreover, the function in brackets is even and $\pi/r$-periodic, so the residues at $i\ell a_s/2+\tau\pi/2r$ and $-i\ell
a_s/2+\tau\pi/2r$, $\tau=0,1$, are either both zero (namely, when the~$G_t$-products both have a~pole), or they have
a~quotient
\begin{gather}
q_{\ell,\tau}=-\prod\limits_{\mu=0}^7\frac{G_t(a_s;i\ell a_s/2-\tau\pi/2r-i\gamma_{\mu})}{G_t(a_s;-i\ell
a_s/2-\tau\pi/2r-i\gamma_{\mu})}
\nonumber\\
\hphantom{q_{\ell,\tau}}{}
= -\prod\limits_{\mu=0}^7\prod\limits_{m=1}^{\ell} \big(1-(-)^{\tau}\exp[2r(\gamma_{\mu}+(\ell +1-2m)a_s/2)]\big),\label{resq}
\end{gather}
where we used the $G_t$-A$\Delta$E~\eqref{Gtade}.

As a~consequence, the poles of the factor $B(\gamma;y+ia_l/2)$ in~\eqref{rewt} that are met when the contour is shifted
down by~$a_l$ can only be located at
\begin{gather}
\label{pind2}
y_{\pm\ell,\tau}:=-ia_l/2\pm i\ell a_s/2 +\tau\pi/2r,\qquad \ell=1,\ldots,L,\qquad \tau=0,1,
\end{gather}
and the residues at $y_{\ell,\tau}$ and~$y_{-\ell,\tau}$ either both vanish or have residue quotient~\eqref{resq}. Now
the remaining two functions in the integrand of~\eqref{rewt} are regular at~\eqref{pind2}, and from~\eqref{cruxid} we
deduce
\begin{gather}
\frac{{\cal S}(\sigma;p,ia_l/2+i\ell a_s/2+\tau\pi/2r)}{{\cal S}(\sigma;p,ia_l/2-i\ell a_s/2+\tau\pi/2r)}=\exp(4\ell
r(\sigma -a)).
\end{gather}
Thanks to the $H_n$-identities~\eqref{Hnident}, we can now conclude that the residue sum for the~$p$-independent
poles~\eqref{pind2} vanishes.

Next, we consider the residue sum at the poles~\eqref{cSp}. Reverting to the representation~\eqref{Qnt}, we see that the
residue sum at the~$p$-dependent poles~\eqref{cSp} is given~by
\begin{gather}
4i\pi \sum\limits_{\tau=+,-}\sum\limits_{\ell=0}^{\nu}V_l(\gamma;-\tau p-i\sigma-i\ell a_s+ia_l)w(\gamma;\tau p
+i\sigma+i\ell a_s-ia_l)
\nonumber\\
\qquad{}\times F_n(\gamma;\tau p -ia_l +i\sigma +i\ell a_s)
 r_{\ell}G(2i\sigma+i\ell a_s-ia)
\nonumber\\
\qquad{}\times
 G(2\tau p+2i\sigma+i\ell a_s-ia)G(-2\tau p -i\ell a_s -ia).
\end{gather}
Comparing this to $Q_n(p)$~\eqref{Qndef}, we deduce that the residue terms cancel pairwise provided we have identities
\begin{gather}
    G(2x-ia-i\ell a_s)G(-2x+2i\sigma-ia+i\ell a_s) V_l(\gamma;x+ia_l-i\sigma-i\ell a_s)
\nonumber
\\
 \qquad{}   =V_l(\gamma';-x)G(2x+2ia_l-ia-i\ell a_s)G(-2x-2ia_l+2i\sigma-ia+i\ell a_s),\label{VGidl}
\end{gather}
with $\ell =0,\ldots,L$.

To prove these identities, we begin by noting that the $\ell =0$ identity amounts to the identity~\eqref{VGid}
with~$a_s$ and~$a_l$ interchanged. Thus the same proof shows that it is valid. As a~result we obtain
\begin{gather}
\frac{V_l(\gamma;x+ia_l-i\sigma)}{V_l(\gamma';-x)}=\frac{G(2x+2ia_l-ia)G(-2x-2ia_l+2i\sigma-ia)}{G(2x-ia)G(-2x+2i\sigma-ia)}\nonumber
\\
\hphantom{\frac{V_l(\gamma;x+ia_l-i\sigma)}{V_l(\gamma';-x)}}{} =\frac{R_s(2x+3ia_l/2-ia)R_s(2x+ia_l/2-ia)}{R_s(2x+3ia_l/2-2i\sigma+ia)R_s(2x+ia_l/2-2i\sigma+ia)}.\label{spid}
\end{gather}
For $\ell>0$ the identity~\eqref{VGidl} amounts to
\begin{gather}
\frac{V_l(\gamma;x+ia_l-i\sigma-i\ell a_s)}{V_l(\gamma';-x)}
\nonumber\\
\qquad{}
=\frac{G(2x+2ia_l-ia-i\ell a_s)G(-2x-2ia_l+2i\sigma-ia+i\ell
a_s)}{G(2x-ia-i\ell a_s)G(-2x+2i\sigma-ia+i\ell a_s)}\nonumber
\\
\qquad{}
=\frac{R_s(2x+3ia_l/2-ia-i\ell a_s)R_s(2x+ia_l/2-ia-i\ell a_s)}{R_s(2x+3ia_l/2-2i\sigma+ia-i\ell
a_s)R_s(2x+ia_l/2-2i\sigma+ia-i\ell a_s)}.\label{spidl}
\end{gather}

Now we use the quasi-periodicity relation~\eqref{Vper} to infer that the ratio of the left-hand sides of~\eqref{spid}
and~\eqref{spidl} equals~$\exp(8\ell r(\sigma -a))$. Thus it remains to verify that the ratio of the right-hand sides
yields the same result. Using the $R_s$-A$\Delta$E~\eqref{Rade} this is routine, and so we have now completed the proof
for the case~\eqref{ii}.

Finally, for the case~\eqref{iii} we can invoke the representation~\eqref{HHnusp}, and prove the pertinent residue
cancellations for the choice~\eqref{qp}. Just as in Appendix~\ref{appendixB}, this involves some minor changes in the above formulas
for case~\eqref{ii}, which we skip for brevity.

\subsection*{Acknowledgments}

We would like to thank M.~Halln\"as for his interest and useful comments. We have also benef\/ited from constructive
criticism of the referees, which helped to improve the exposition of the paper.

\LastPageEnding

\end{document}